\pdfoutput=1
\documentclass[11pt,a4paper]{article}

\usepackage{jheppub}
\usepackage{amsmath}
\usepackage{amssymb}
\usepackage{appendix}
\usepackage{multirow}
\usepackage{rotating}
\usepackage{diagbox}
\usepackage{float}
\usepackage{xcolor}
\usepackage{cancel}

\usepackage{color}

\usepackage{color}
\definecolor{nicered}{rgb}{0.7,0.1,0.1}
\definecolor{nicegreen}{rgb}{0.1,0.5,.1}

\usepackage{graphicx}
\usepackage{subcaption}
\graphicspath{{figures/}{./}}

\title{Symmetries in $B\to D^*\ell\nu$ angular observables}

\author[a,b]{Marcel Alguer\'o,}
\author[c]{S\'ebastien Descotes-Genon,}
\author[a,b]{Joaquim Matias,}
\author[c]{and Mart\'in Novoa-Brunet}

\affiliation[a]{Grup de F\'{\i}sica Te\`orica (Departament de F\'{\i}sica),\\
Universitat Aut\`onoma de Barcelona, E-08193 Bellaterra (Barcelona), Spain}
\affiliation[b]{Institut de F\'{\i}sica d'Altes Energies (IFAE),
The Barcelona Institute of Science and Technology,\\ Campus UAB, E-08193 Bellaterra (Barcelona), Spain}
\affiliation[c]{Universit\'e Paris-Saclay, CNRS/IN2P3, IJCLab, 91405 Orsay, France}

\abstract{We apply the formalism of amplitude symmetries to the angular distribution of the decays $B \to D^* \ell \nu$ for $\ell=e,\mu,\tau$. We show that the angular observables used to describe the distribution of this class of decays are not independent in absence of New Physics contributing to tensor operators. We derive sets of relations among the angular coefficients of the decay distribution for the massless and massive lepton cases which can be used to probe in a very general way the consistency among the angular observables and the underlying New Physics at work. We use these relations to access the longitudinal polarisation fraction of the $D^*$ using different angular coefficients from the ones used by Belle experiment. This in the near future can provide an alternative strategy to measure $F_L^{D^*}$ in $B \to D^* \tau \nu$ and to understand the relatively high value measured by the Belle experiment. Using the same symmetries, we identify three observables which may exhibit a tension if the experimental value of $F_L^{D^*}$ remains high.
We discuss how these relations can be exploited for binned measurements. We also propose a new observable that could test for specific scenarios of New Physics  generated by light right-handed neutrinos.  Finally we study the prospects of testing these relations based on the projected experimental sensitivity of new experiments.}

\emailAdd{malguero@ifae.es}
\emailAdd{sebastien.descotes-genon@ijclab.in2p3.fr}
\emailAdd{matias@ifae.es}
\emailAdd{martin.novoa@ijclab.in2p3.fr}

\arxivnumber{2003.02533}

\begin{document}

\maketitle

\section{Motivation}

Over the last six years, the hints of a tension with respect to Standard Model (SM) expectations have been growing concerning two different classes of $b$-quark decays, generically described as $b$-anomalies.

On the one hand, the interest of neutral-current $b\to s\mu\mu$ transitions was highlighted by the measurement of $B\to K^*\mu\mu$ angular observables, and in particular the observable called $P_5^\prime$~\cite{Descotes-Genon:2013vna} exhibiting discrepancies with respect to the SM at the level of $3.7\sigma$~\cite{Aaij:2013qta,Aaij:2015oid,ATLAS:2017dlm,CMS:2017ivg,Abdesselam:2016llu}. Consistent deviations appeared in other channels such as $B\to K\mu\mu$ and $B_s \to \phi\mu\mu$ (mainly for the branching ratios), but also in a different type of observable, namely Lepton Flavour Universality Violating (LFUV) observables probing the universality of the lepton coupling  in $b\to s\ell\ell$ comparing $\ell=e$ and $\ell=\mu$. Recent experimental updates have confirmed the presence of these deviations at the level of $2.5\sigma$~\cite{Aaij:2014ora,Aaij:2017vbb,Wehle:2016yoi}. Global fits within an Effective Field Theory (EFT) approach  performed on the large set of observables available have shown the remarkable consistency of the deviations observed, which can be explained through various New Physics (NP) scenarios affecting only a limited number of operators by shifting the short-distance physics encoded in Wilson coefficients. For instance, in Refs.~\cite{Capdevila:2017bsm,Alguero:2019ptt}, it was shown that adding NP to one or two Wilson coefficients is sufficient to obtain an improvement of the fit with respect to the SM (measured by the corresponding pull) by more than $5\sigma$.

On the other hand, charged-current $b\to c\ell\nu$ transitions have also exhibited deviations in LFUV observables comparing $\ell=\tau$ and lighter leptons. First measured as deviating significantly from the SM in 2012~\cite{Lees:2012xj,Lees:2013uzd}, the relevant ratios $R_D$ and $R_{D^*}$ have been updated regularly, leading to a recent decrease of the deviation with respect to the Standard Model down to $3.1\sigma$~\cite{Huschle:2015rga,Aaij:2015yra,Aaij:2017uff,Abdesselam:2019dgh,Amhis:2019ckw}. Additional observables have been considered for $B\to D^*\tau\nu$ concerning the polarisation of both the $D^*$ meson~\cite{Abdesselam:2019wbt} and the $\tau$ lepton~\cite{Hirose:2016wfn,Hirose:2017dxl}. If the latter agrees with the SM within large uncertainties, the precise Belle measurement of the integrated $F_L^{D^*}$ yields a relatively high value  compared to the SM prediction, which appears difficult to accommodate with NP scenarios, as can be seen in Refs.~\cite{Blanke:2018yud,Blanke:2019qrx,Becirevic:2019tpx,Murgui:2019czp,Asadi:2019xrc,Shi:2019gxi} which considered a wide set of NP benchmark points.

While neutral-current anomalies hinted at in a large set of channels and observables can be caused by small NP contributions competing with the SM ones generated at the loop level, charged-current anomalies seen in two LFUV ratios should correspond to much larger NP contributions able to compete with tree-level SM processes. In this sense, the latter were much more unexpected and should be scrutinised in more detail, in order to confirm their existence.

In this note we pay  close attention to the decay $B \to D^* \ell \nu$ governed by the quark level transition $b \to c \ell \bar{\nu}$ with $\ell=\tau$ and $\ell=e,\mu$, and more specifically to its angular distribution. Depending on the NP hypotheses chosen, we will identify a set of symmetries for the massless (electron and muon) and massive (tau) distributions that will lead us to find a set of dependencies or relations among the angular coefficients of the distribution.
A similar exercise  was done in Refs.~\cite{Egede:2010zc,Matias:2014jua,Hofer:2015kka} for the case of  the decay mode $B \to K^*\mu\mu$. Here we will follow closely the  detailed work in Ref.~\cite{Egede:2010zc} to use the symmetries of the distribution in order to show that depending on the assumptions of the type of NP at work and the mass of the leptons, not all angular coefficients are independent. These relations can be used in the case of the $B \to D^* \ell \nu$ decay as a way of cross-checking the consistency of the measurements of angular observables~\footnote{An alternative approach is illustrated in Ref.~\cite{Colangelo:2019axi} in the case of  $B \to \rho (a_1)\ell\nu$  semileptonic decays where the study of specific NP operators extending the SM effective hamiltonian and the large-energy limit of form factors allows one to disentangle the role of the possible new structures in the differential 4-body distribution.}, but also to provide orientation on which kind of NP can be responsible for deviations with respect to the SM observed in these observables.

These relations among the observables, based on the symmetries of the angular distribution, lead to a new way of measuring $F_L^{D^*}$ for $B \to D^* \tau \nu$, relying on different coefficients of the distribution compared to the direct measurement performed by the Belle experiment. This can provide a different handle for experimentalists to cross-check the polarisation fraction and confirm or not its high value.  Such an alternative extraction of the longitudinal $D^*$ polarisation can also be useful if instabilities occur when extracting the p.d.f. of angular observables due to values of $F_L^{D^*}$ beyond physical boundaries for instance~\footnote{This problem has already occurred in the case of the angular analysis of $B \to K^*\mu^+\mu^-$: the fit to CMS data~\cite{CMS:2017ivg}
 used to extract $P_1$, $P_5^\prime$ and $F_L$ altogether from the data has exhibited instabilities that forced the authors of Ref.~\cite{CMS:2017ivg} to include additional information on $F_L$ rather than leave it free in the fit.}.
We will provide general expressions for the relations among observables but we will focus mainly on a baseline case without tensor contributions~\footnote{See Ref.~\cite{Biancofiore:2013ki} for  the impact of tensor operators on $R_{D^*}$ and other observables.} (for the benchmark points analysed in Ref.~\cite{Becirevic:2019tpx}, the presence of tensor operators decreases the value of $F_L^{D^*}$ for $B \to D^* \tau \nu$ substantially,
increasing the discrepancy with the measured value). On the other hand, we will consider the contribution of the pseudoscalar operator that can help to increase $F_L^{D^*}$ and bring it closer to the Belle measurement, as found in Ref.~\cite{Becirevic:2019tpx}. We will also discuss the simplified case where there are no large NP phases in the Wilson coefficients, i.e. when we assume the coefficients are real or the NP phases are small.

In Section~\ref{sec:angdist} we recall the structure of the angular distribution and define the most relevant observables following Ref.~\cite{Becirevic:2019tpx}. In Section~\ref{sec:relationsangular} we describe the formalism and explain how to count the number of  symmetries and dependencies for each particular case and we work out the dependencies in the massless and massive cases, paying special attention to the presence of pseudoscalar operators. In Section~\ref{sec:Dstarpol} these dependencies are used to determine $F_L^{D^*}$ (or equivalently $F_T^{D^*}$) in terms of the other observables in various ways and we discuss the impact of binning when using these relations.
In Section~\ref{rhn} we discuss a possible signature of the presence of light right-handed neutrinos in the absence of tensors and imaginary contributions using the different determinations of $F_L^{D^*}$. And in Section~\ref{expsensitivity}  the expected experimental sensitivity of forthcoming experiments is discussed.
We give our conclusions in Section~\ref{sec:conclusions}. In App.~\ref{app:massivecase} some details on the derivation of the exact massive dependencies are provided. Finally, illustrations of the binning effects for the relations discussed in this article are given in App.~\ref{app:allplots}.

\section{$\bar{B}\to D^*\ell\bar\nu$ angular distribution}\label{sec:angdist}

\subsection{Effective Hamiltonian and angular observables}

The angular distribution for $B\to D^*\ell\nu$ has been extensively studied in the literature~\cite{Duraisamy:2013pia,Duraisamy:2014sna,Becirevic:2016hea,Alonso:2016gym,Ligeti:2016npd,Hill:2019zja,Aebischer:2019zoe}.
We will base our studies on the studies in Ref.~\cite{Becirevic:2019tpx}.
Assuming that there are no light right-handed neutrinos, the distribution can be computed using the effective Hamiltonian:
\begin{eqnarray}
 H_{\rm eff} &=& \sqrt2 G_F V_{cb}  \left[
  (1+g_V) (\bar{c} \gamma_\mu b) (\bar{\ell}_L \gamma^\mu \nu_L) +
  (-1+g_A) (\bar{c} \gamma_\mu\gamma_5 b) (\bar{\ell}_L \gamma^\mu \nu_L) \right. \\
  \nonumber&&\quad + g_S (\bar{c} b) (\bar{\ell}_R \nu_L)
  + g_P (\bar{c} \gamma_5 b) (\bar{\ell}_R \nu_L) \\
  \nonumber&&\quad + g_T (\bar{c} \sigma_{\mu\nu} b) (\bar{\ell}_R \sigma^{\mu\nu} \nu_L)
  + \left. g_{T5} (\bar{c} \sigma_{\mu\nu} \gamma_5 b) (\bar{\ell}_R \sigma^{\mu\nu} \nu_L) \right]  + \mathrm{h.c.}
\end{eqnarray}
As it can be seen, we do not include right-handed neutrinos at this stage, which will be discussed later on.
One may also use the equivalent notation of Refs.~\cite{Blanke:2018yud,Blanke:2019qrx} (for instance)
\begin{eqnarray}
 H_{\rm eff} &=&{4\frac{G_F}{\sqrt2}} V_{cb}  \left[
 (1+g_{V_L}) (\bar{c}_L \gamma_\mu b_L) (\bar{\ell}_L \gamma^\mu \nu_L) +
 g_{V_R} (\bar{c}_R \gamma_\mu b_R) (\bar{\ell}_L \gamma^\mu \nu_L) \right. \\
 &&\qquad + g_{S_L} (\bar{c}_R b_L) (\bar{\ell}_R \nu_L)
 + g_{S_R} (\bar{c}_L b_R) (\bar{\ell}_R \nu_L) + \left. g_{T_L} (\bar{c}_R \sigma_{\mu\nu} b_L) (\bar{\ell}_R \sigma^{\mu\nu} \nu_L) \right] + \mathrm{h.c.} \nonumber
\end{eqnarray}
with the corresponding effective coefficients
\begin{equation}
 g_{V,\,A} = g_{V_R} \pm g_{V_L} \,, \quad\qquad
 g_{S,\,P} = g_{S_R} \pm g_{S_L} \,, \quad\qquad
 g_T = -g_{T5} = g_{T_L} \,.
 \label{eq:coeff_relations}
\end{equation}

The resulting angular distribution is
\begin{eqnarray}
 {\frac{d^4\Gamma}{dq^2 d\cos\theta_D d\cos\theta_\ell d\chi}} &\!\!\!\!\!=\!\!\!\!\!& \frac{9}{32\pi} \biggl\{ \biggr. I_{1c}\cos^2\theta_D + I_{1s}\sin^2\theta_D + \bigl[ I_{2c}\cos^2\theta_D + I_{2s}\sin^2\theta_D \bigr] \cos2\theta_\ell \\
 &&+ \bigl[ I_{6c}\cos^2\theta_D + I_{6s}\sin^2\theta_D \bigr] \cos\theta_\ell + \bigl[ I_3\cos2\chi + I_9\sin2\chi \bigr] \sin^2\theta_\ell \sin^2\theta_D \nonumber \\
 && + \bigl[ I_4\cos\chi + I_8\sin\chi \bigr] \sin2\theta_\ell \sin2\theta_D + \bigl[ I_5\cos\chi + I_7\sin\chi \bigr] \sin\theta_\ell \sin2\theta_D \biggl.\biggr\} \,, \nonumber
 \label{eq:fullangdist}
\end{eqnarray}
where the angular coefficients $I_i\equiv I_i(q^2)$ are given in Ref.~\cite{Becirevic:2019tpx}:
\begin{eqnarray}
 I_{1c} &=& 2N \biggl[ |\tilde{H}_0^-|^2 +  \frac{m_\ell^2}{q^2} |\tilde{H}_0^+|^2 + 2\frac{m_\ell^2}{q^2} |\tilde{H}_t|^2 \biggr] \,, \\
 I_{1s} &=& \frac{N}{2} \biggl[ 3\bigl( |\tilde{H}_+^-|^2 + |\tilde{H}_-^-|^2 \bigr) + \frac{m_\ell^2}{q^2} \bigl( |\tilde{H}_+^+|^2 + |\tilde{H}_-^+|^2 \bigr) \biggr] \,, \\
 I_{2c} &=& 2N \biggl[ -|\tilde{H}_0^-|^2 + \frac{m_\ell^2}{q^2} |\tilde{H}_0^+|^2 \biggr] \,, \\
 I_{2s} &=& \frac{N}{2} \biggr[ |\tilde{H}_+^-|^2 + |\tilde{H}_-^-|^2 - \frac{m_\ell^2}{q^2} \bigl( |\tilde{H}_+^+|^2 + |\tilde{H}_-^+|^2 \bigr) \biggr] \,, \\
 I_3 &=& -2N \, {\rm Re}\biggl[ \tilde{H}_+^- \tilde{H}_-^{-*} - \frac{m_\ell^2}{q^2} \tilde{H}_+^+ \tilde{H}_-^{+*} \biggr]\\
 I_4 &=& N \, {\rm Re}\biggl[ (\tilde{H}_+^- + \tilde{H}_-^-) \tilde{H}_0^{-*} - \frac{m_\ell^2}{q^2} (\tilde{H}_+^+ + \tilde{H}_-^+) \tilde{H}_0^{+*} \biggr] \\
 I_5 &=& 2N \, {\rm Re}\biggl[ (\tilde{H}_+^- - \tilde{H}_-^-) \tilde{H}_0^{-*} - \frac{m_\ell^2}{q^2} (\tilde{H}_+^+ + \tilde{H}_-^+) \tilde{H}_t^* \biggr] \,, \\
 I_{6c} &=& 8N \frac{m_\ell^2}{q^2} \, {\rm Re}\bigl[ \tilde{H}_0^+ \tilde{H}_t^* \bigr] \,, \\
 I_{6s} &=& 2N \bigl( |\tilde{H}_+^-|^2 - |\tilde{H}_-^-|^2 \bigr) 
 \end{eqnarray}
\begin{eqnarray}
 I_7 &=& 2N \, {\rm Im}\biggl[ (\tilde{H}_+^- + \tilde{H}_-^-) \tilde{H}_0^{-*} - \frac{m_\ell^2}{q^2} (\tilde{H}_+^+ - \tilde{H}_-^+) \tilde{H}_t^* \biggr] \,, \\
 I_8 &=& N \, {\rm Im}\biggl[ (\tilde{H}_+^- - \tilde{H}_-^-) \tilde{H}_0^{-*} - \frac{m_\ell^2}{q^2} (\tilde{H}_+^+ - \tilde{H}_-^+) \tilde{H}_0^{+*} \biggr] \\
 I_9 &=& -2N \, {\rm Im}\biggl[ \tilde{H}_+^- \tilde{H}_-^{-*} - \frac{m_\ell^2}{q^2} \tilde{H}_+^+ \tilde{H}_-^{+*} \biggr]
\end{eqnarray}
where $N$ is a normalisation
\begin{equation}
 N={\cal B}_{D^* \to D \pi} \frac{G_F^2 |V_{cb}|^2}{48 (2\pi)^3 m_B^3} q^2 \lambda^{1/2}_{BD^*}(q^2) \left(1-\frac{m_\ell^2}{q^2}\right)^2,
\end{equation}
with $\lambda_{BD^*}(q^2)=m_B^4 +m_{D^*}^4+q^4-2(m_B^2 m_{D^*}^2+m_B^2 q^2 + m_{D^*}^2 q^2)$
and the amplitudes $\tilde{H}$ correspond to linear combinations of transversity amplitudes for various currents. We can write them in the following way to make the dependence on $m_\ell$ explicit:
\begin{equation} \label{definitionsH}
 \tilde{H}_i^+=H_i-2\frac{\sqrt{q^2}}{m_\ell} H_{T,i}
 \qquad
 \tilde{H}_i^-=H_i-2\frac{m_\ell}{\sqrt{q^2}} H_{T,i}
 \qquad
 \tilde{H}_t=\frac{\sqrt{q^2}}{m_\ell} \tilde{H}_P
\end{equation}
where $i=0,+,-$ and $H_i$ correspond to vector and axial currents whereas $H_{T,i}$ correspond to tensor currents, and $\tilde{H}_P$ combines two amplitudes $H_t$ and $H_P$:
\begin{equation} \label{pseudo}
 \tilde{H}_P=\frac{m_\ell}{\sqrt{q^2}}H_t + H_P
\end{equation}
The $H_i$ amplitudes depend on form factors and on $q^2$, but not on the lepton mass. In particular, the presence of $1/m_\ell$ in $\tilde{H}^+_i$ means that the discussion of the limit $m_\ell\to 0$ should be considered after expressing all the angular coefficients in terms of $H_i$.

\subsection{Observables}\label{observablessec}

Contrary to $B\to K^*\ell\ell$~\cite{Matias:2012xw,Descotes-Genon:2013vna}, there are no specific discussions to consider concerning the possibility of optimised observables, since all $B\to D^*$ form factors  either vanish or yield the same Isgur-Wise function $\xi$ in the heavy quark limit, so any ratio of angular observables is appropriate to reduce uncertainties from form factors.
We thus take almost the same list as
Ref.~\cite{Becirevic:2019tpx}  for the 12 observables that form  a basis\footnote{Further discussions of
 this differential decay rate can be found in Ref.~\cite{Bhattacharya:2019olg} including CP-violating observables and in Ref.~\cite{Colangelo:2018cnj} when $D^*$ subsequently decays either to $D\pi$ or to $D\gamma$.}:
\begin{equation}
 O_i=\left\{A_0, A_3,A_4,A_5,A_{6s},A_7,A_8,A_9,A_{\rm FB}, R_{A,B}, F_L^{D^*},d \Gamma/dq^2\ \right\}
\end{equation}
Compared to Ref.~\cite{Becirevic:2019tpx}, we do not include the observable $A_{\lambda_\ell}$  in this list
because it is related to the $\tau$ polarisation and requires one coefficient not included in the angular distribution. Instead we must
introduce an additional observable (not included in Ref.~\cite{Becirevic:2019tpx}) so that the numbers of angular coefficients and observables match. We may choose for instance:
\begin{equation}
 A_0=\frac{1}{d\Gamma/dq^2} (I_{1c} + I_{1s})
\end{equation}
We recall here the definition of the
observables defined in Ref.~\cite{Becirevic:2019tpx} that will play an important role in this article:
\begin{itemize}
 \item The differential decay rate
       \begin{equation} \label{dgstandard}
        \frac{d\Gamma}{dq^2}=\frac{1}{4}(3I_{1c}+6I_{1s}-I_{2c}-2I_{2s})
       \end{equation}
 \item The longitudinal and transverse $D^*$ polarisation decay rates:
       \begin{eqnarray} \label{deffLstandard}
        F_L^{D^*}&=&\frac{d\Gamma_L/dq^2}{d\Gamma/dq^2}=\frac{1}{d\Gamma/dq^2}\frac{1}{4} (3 I_{1c}-I_{2c}) \\  \label{deffTstandard}
        F_T^{D^*}&=&1-F_L^{D^*}=\frac{d\Gamma_T/dq^2}{d\Gamma/dq^2}=\frac{1}{d\Gamma/dq^2}\frac{1}{2} (3 I_{1s}-I_{2s})
       \end{eqnarray}
       In order to make a more explicit contact with the integrated longitudinal polarisation we also introduce $\tilde{F}_L^{D^*}=(d\Gamma_L/dq^2)/\Gamma$ and $\tilde{F}_T^{D^*}=(d\Gamma_T/dq^2)/\Gamma$, where $\Gamma=\Gamma(B\to D^*\ell\nu)$ with $\ell=\tau,\mu,e$.
 \item The ratio $R_{A,B}$ describing the relative weight of the various angular coefficients in the partial differential decay rate with respect to $\theta_\ell$, in analogy with the longitudinal polarisation fraction
       \begin{equation}
        R_{A,B}(q^2)=\frac{d\Gamma_A/dq^2}{d\Gamma_B/dq^2}=\frac{1}{2} \frac{(I_{1c}+2I_{1s}-3 I_{2c}-6 I_{2s})}{(I_{1c}+2 I_{1s}+I_{2c}+ 2 I_{2s})}
       \end{equation}
\end{itemize}
Eqs.~(\ref{dgstandard}), (\ref{deffLstandard}) and (\ref{deffTstandard}) are the ``standard definitions" of $d\Gamma/dq^2$, $F_L^{D^*}$ and $F_T^{D^*}$ respectively, and they are used to determine these observables with this particular functional dependence of the angular coefficients $I$.

Similarly to the discussion in Ref.~\cite{DescotesGenon:2012zf}, the definition of observables integrated over a bin (or over the whole phase space) requires some care. Experimentally, the measurement yields the integrated angular coefficients  $\langle I_k \rangle_\ell$ with the definition\footnote{Notice that the definition of $\langle I_i \rangle$ in Ref.~\cite{Becirevic:2019tpx} is normalised with $\Gamma(B \to D^* \ell \nu)$, while we prefer to keep the dependence on $\Gamma(B \to D^* \ell \nu)$  explicit.}
\begin{equation} \label{idef}
 \langle X \rangle_\ell = \int_{m_\ell^2}^{(m_B-m_{D^*})^2} dq^2\, X
\end{equation}
where the subscripts $\ell$ and $0$ indicate the massive case (with $m_\ell$) and the massless case respectively.
We can then define the ``standard" integrated longitudinal and transverse polarisations
\begin{eqnarray} \label{deffL}
 \langle \tilde{F}_L^{D^*} \rangle_\ell&=&\frac{1}{4\Gamma} \left(3 \langle I_{1c} \rangle_\ell - \langle I_{2c}  \rangle_\ell \right)\\
 \langle \tilde{F}_T^{D^*} \rangle_\ell&=&\frac{1}{2\Gamma} ( 3 \langle I_{1s}\rangle_\ell-\langle I_{2s}\rangle_\ell) \label{deffT}
\end{eqnarray}
The Belle measurement is actually $\langle \tilde{F}_L^{D^*} \rangle_\tau^{\rm Belle}=0.60\pm 0.09$.

\subsection{Global fits}

At this stage, a brief overview of our current understanding of the possible NP contributions
is useful. Global fits to $b\to c\tau\nu$ favour overwhelmingly a NP contribution through a real $g_{V_L}$ for $b\to c\tau\nu$, as it allows one to modify the tauonic branching ratios involved in $R_D$ and $R_{D^*}$ by the same amount without altering the angular observables, in agreement with the current data (apart from $F^{D^*}_L$ already discussed)~\cite{Blanke:2018yud,Blanke:2019qrx,Becirevic:2019tpx}.
For real contributions, scenarios based purely on scalar and pseudoscalar contributions exhibit some tension with the $B_c$ lifetime, depending on the relative size of the contribution allowed for $B_c\to\tau\nu$ in the total lifetime, which requires the pseudoscalar contribution to be somewhat small~\cite{Li:2016vvp,Alonso:2016oyd,Akeroyd:2017mhr}. Similarly, real tensor contributions are disfavoured, as  they tend to decrease the longitudinal polarisation of the $D^*$ meson compared to the SM~\cite{Becirevic:2019tpx}, when the first measurement from the Belle experiment indicated a value higher than SM expectations~\cite{Abdesselam:2019wbt}.
If $g_{V_L}$ is allowed as well as contributions of other operators, the former is dominant and the other operators (scalar, pseudoscalar, tensors) are subleading. Other constraints on $b\to c\tau\nu$ come from direct searches at LHC involving mono-$\tau$ jets~\cite{Greljo:2018tzh}. The corresponding bounds are again much tighter on tensor operators than on vector or scalar operators.

Some of these scenarios allow large imaginary parts~\cite{Blanke:2018yud,Blanke:2019qrx,Becirevic:2019tpx}, with a similar hierarchy of scenarios as in the real case. However, one must take into account that such large imaginary parts are allowed due to the limited number of observables. Additional observables could bring a dramatic modification of the landscape of the allowed scenarios, restricting the possible size of imaginary parts and the applicability of scenarios currently viable severely. Indeed some of the NP scenarios favour large imaginary parts so that there are no interferences between the SM and NP contributions, which add up in quadrature only (see for instance the scenario of a purely imaginary $g_{S_L}$ discussed in Ref.~\cite{Becirevic:2018afm}).
Restricting the size of these imaginary parts would enhance the interferences between SM and NP parts and would restrict the viability of the NP models where these interferences are negative.

This trend is confirmed by model-dependent analyses.
Most of the models with a single-particle exchange aiming at reproducing the data in $b\to c\ell\nu$ do not generate tensor contributions, apart from the scalar $SU(2)_L$-doublet leptoquark $R_2$ (as illustrated, for instance, in Ref.~\cite{Tanaka:2012nw}) which however generates much larger contributions to  $g_{S_L}$ (i.e. $g_S$ and $g_P$) than to $g_{T_L}$ (i.e. $g_T$ and $g_{T5}$). This effect is enhanced by the running from the NP scale (1 TeV) down to the $m_b$ scale (reducing the tensor contribution by $\sim 20\%$ and increasing the scalar contribution by $\sim 80\%$), so that scalar contributions are likely to be larger than the tensor contributions if the latter are present~\cite{Blanke:2018yud}. In Ref.~\cite{Blanke:2018yud, Blanke:2019qrx}, a model with a single $R_2$ leptoquark with complex couplings was shown to have a lower SM-pull than other NP scenarios once the constraint from the $B_c$ lifetime was taken into account. In Ref.~\cite{Becirevic:2018afm}, a viable model with the $R_2$ leptoquark was proposed in combination with the $S_1$ leptoquark, leading to (large real) vector couplings as well as (large imaginary) scalar and (smaller imaginary) tensor couplings.

We will thus consider as a baseline scenario that tensor contributions are subleading compared to other operators. We will also consider that  the imaginary parts of the amplitudes can be neglected. In the SM as well as in the case of real NP, the only phase comes from the CKM matrix element, and it is actually the same for all the amplitudes. Under our baseline scenario,  for instance, the angular coefficients corresponding to imaginary parts ($I_{7,8,9}$) are either small or vanishing, as well as any imaginary contribution. For completeness we will provide full expressions for the relations among the coefficients including these terms (see App.~\ref{app:massivecase} for the  general expressions in the massive case).

\section{Relations among angular coefficients}\label{sec:relationsangular}

\subsection{Symmetries and dependencies}

The decay $B\to D^*\ell\nu$ has a rich angular structure, and it is interesting to investigate whether all the angular observables defined in the previous section are independent, following the same steps as in Refs.~\cite{Egede:2010zc,Matias:2014jua,Hofer:2015kka,
 Matias:2012xw} for $B\to K^*\ell\ell$. We can consider the angular coefficients as being bilinears in
\begin{eqnarray}
 \vec{A}&=&\{ {\rm Re}[H_0], {\rm Im}[H_0],{\rm Re}[H_+], {\rm Im}[H_+],
 {\rm Re}[H_-], {\rm Im}[H_-], \\
 && \qquad {\rm Re}[H_{T,0}], {\rm Im}[H_{T,0}],{\rm Re}[H_{T,+}], {\rm Im}[H_{T,+}],
 {\rm Re}[H_{T,-}], {\rm Im}[H_{T,-}],
 {\rm Re}[\tilde{H}_P], {\rm Im}[\tilde{H}_P]\} \nonumber
\end{eqnarray}
An infinitesimal transformation will be given by
\begin{equation}
 \vec{A}'=\vec{A}+\vec{\delta}
\end{equation}
For the infinitesimal transformation to leave the coefficients $I$ unchanged, the vector $\vec\delta$ has to be perpendicular to the hyperplane spanned by the set of gradient vectors $\vec\nabla I_i$ (with the derivatives taken with respect to the various elements of $\vec{A}$). If the $I_i$ are all independent, the gradient vectors should span the whole space available for the coefficients, i.e. the dimension of the space for the gradient vectors should be identical to the number of angular coefficients.

\begin{table}[t]
 \begin{center}
  \begin{tabular}{c|c|c|c|c|c|c}
   $m_\ell$ & Tensor ops. & Pseudoscalar op. & Coefficients & Dependencies & Amplitudes & Symmetries \\
   \hline
   0        & No          & No               & 11           & 6            & 3          & 1          \\
   0        & No          & Yes              & 11           & 5            & 4          & 2          \\
   0        & Yes         & No               & 11           & 0            & 6          & 1          \\
   0        & Yes         & Yes              & 12           & 0            & 7          & 2          \\
   $\neq 0$ & No          & No               & 12           & 5            & 4          & 1          \\
   $\neq 0$ & No          & Yes              & 12           & 5            & 4          & 1          \\
   $\neq 0$ & Yes         & No               & 12           & 0            & 7          & 2          \\
   $\neq 0$ & Yes         & Yes              & 12           & 0            & 7          & 2
  \end{tabular}
  \caption{Symmetries and dependencies among the $B\to D^*\ell\nu$ angular observables depending on the mass of the lepton and the contribution of tensor and pseudoscalar operators.}
  \label{tab:symmetries}
 \end{center}
\end{table}

One can define:
\begin{itemize}
 \item The number of coefficients $n_c$, given directly by the angular distribution
 \item The number of dependencies $n_d$, given by the difference between the number of angular coefficients $I_i$ and the dimension of the space given by the gradient vectors (provided by the rank of the matrix $M_{ij}=\nabla_i I_j$)
 \item The number of helicity/transversity amplitudes $n_A$, leading to $2n_A$ real degrees of freedom
 \item The number of continuous symmetries  $n_s$ explaining the degeneracies among angular coefficients
\end{itemize}
One has the following relation
\begin{equation}
 n_c-n_d=2n_A-n_s
\end{equation}
which we can investigate in various cases for $B\to D^*\ell\nu$ summarised in Table~\ref{tab:symmetries}.

As discussed above, the assumption of no tensor contributions seems favoured by
the current global fits and we will stick to this assumption. In this case it is expected according to Table~\ref{tab:symmetries} the existence of 5 or 6 relations.
The presence or absence of the pseudoscalar operator  does not modify the
outcome of the analysis and the number of dependencies in the massive case due to Eq.~(\ref{pseudo}). However,
we find interesting to discuss its effect separately as it was found in Ref.~\cite{Becirevic:2019tpx} that such a pseudoscalar contribution can help to alleviate the tension in $F_L^{D^*}$ for $B \to D^* \tau \nu$.

We can now explore the dependence relations between angular coefficients, depending on the lepton mass, the presence of pseudoscalar and tensor operators.
These relations can be used as a consistency test among the observables if all of these observables are measured
in order to check the very general assumptions made to derive them. If these relations are not fulfilled, it means that there is an issue with one or more of the measurements or some of the underlying assumptions (negligible NP in tensor operator, negligible imaginary parts) are not correct. Such tests are completely independent of any assumption on the details of the NP model or the hadronic inputs.

\subsection{Massless case with no pseudoscalar operator and no tensor operators}

The expressions for the angular observables become
in terms of the amplitudes themselves
\begin{eqnarray}
 I_{1c}&=&2N \times |H_0|^2\\
 I_{1s}&=&\frac{N}{2}\times 3\left[|H_+|^2+|H_-|^2\right]\\
 I_{2c}&=&2N\times(-1)|H_0|^2\\
 I_{2s}&=&\frac{N}{2}\left[|H_+|^2+|H_-|^2\right]\\
 I_{3}&=&-2N\times{\rm Re}[H_+H_-^*]\\
 I_{4}&=&N\left[{\rm Re}[H_0H_+^*+{\rm Re}[H_0H_-^*]\right]\\
 I_{5}&=&2N\left[{\rm Re}[H_0H_+^*-{\rm Re}[H_0H_-^*]\right]\\
 I_{6c}&=&0\\
 I_{6s}&=&2N\left[|H_+|^2-|H_-|^2\right]
\end{eqnarray}
\begin{eqnarray}
 I_{7}&=&2N\left[-{\rm Im}[H_0H_+^*]-{\rm Im}[H_0H_-^*]\right]\\
 I_{8}&=&N\left[-{\rm Im}[H_0H_+^*]+{\rm Im}[H_0H_-^*]\right]\\
 I_{9}&=&-2N\times{\rm Im}[H_+H_-^*]
\end{eqnarray}

In this case, the only continuous symmetry that can be found is simply
\begin{equation}
 H_0\to e^{i\alpha} H_0\,, \qquad
 H_-\to e^{i\alpha} H_-\,, \qquad
 H_+\to e^{i\alpha} H_+
\end{equation}
and only 5 of the 11 observables~\footnote{Notice that there are 11 coefficients in this case: $I_{6c}=0$ and consequently there are 11 observables since  $A_{FB}$ and $A_{6s}$ are proportional.} are independent and 6 dependencies are found. Consequently, one can invert the system to determine the value of the real and imaginary parts of the amplitudes in terms of some of the angular coefficients, and re-express the other ones in terms of the same angular coefficients leading to the following relations:
\begin{eqnarray}
 I_{1c}&=&-I_{2c} \label{eq:massless0}\\
 I_{1s}&=&3 I_{2s} \label{eq:masslessm1}\\
 -4 I_3 I_{2c}&=& -4 I_{4}^2+I_5^2-I_7^2+4 I_8^2 \label{eq:massless1} \\
 -2 I_9 I_{2c}&=& I_5 I_7 - 4 I_4 I_8 \label{eq:massless2}\\
 -4 I_{2c} \left(\frac{1}{2} I_{6s}+ \frac{2}{3} I_{1s} \right)&=& (2 I_4 +I_5)^2 + (I_7+2 I_8)^2  \label{eq:massless3}\\
 -4 I_{2c} \left(-\frac{1}{2} I_{6s}+ \frac{2}{3} I_{1s} \right)&=& (-2 I_4 +I_5)^2 + (I_7-2 I_8)^2
 \label{eq:massless4}
\end{eqnarray}
These relations can be used as a consistency test among the observables if all of these observables are measured, under the hypothesis that we have outlined (negligible lepton mass, negligible pseudoscalar and tensor operators).

Another way of exploiting these equations consists in combining  the non-trivial relations  Eqs.~(\ref{eq:massless1})-(\ref{eq:massless4}) under the assumption that $I_{7,8,9}=0$ (taking all imaginary parts to be zero). For future use under this assumption we reorganise these equations, allowing us to make contact with the massive ones later on:
\begin{eqnarray} \label{eqmasszero}
 I_3^2&=&\frac{4}{9} I_{1s}^2-\frac{1}{4} I_{6s}^2   \\  \label{eqmasszero1}
 I_4^2&=&-\frac{1}{3} I_{1s} I_{2c}+ \frac{1}{2} I_{2c} I_3   \\  \label{eqmasszero2}
 I_5^2&=&-\frac{2}{3} I_{2c} (2 I_{1s}+ 3 I_3)  \label{eqmasszero3}
\end{eqnarray}
One of the dependencies disappears once  $I_{7,8,9}=0$ is taken.

\subsection{Massless case with pseudoscalar operator but no tensor operators}

The same relations between angular observables and amplitudes hold as in the previous case, apart from
\begin{equation}
 I_{1c}=2N\left[|H_0|^2+2|H_P|^2\right]
\end{equation}
One can see that the two symmetries are
\begin{equation}
 H_0\to e^{i\alpha} H_0\,, \qquad
 H_-\to e^{i\alpha} H_-\,, \qquad
 H_+\to e^{i\alpha} H_+\,, \qquad
 H_P\to e^{i\beta} H_P\,, \qquad
\end{equation}

Again, by inverting the system one can obtain the same relations as in the massless case without pseudoscalar contributions, see Eqs.~(\ref{eq:masslessm1})-(\ref{eq:massless4}), except for Eq.~(\ref{eq:massless0}) which is not fulfilled.

Like in the previous case, these relations can be used as a consistency test among the observables if all of these observables are measured, under the hypothesis that we have outlined (negligible lepton mass, negligible tensor operators).

\subsection{Massive case with pseudoscalar operator but no tensor operators}

The symmetries in the massive case with pseudoscalar operator but no tensors are in principle a simple extension of the analogous massless case. However, obtaining the expression of the dependencies in the massive case is a rather non-trivial task. The absence of  tensors implies that there is no distinction between ``+'' and ``-'' components of $\tilde H_i^+$ and $\tilde H_i^-$ (see Eq.~(\ref{definitionsH})) and the only surviving symmetry in this case is
\begin{equation}
 H_0\to e^{i\alpha} H_0\,, \quad
 H_-\to e^{i\alpha} H_-\,, \quad
 H_+\to e^{i\alpha} H_+\,, \quad
 H_t \to e^{i\alpha} H_t\,, \quad
 H_P\to e^{i\alpha} H_P
\end{equation}
One finds five dependencies in this case, which are identified by solving the system of non-linear equations. The first one is trivial:
\begin{eqnarray}\label{eq:trivial}
 0&=&I_{1s} \left(1 - \frac{m_\ell^2}{q^2}\right) -  I_{2s} \left(3 + \frac{m_\ell^2}{q^2}\right) \end{eqnarray}
and the other exact four non-trivial dependencies are detailed in App.~\ref{app:massivecase}.

We will consider the simplifying case where all Wilson coefficients are real so that $I_{7,8,9}$ and all imaginary contributions can be neglected (see App.~\ref{app:massivecase} for the general case without these assumptions). The remaining four dependencies are then simplified substantially
\begin{eqnarray}
 I_3^2&=&\left(1-\frac{m_\ell^2}{q^2}\right)^2 \left[ \left(\frac{2 I_{1s}}{3 + m_\ell^2/q^2} \right)^2-\frac{I_{6s}^2}{4}
  \right]\label{eq:dependency3}
 \\[1.5mm] \label{eq:dependencym1} I_4^2&=& \frac{I_{2c} (2 I_{1s} (m_\ell^2-q^2)+I_3 (m_\ell^2+3 q^2))}{2 (m_\ell^2+3 q^2)}
 \\[1.5mm] \label{eq:dependencysm2}
 I_5^2&=& \left[-4 I_{2c}  I_{6c} I_{6s} (m_\ell^2 - q^2)^2 (m_\ell^2 + 3 q^2) +
  I_{6c}^2 (m_\ell^2 - q^2)^2 \left[ 2 I_{1s} (m_\ell^2 - q^2) + I_{3} (m_\ell^2 + 3 q^2) \right]  \right. \nonumber \\[1.5mm]
 &&\qquad \left.
 \!  \!\!\!\!\!\! \!\!\!-16 I_{2c}^2 q^4 \left[2 I_{1s} (-m_\ell^2 + q^2) + I_3 (m_\ell^2 + 3 q^2)\right]\right] /\left[
  8 I_{2c} (m_\ell^2 - q^2)^2 (m_\ell^2 + 3 q^2)\right] \quad \, \,
 \label{eq:dependency4} \\[1.5mm]
 I_{6c}^2  &=&  -8 m_\ell^2 \left[
  I_{1c} I_{2c}  (-m_\ell^2 + q^2) +  I_{2c}^2  (m_\ell^2 + q^2) \right]/\left[(m_\ell^2 - q^2)^2\right]    \label{eq:dependency5}
\end{eqnarray}
The first three equations above are the generalisation of Eqs.~(\ref{eqmasszero})-(\ref{eqmasszero2}) in the massive case while the last equation is new: it would vanish in the massless limit with no tensors.  These relations can be used as a consistency test among the observables if all of these observables are measured, under the hypothesis that we have outlined (no tensor operators, imaginary contributions negligible).

The last two equations can be combined  to get rid of the $I_{6c}^2$ term and obtain the massive counterpart of Eq.~(\ref{eqmasszero2}):
\begin{eqnarray}
 \!\!\!\!\!\!\!\!\! I_5^2&=&\left[ 4 (m_\ell^2 - q^2)^2 I_{1s} (m_\ell^2 (I_{1c} - I_{2c}) - 2 q^2 I_{2c}) +
  2 (m_\ell^2 + 3 q^2) (m_\ell^4 (I_{1c} - I_{2c}) - 2 q^4 I_{2c} \right. \nonumber \\ && \quad \left.
  \!\!\!\! \!\!\!\!\!\! -
  m_\ell^2 q^2 (I_{1c} + I_{2c})) I_3
  - (m_\ell^2 - q^2)^2 (m_\ell^2 +
  3 q^2) I_{6c} I_{6s}\right]/\left[2 (m_\ell^2-q^2)^2 \, (m_\ell^2+3 q^2)\right]
\end{eqnarray}
Eq.~(\ref{eq:dependency5}) has obviously no counterpart in the massless case, as it vanishes then~\footnote{In the massive case, this relation provides access to a sum of two related observables $A_{6s}$ and $A_{FB}$:
 \begin{equation}
  2 \langle A_{6s}\rangle_\ell + 9 \langle A_{\rm FB} \rangle_\ell  = \frac{27}{2\sqrt{2}}\frac{1}{\Gamma}\, m_\ell\,  \langle
  \frac{1}{q^2-m_\ell^2}
  \sqrt{
   I_{1c} I_{2c}  (m_\ell^2 - q^2) -  I_{2c}^2  (m_\ell^2 + q^2)  }\rangle_\ell \nonumber
 \end{equation}}.

\subsection{Cases with tensor operators}

In the massive case with tensors the degeneracy between the $\tilde H_i^+$ and $\tilde H_i^-$  is broken and two symmetries are identified.  The symmetries are better described in terms of the tilde-fields:
\begin{equation}
 \tilde{H}_i^{-}\to e^{i\alpha} \tilde{H}_i^{-}\,, \qquad
 \tilde{H}_i^{+}\to e^{i\beta} \tilde{H}_i^{+}\,, \qquad
 \tilde{H}_t\to e^{i\beta} \tilde{H}_t\,.
\end{equation}
Unfortunately there are no dependencies in this case. The same is true in the massless case.

\section{Expressions of the $D^*$ polarisation} \label{sec:Dstarpol}

In the previous section, we have obtained several relationships between the angular coefficients under various hypotheses, assuming that tensor contributions are negligible. We can use these relations in order to obtain alternative determinations of the longitudinal polarisation $F_L^{D^*}$. From Sec.~\ref{sec:masslessreal} to Sec.~\ref{sec:imaginary}, we will provide these exact relationships in their binned form, but the corresponding unbinned versions have exactly the same form.

\subsection{Massless case without pseudoscalar operator}

For completeness we discuss the case with zero mass and no pseudoscalar operator, but still including all imaginary terms. Eqs~(\ref{eq:massless0})-(\ref{eq:masslessm1}) are trivial. Eqs.~(\ref{eq:massless1})-(\ref{eq:massless4}) can be rewritten in terms of observables providing different determinations of $F_L^{D^*}$:
\begin{eqnarray}
 \pi A_3 F_L^{D^*}&=&\frac{2}{9} (A_5^2-A_7^2) - \frac{1}{8} \pi^2 (A_4^2-A_8^2)
 \\
 \pi A_9 F_L^{D^*}&=&\frac{4}{9} A_5 A_7 +  \frac{1}{4} \pi^2 A_4 A_8
 \\
 (F_L^{D^*})^2
 &=&\left[ \frac{8}{9} (A_5^2+ A_7^2)  + \frac{1}{2} \pi^2 \left(A_4^2+A_8^2\right)\right] R_{A,B} \label{eq:FLD2}\\
 A_{FB} F_L^{D^*}&=& \pi \left( A_4 A_5-A_7 A_8 \right)
\end{eqnarray}
We recall that $A_i$ are defined from the angular observables up to a numerical normalisation given in Ref.~\cite{Becirevic:2019tpx}.
A similar set of expressions can be written for $\tilde{F}_L^{D^*}$, $\tilde{A}_i$ and $\tilde{A}_{FB}$ rather than $F_L^{D^*}$, $A_i$ and $A_{FB}$, respectively, by
substituting the normalization in terms of  $d\Gamma/dq^2$ by the integrated decay rate $\Gamma$.
These expressions can then be binned  trivially, however they are rather cumbersome to use. In the following two subsections we will restrict to the case of removing any imaginary contribution corresponding to our baseline scenario  that will be relevant to the extraction of $F_L^{D^*}$.

\subsection{Massless case without imaginary contributions}\label{sec:masslessreal}

Using Eqs.~(\ref{eq:masslessm1}) and (\ref{eqmasszero}) we obtain one of the important results of this article:
\begin{eqnarray} \label{first}
 \langle \tilde{F}_T^{D^*}\rangle_0 &=&   { \frac{1}{\Gamma}} \langle  2 \sqrt{  I_3^2 + \frac{1}{4} I_{6s}^2} \rangle_0 \qquad\qquad {\rm where\ }\langle \tilde{F}_T^{D^*}\rangle_0=1- \langle \tilde{F}_L^{D^*}\rangle_0
\end{eqnarray}

This expression can be used as an alternative way to determine the integrated $F_L^{D^*}$ in the massless case (without imaginary contributions but allowing for the presence of pseudoscalars) from experiment instead of the traditional determination in terms of $I_{1s}$ and $I_{2s}$ in Eq.~(\ref{deffL}) and Eq.~(\ref{deffT}).

This expression can be generalised to the case of smaller bins spanning only part of the whole kinematic range, leading to
\begin{eqnarray} \label{firstsmallbins}
 \langle \tilde{F}_T^{D^*}\rangle_0^i &=&  { \frac{1}{\Gamma}} \langle  2 \sqrt{  I_3^2 + \frac{1}{4} I_{6s}^2} \rangle_0^i
\end{eqnarray}
where $i$ means that the integral in Eq.(\ref{idef}) is taken over the bin $i$ with a narrower $[q^2_{i,{\rm min}},q^2_{i,{\rm max}}]$ range~\footnote{Notice that  $\langle \tilde{F}_L^{D^*}\rangle_0 + \langle \tilde{F}_T^{D^*}\rangle_0=1$ holds because the integration is performed over the whole kinematic range.
For the observables $\langle \tilde{F}_L^{D^*}\rangle_0^i$ and $\langle \tilde{F}_T^{D^*}\rangle_0^i$ shown in Figs.~\ref{fig:binning}-\ref{fig:appendix6}, this is no longer the case due to the normalisation of $\tilde{F}_L^{D^*}$ and $\tilde{F}_T^{D^*}$:
$\langle \tilde{F}_L^{D^*}\rangle_0^i+\langle \tilde{F}_T^{D^*}\rangle_0^i=\langle d\Gamma/dq^2\rangle_0^i/\Gamma<1$.
It is trivial to check that a different normalisation for $\tilde{F}_L^{D^*}$ and $\tilde{F}_T^{D^*}$ would only affect the normalisation $1/\Gamma$ appearing in the binned expressions.
}.

If we restrict further to the case without pseudoscalars (in this case $I_{1c}=- I_{2c}$ is fulfilled), we obtain further expressions using Eqs.~(\ref{eqmasszero1}) and (\ref{eqmasszero3}):
\begin{eqnarray} \label{firstbis}
 \langle \tilde{F}_L^{D^*}\rangle_0 &=&{ \frac{1}{\Gamma}}
 \langle \frac{I_5^2- 4 I_4^2}{4 I_3} \rangle_0
 \\
 &=&{ \frac{1}{\Gamma}}\langle R_{A,B}\left( I_3 +\sqrt{4 \frac{I_4^2}{R_{A,B}}+ I_3^2}\right) \rangle_0={ \frac{1}{\Gamma}}\langle R_{A,B}\left( -I_3 +\sqrt{\frac{I_5^2}{R_{A,B}}+ I_3^2}\right)  \rangle_0
\end{eqnarray}
where $R_{A,B}$ is positive and non-vanishing by construction.

\subsection{Massive case with pseudoscalar operator but without imaginary contributions}\label{sec:massivereal}

In this case, we focus on Eqs.~(\ref{eq:trivial}),(\ref{eq:dependency3}) and (\ref{eq:dependencym1}) to derive new descriptions of $F_L^{D^*}$  since
Eq.~(\ref{eq:dependencysm2}) is too involved to provide a useful alternative approach to $F_L^{D^*}$.
Eqs.~(\ref{eq:trivial}) and (\ref{eq:dependency3}) yield:
\begin{equation} \label{second}
 \langle \tilde{F}_T^{D^*}\rangle_\ell ={\frac{1}{\Gamma}} \langle \sqrt{ {(A\, I_3)}^2 + \frac{1}{4} {(B\, I_{6s})}^2} \,\rangle_\ell \qquad\qquad {\rm where\ }\langle \tilde{F}_T^{D^*}\rangle_\ell=1- \langle \tilde{F}_L^{D^*}\rangle_\ell
\end{equation}
where we define the auxiliary kinematic quantities (whose value in the massless case  is two)
\begin{equation}\label{eq:kinematicalWeights}
 A=\frac{m_\ell^2+2 q^2}{q^2-m_\ell^2} \qquad\qquad B=2+\frac{m_\ell^2}{q^2}
\end{equation}
One can write an equivalent equation to Eq.~(\ref{second}) for narrower $q^2$ bins similary to the previous section.
In the case of Eq.~(\ref{eq:dependencym1}) we do not substitute $I_{2c}$, leading to:
\begin{equation} \label{fLmass}
 \langle \tilde{F}_T^{D^*}\rangle_\ell =
 1- \langle \tilde{F}_L^{D^*}\rangle_\ell ={\frac{1}{\Gamma}} \langle  A\left(I_3 - 2 \frac{{ I_4}^2}{I_{2c}} \right)\rangle_\ell
\end{equation}
Relating this equation with the massless case is not straightforward given that in the massless case $I_{2c}$ was substituted (before integrating) in terms of $F_L^{D^*}$ and $R_{A,B}$.

\subsection{Cases with pseudoscalar operator and imaginary contributions}\label{sec:imaginary}

This corresponds to the most complete expression allowing for the presence of pseudoscalars and also imaginary parts, but no tensors. This can be achieved, as in the previous section, by using $I_{1s}$ and $I_{2s}$ instead of $I_{1c}$ and $I_{2c}$ as a starting point. The corresponding expression in the massless case is:
\begin{eqnarray} \label{firstimag}
 \langle \tilde{F}_T^{D^*}\rangle_0 &=&   { \frac{1}{\Gamma}} \langle  2 \sqrt{  I_3^2 + I_9^2 + \frac{1}{4} I_{6s}^2} \rangle_0
 \qquad\qquad {\rm where\ }\langle \tilde{F}_T^{D^*}\rangle_0=1- \langle \tilde{F}_L^{D^*}\rangle_0
\end{eqnarray}
and in the massive case
\begin{equation} \label{secondimag}
 \langle \tilde{F}_T^{D^*}\rangle_\ell ={\frac{1}{\Gamma}} \langle \sqrt{ {(A\, I_3)}^2 + {(A\, I_9)}^2+ \frac{1}{4} {(B\, I_{6s})}^2} \,\rangle_\ell
 \qquad\qquad {\rm where\ }\langle \tilde{F}_T^{D^*}\rangle_\ell=1- \langle \tilde{F}_L^{D^*}\rangle_\ell
\end{equation}

Similar expressions can be written for $\langle \tilde{F}_T^{D^*}\rangle_\ell^i$ defined for narrower $q^2$ bins.
These expressions represent the most general alternative ways to determine the massless and massive polarisation fractions. {Compared to the previous case, one can see that the presence of imaginary contributions comes simply from the additional $I_9$ term in Eqs.~(\ref{firstimag}) and (\ref{secondimag}), see also Eq.~(\ref{seconddep}) in App.~\ref{app:massivecase}.}

Within this more general framework, Eqs.~(\ref{eq:trivial}) and (\ref{seconddep})
yield the following simple relation among the observables defined in Section \ref{observablessec}:
\begin{equation}\label{new}
 \langle  x_1 (\tilde{F}_T^{D^*})^2\rangle_\ell =\langle x_2 \left( \tilde{A}_3^2+\tilde{A}_9^2\right)+ x_3 \left(\tilde{A}_{6s}\right)^2\rangle_\ell
\end{equation}
where $\tilde{A_i}$ stands for the observables $A_i$ normalised to $\Gamma$ rather than $d\Gamma/dq^2$, $x_1=(m_\ell^2-q^2)^2$, $x_2=4 \pi^2 (m_\ell^2+2 q^2)^2$ and $x_3=4 x_1 x_2/(729 \pi^2 q^4)$ ($A_9$ vanishes in the absence of large imaginary contributions).
This relation implies that the large (small) value of $F_L^{D^*}$ ($F_T^{D^*}$) requires a corresponding suppression in $A_3^2+A_9^2$, in $A_{6s}$ or both. For this reason it would be particularly interesting to have available predictions in specific models for this couple of observables in case  that the unexpectedly large value of this polarisation fraction remains.

\subsection{Binning}\label{sec:binning}

We have obtained these alternative expressions for $\langle \tilde{F}_L^{D^*} \rangle_\ell$ (or $\langle \tilde{F}_T^{D^*} \rangle_\ell$) assuming that there are no tensors and (in some cases)  no large imaginary contributions at short distances. From now on we introduce  the notation  $\langle\tilde{F}_T^{D^* \, \rm alt}\rangle_\ell$ (or $\langle\tilde{F}_L^{D^* \, \rm alt}\rangle_\ell$) to refer to Eq.~(\ref{secondimag}) as the alternative way to extract $F_T^{D^*}$ (or $F_L^{D^*}$).  In the absence of imaginary contributions we will use the notation $\langle\tilde{F}_T^{D^* \, \rm alt}\rangle^{\rm I_9=0}_\ell$ corresponding to Eq.~(\ref{second}). In the massless case we denote $\langle\tilde{F}_T^{D^* \, \rm alt}\rangle_0$ for Eq.~(\ref{firstimag}) and $\langle\tilde{F}_T^{D^* \, \rm alt}\rangle^{I_9=0}_0$ for Eq.~(\ref{first}).

Experimentally we have to consider binned versions of these expressions, which are nonlinear functions of the angular coefficients. Since the binned angular coefficients are the only quantities measured, we should be careful that $f(\langle I_k\rangle_\ell)\neq \langle f(I_k)\rangle_\ell$ when $f$ is non-linear.
From an experimental perspective there are two ways to proceed: \emph{i)} measure the coefficients $I_3$ and $I_{6s}$ of the massless or massive distribution in very small bins  in order
to reconstruct a $q^2$ dependence  of these functions, so that we can perform the integration  in Eq.~(\ref{first}) for the massless case or in  Eq.~(\ref{second}) in the massive case (or their counterparts including imaginary parts Eq.~(\ref{firstimag}) and Eq.~(\ref{secondimag})); \emph{ii)} use an unbinned measurement method
(as was done for $B \to K^*\mu\mu$ \cite{Egede:2015kha})  to determine the $q^2$ dependence of the coefficients and introduce the obtained expressions inside Eq.~(\ref{first}) or Eq.~(\ref{second}) as explained above.

Both approaches are however difficult to implement when the statistics is low, and one has to choose between the extraction of the whole angular distribution and the study of the $q^2$ dependence of simpler observables like the decay rate. Currently, the measurements are integrated over the whole kinematic range, which constitutes a single bin for the analysis.

By comparing with our exact results, we will thus investigate the accuracy of the approximation $f(\langle I_k\rangle_\ell)= \langle f(I_k)\rangle_\ell$,
which requires the following transformation on the unbinned expressions:
\begin{equation}\label{errorbin}
 d\Gamma_X/dq^2 \to \langle d\Gamma_X/dq^2 \rangle \qquad I_i \to \langle I_i \rangle \qquad w I_i \to \langle w I_i \rangle \qquad w I_i^2 \to \langle \sqrt{|w|} I_i \rangle^2
\end{equation}
where $w$ stands for any positive weight depending on $m$ and $q^2$.  This
leads to the following ``approximate formulae'' in the massless case, starting from Eq.~(\ref{firstimag}):
\begin{equation}\label{eq:FLbinnedmassless}
 \langle \tilde{F}_T^{D^* \, \rm alt}\rangle_0 \simeq    { \frac{1}{\Gamma}}   2 \sqrt{  \langle I_3 \rangle_0^2 +  \langle I_9 \rangle_0^2+  \frac{1}{4} \langle I_{6s} \rangle_0^2}
\end{equation}
and in the massive case, starting from Eq.~(\ref{secondimag}):
\begin{equation}\label{eq:FLbinnedmassive}
 \langle \tilde{F}_T^{D^* \, \rm alt}\rangle_\ell \simeq {\frac{1}{\Gamma}}  \sqrt{ \langle A\, I_3 \rangle_\ell^2 +\langle A\, I_9 \rangle_\ell^2 + \frac{1}{4} \langle  B\, I_{6s}\rangle_\ell ^2}
\end{equation}
In the massive case, one should measure the $I_i$ and multiply each event by a numerical factor $A$ for $I_3$, $I_9$ and $B$ for $I_{6s}$.

Similarly, in the absence of imaginary parts, we obtain the approximate binned expression, starting from Eq.~(\ref{second}):
\begin{equation} \label{binnedsecond}
 \langle \tilde{F}_T^{D^* \, \rm alt}\rangle^{I_9=0}_\ell \simeq {\frac{1}{\Gamma}} \sqrt{ \langle A\, I_3\rangle_\ell^2 + \frac{1}{4} \langle B\, I_{6s} \rangle_\ell ^2 }
\end{equation}
and the approximate expression for $ \langle \tilde{F}_T^{D^*}\rangle_\ell$ starting from Eq.~(\ref{fLmass})
\begin{equation} \label{binnedfLmass}
 {\frac{1}{\Gamma}} \langle  A\left(I_3 - 2 \frac{{ I_4}^2}{I_{2c}} \right)\rangle_\ell  \simeq  {\frac{1}{\Gamma}}\left[ \langle  A I_3\rangle_\ell - 2 \frac{\langle A I_4\rangle_\ell^2}{\langle AI_{2c}\rangle_\ell}\right]
\end{equation}
All these expressions have a corresponding expression for $\langle \tilde{F}_T^{D^*}\rangle_\ell^i$
for narrower bins where $\langle \rangle_\ell$ is transformed into $\langle \rangle_\ell^i$ corresponding to the integration over the narrow bin $i$.

In order to get an idea of the accuracy of these approximate relations, we perform the following numerical exercise.
We consider a set of  benchmark points corresponding to the best-fit-points of the 1D and 2D NP hypotheses in Ref.~\cite{Blanke:2018yud,Blanke:2019qrx}. Among the 1D hypotheses, the most favoured one is assuming NP in $g_{V_L}$, followed by NP in $g_{S_R}$.
Specifically we will take for this numerical analysis as benchmark points the best-fit-points of the following four different NP hypotheses (in each case, the remaining couplings are set to zero):
\begin{eqnarray}
 (R1)&:& \qquad g_{V_L}=0.07\\
 (R2)&:&\qquad g_{S_R}=0.09\\
 (R3)&:& \qquad g_{S_L}=0.07\\
 (R4)&:& \qquad g_{S_L}=4g_T=-0.03
\end{eqnarray}
where the values are given at the scale $\mu=1$ TeV, and we run them down to the scale $\mu=m_b$~\cite{Blanke:2018yud,Blanke:2019qrx,Gonzalez-Alonso:2017iyc}.
For 2D hypotheses, there is a wider range of relevant possibilities, and we select the following ones~\footnote{Even though $(C0)$ and $(C0)^*$ are formally different scenarios corresponding to opposite imaginary parts, they yield the same results for our observables which are not sensitive to the sign of the imaginary part.}:
\begin{eqnarray}
 (R5)&:&  (g_{V_L},g_{S_L}=-4g_T)=(0.10,-0.04)\\
 (R6)-(R7)&:&(g_{S_R},g_{S_L})=(0.21,-0.15)\ {\rm or}\ (-0.26,-0.61)\\
 (R8)&:&(g_{V_L},g_{S_R})=(0.08,-0.01)\\
 (C0)-(C0)^*&:& g_{S_L}=4g_T=-0.06\pm i\, 0.31
\end{eqnarray}
where once again we run these coefficients down to $\mu=m_b$.

\begin{figure}
 \centering
 \begin{subfigure}{\linewidth}
  \centering
  \includegraphics[height=.27\linewidth]{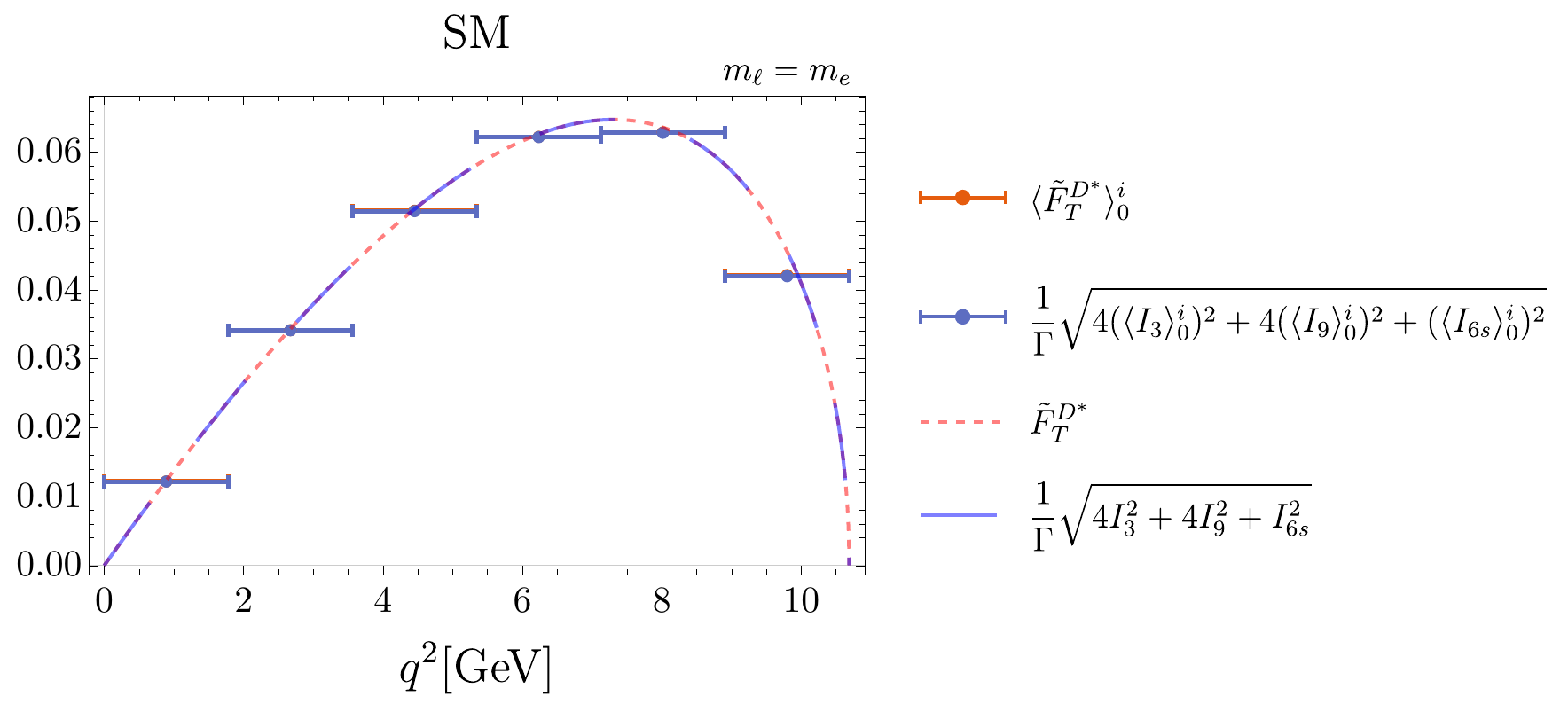}
  \label{fig:binning2Massless}
 \end{subfigure}\\
 \begin{subfigure}{\linewidth}
  \centering
  \includegraphics[height=.27\linewidth]{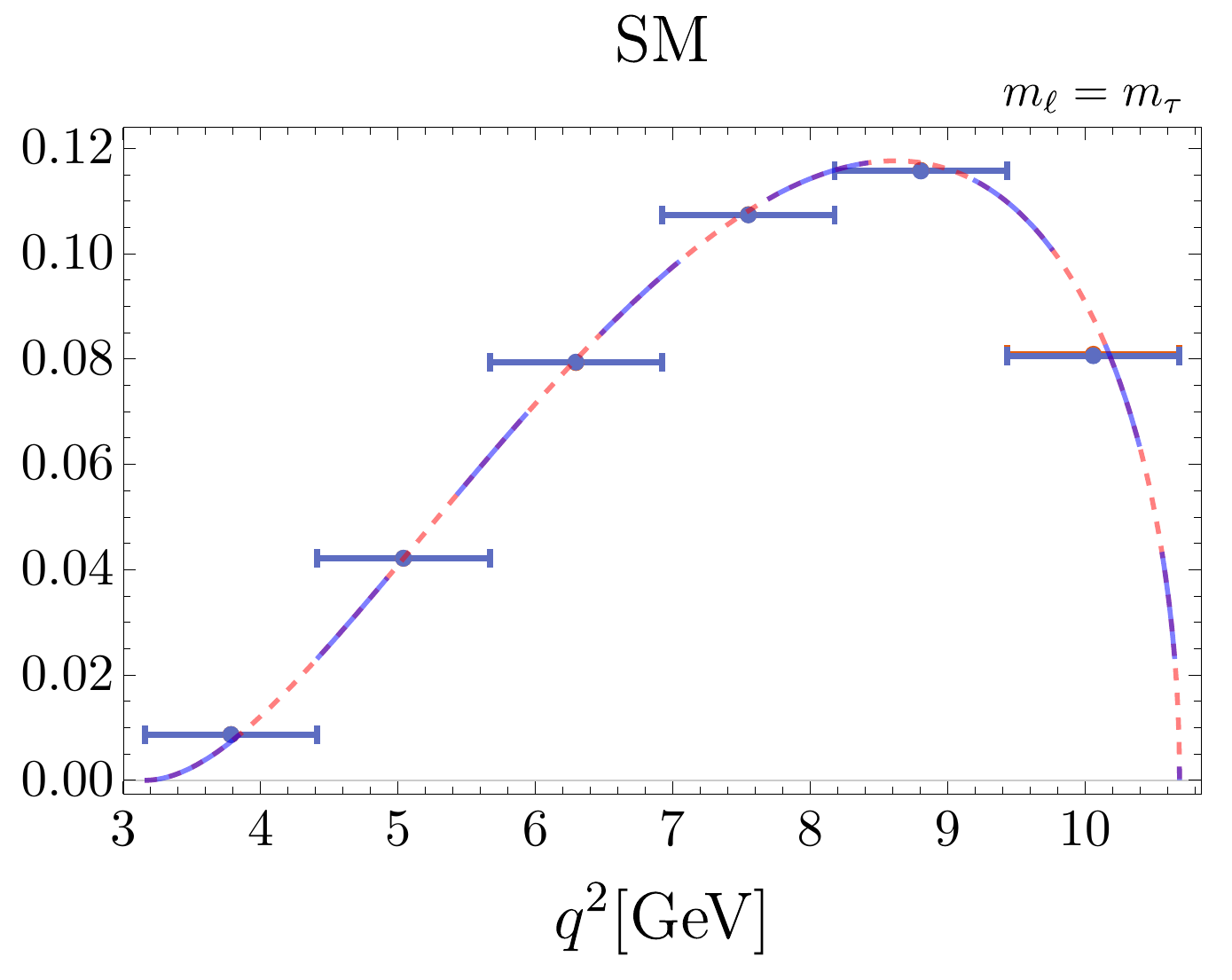}
  \hspace{10pt}
  \includegraphics[height=.27\linewidth]{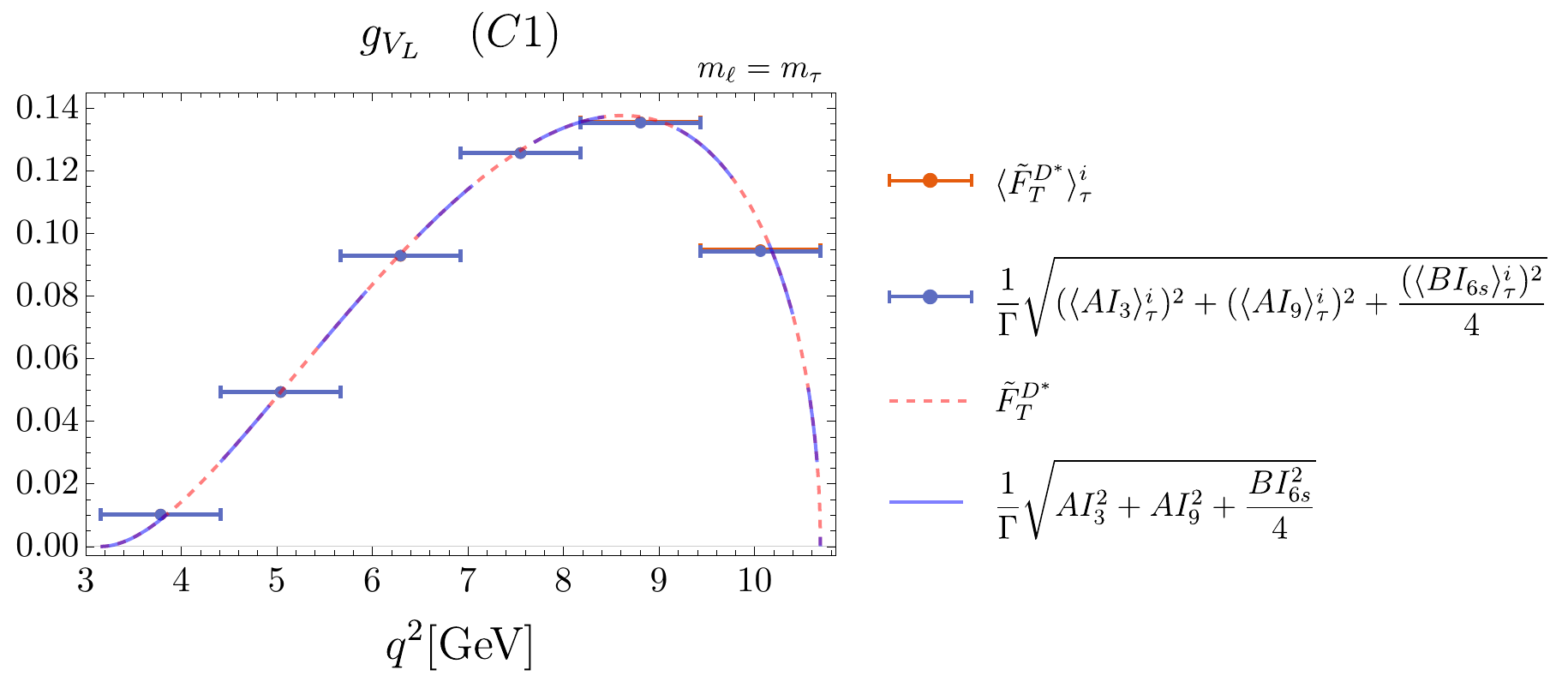}
  \label{fig:binning2Vector}
 \end{subfigure}

 \begin{subfigure}{\linewidth}
  \centering
  \includegraphics[height=.27\linewidth]{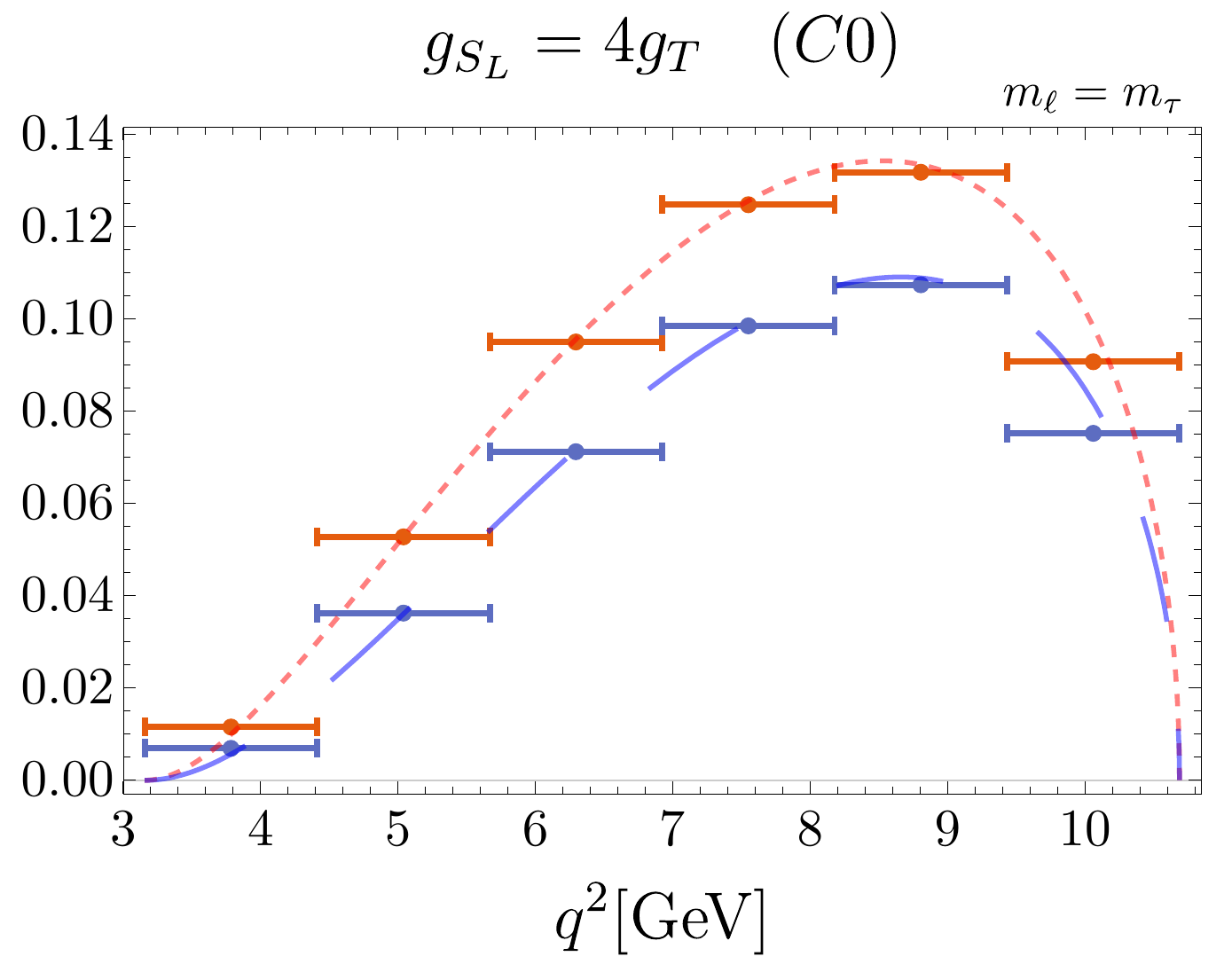}
  \hspace{10pt}
  \includegraphics[height=.27\linewidth]{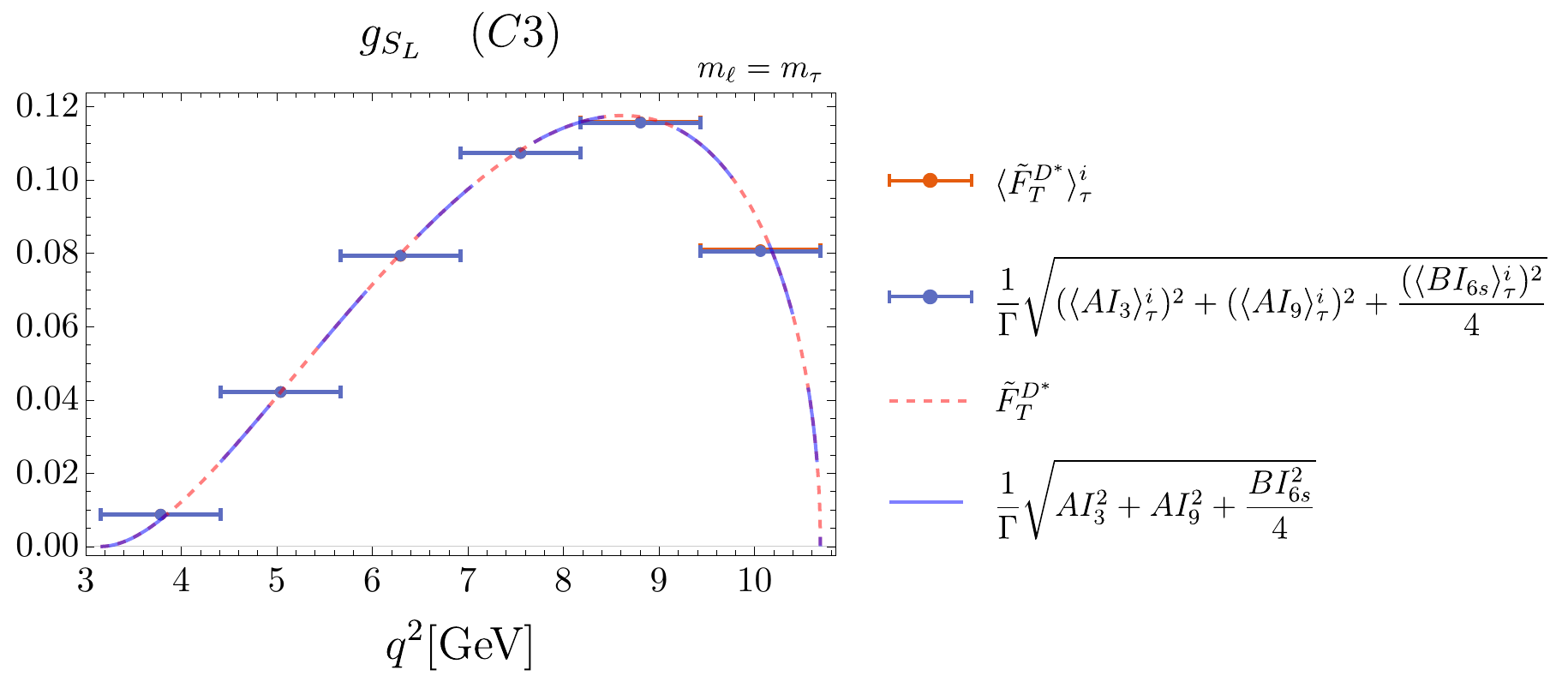}
  \label{fig:binning2Scalar}
 \end{subfigure}
 \caption{Illustration of the errors induced by binning on the relation in Eq.~(\ref{eq:FLbinnedmassive}). The orange dashed curve corresponds to the standard definition of $\tilde{F}_T^{D^*}$, whereas the blue one corresponds to $\tilde{F}_T^{D^* \, \rm alt}$.
 The orange bins in this plot are obtained using the binning form of the ``standard'' expression for  $\tilde{F}_T^{D^*}$  while the blue ones are obtained using the approximate binned expression of $\tilde{F}_T^{D^* \, \rm alt}$ in Eq.~(\ref{eq:FLbinnedmassive}).
 The  plots labelled SM correspond to the case  $m_\ell=m_e$ and $m_\ell=m_\tau$ in the SM and the  other plots correspond to $\tilde{F}_T^{D^*}$ in $B\to D^*\tau\nu$  in different NP scenarios described in the text. The differences come from the presence of tensor currents for $(C0)$ or from binning effects for the SM case. }
 \label{fig:binning}
\end{figure}

In Ref.~\cite{Becirevic:2019tpx}, a set of benchmark points is determined by considering the best-fit points of different scenarios with one free complex parameter. The resulting 2D benchmark points  (in each case, the remaining couplings are set to zero) at the scale $\mu=m_b$ are:
\begin{eqnarray}
 (C1)&:&\qquad g_{V_L}=0.07-i 0.16\\
 (C2)&:&\qquad g_{V_R}=-0.01-i0.39\\
 (C3)&:&\qquad g_{S_L}=0.29-i0.67\\
 (C4)&:& \qquad g_{S_R}=0.19+i0.08\\
 (C5)&:&\qquad g_{T}=0.11-i0.18
\end{eqnarray}
Using a different operator basis, alternative benchmark points are found to be~\footnote{For completeness, we quote $(C8)$ although this NP scenario has no impact on $B\to D^*\ell\nu$ and is thus equivalent to the SM for our purposes.}:
\begin{eqnarray}
 (C6)&:&\qquad g_{V}=0.20+i 0.19\\
 (C7)&:&\qquad g_{A}=0.69+i1.04\\
 (C8)&:&\qquad g_{S}=0.17-i0.16\\
 (C9)&:&\qquad g_{P}=0.58+i0.21
\end{eqnarray}
In the following we will check the relations given in the previous sections against these benchmark scenarios. We have used the binned approximation of the relations using 6 bins of equal length as shown in Fig.~\ref{fig:binning}. On the one hand, this allows us to test the quality of the binned approximation. On the other hand, we can check the impact of the assumptions used in order to derive the various relations: for instance, checking the expressions obtained for real NP contributions in Sec.~\ref{sec:massivereal} in the case of the scenarios $(C0)-(C9)$ with complex parameters  provides an estimate of the impact of realistic NP imaginary contributions on these expressions.

We need to choose a set of  form factors to evaluate the hadronic contributions and to be able to test how accurate the relations remain within the binned approximation discussed above, taking into account possible unexpected NP contributions (imaginary parts, tensor contributions). Since our goal is only to check the accuracy of this approximation for the various NP benchmark points it is enough to work using a simplified setting. For this reason, we refrain from using form factors obtained by elaborate combinations of heavy-quark effective theory~\cite{Isgur:1989vq,Isgur:1989ed,Falk:1992wt,Grinstein:2001yg} sum rules and lattice simulations~\cite{Boyd:1995cf,
 Caprini:1997mu,Gambino:2010bp,Gambino:2012rd,Gubernari:2018wyi,Bernard:2008dn,Bailey:2014tva,Harrison:2016gup,Bernlochner:2017jka,Blanke:2018yud,Becirevic:2019tpx} and we stick to the simpler quark model in Ref.~\cite{Melikhov:2000yu} without attempting to assign uncertainties to these computations.

\begin{figure}
 \centering
 \begin{subfigure}{\linewidth}
  \centering
  \includegraphics[height=.285\linewidth]{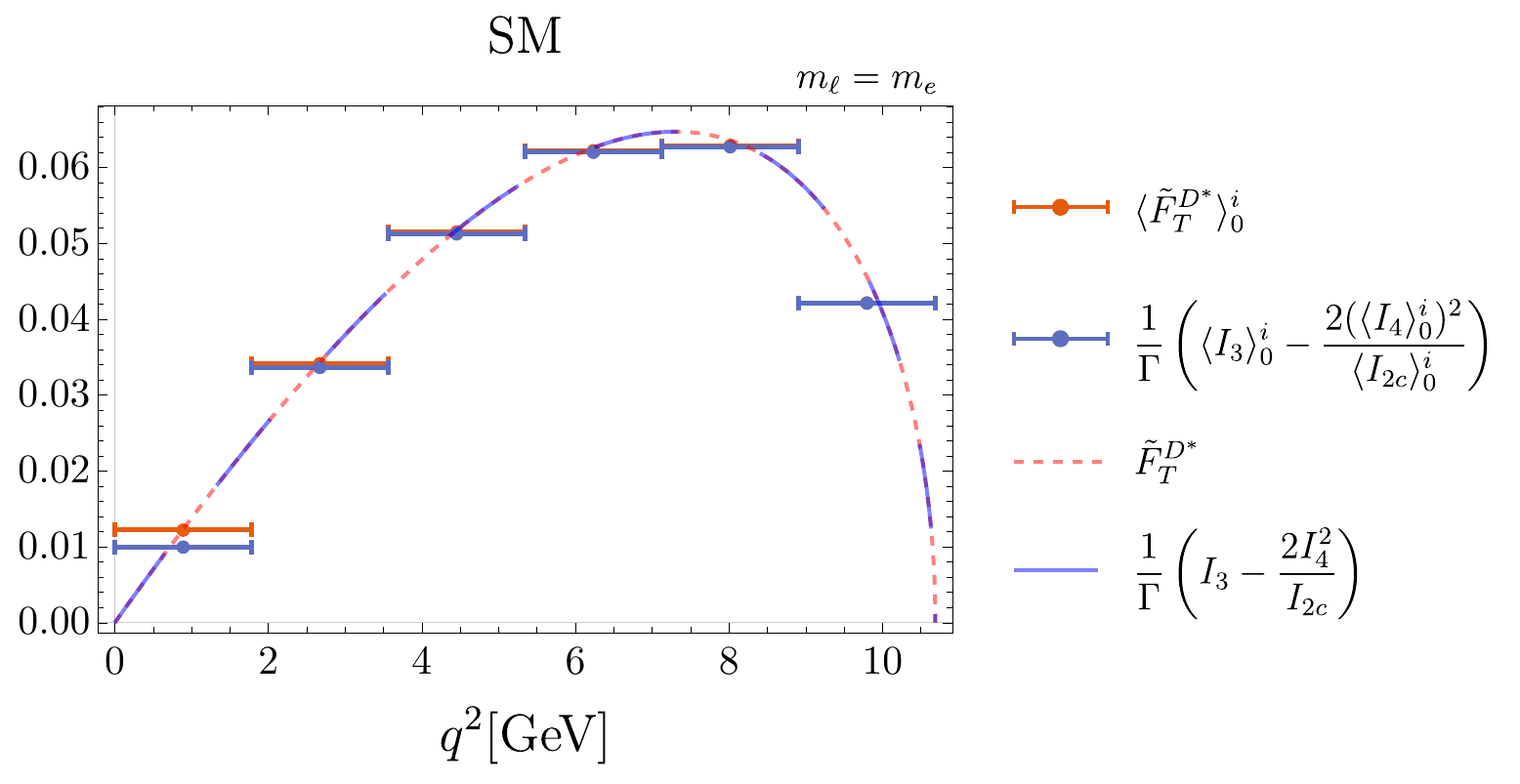}
  \label{fig:binningMassless}
 \end{subfigure}

 \begin{subfigure}{\linewidth}
  \centering
  \includegraphics[height=.285\linewidth]{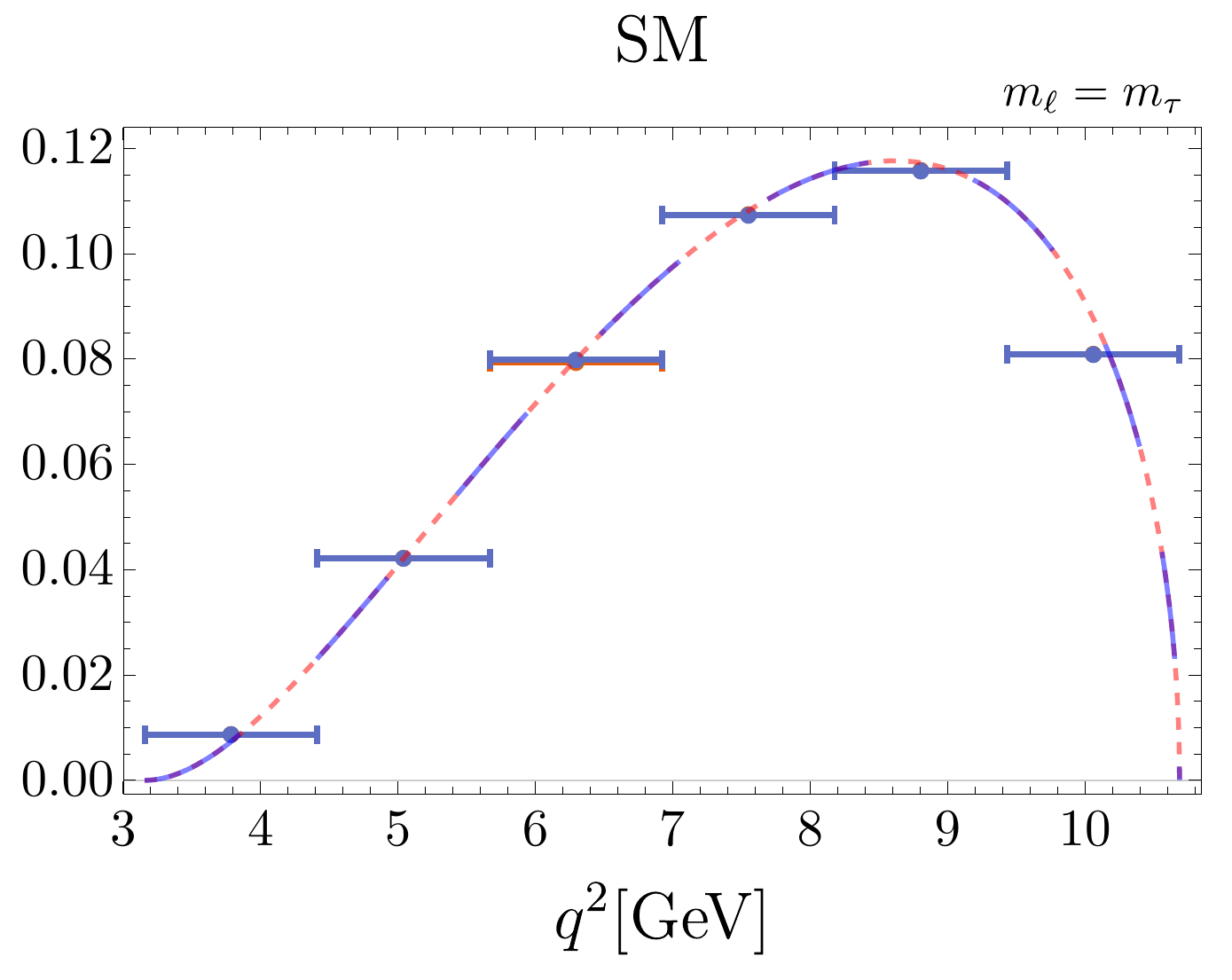}
  \hspace{10pt}
  \includegraphics[height=.285\linewidth]{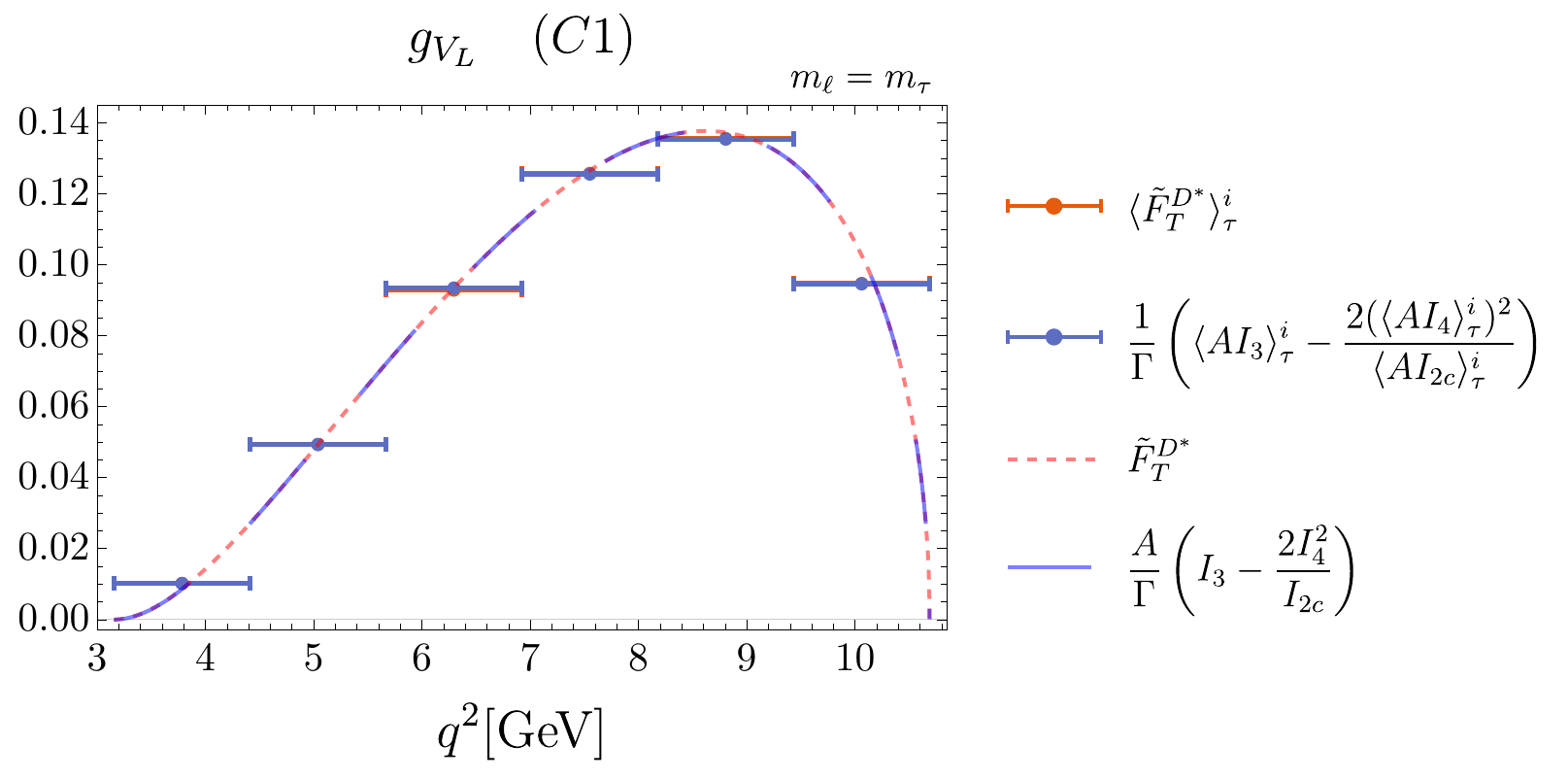}
  \label{fig:binningVector}
 \end{subfigure}

 \begin{subfigure}{\linewidth}
  \centering
  \includegraphics[height=.285\linewidth]{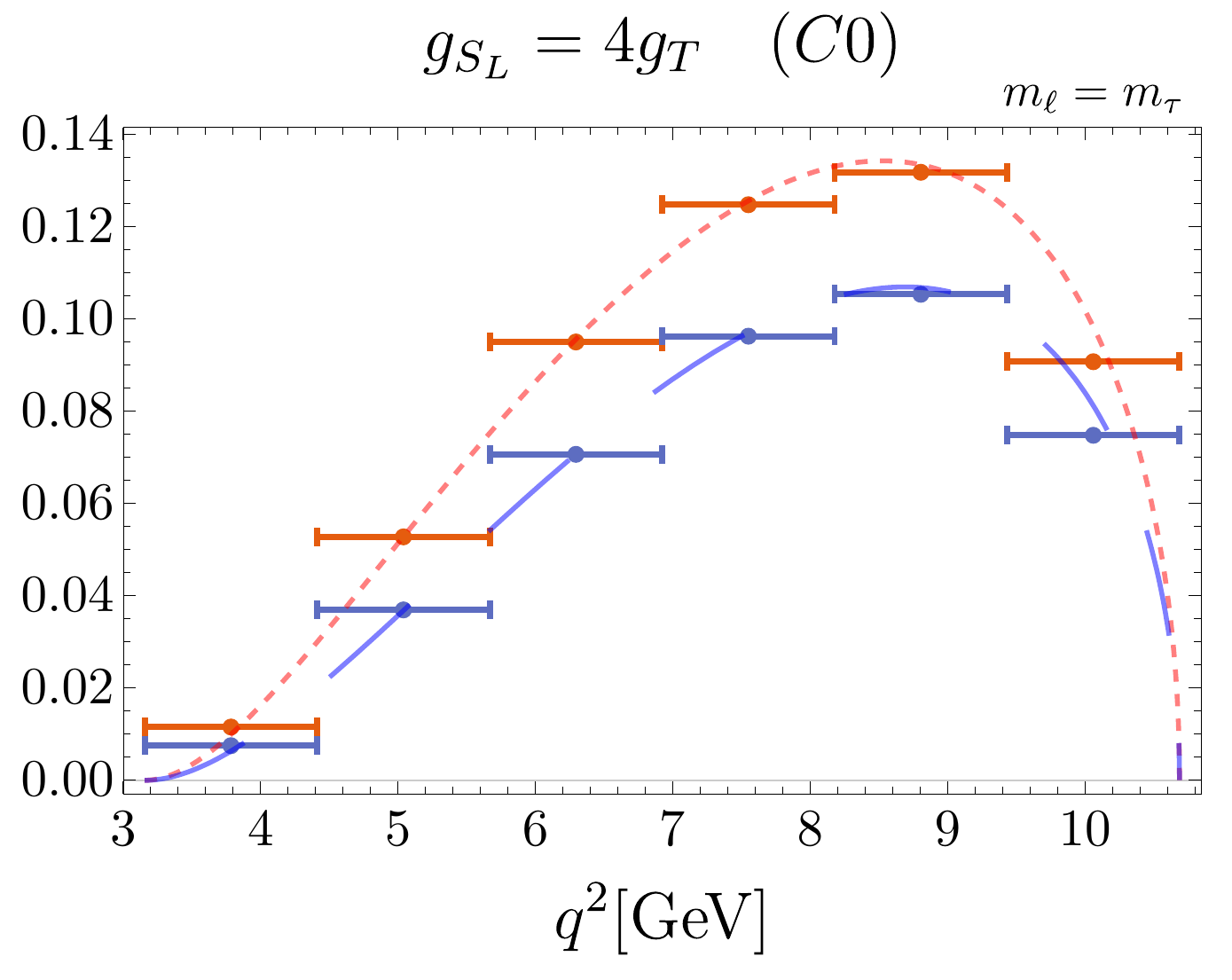}
  \hspace{10pt}
  \includegraphics[height=.285\linewidth]{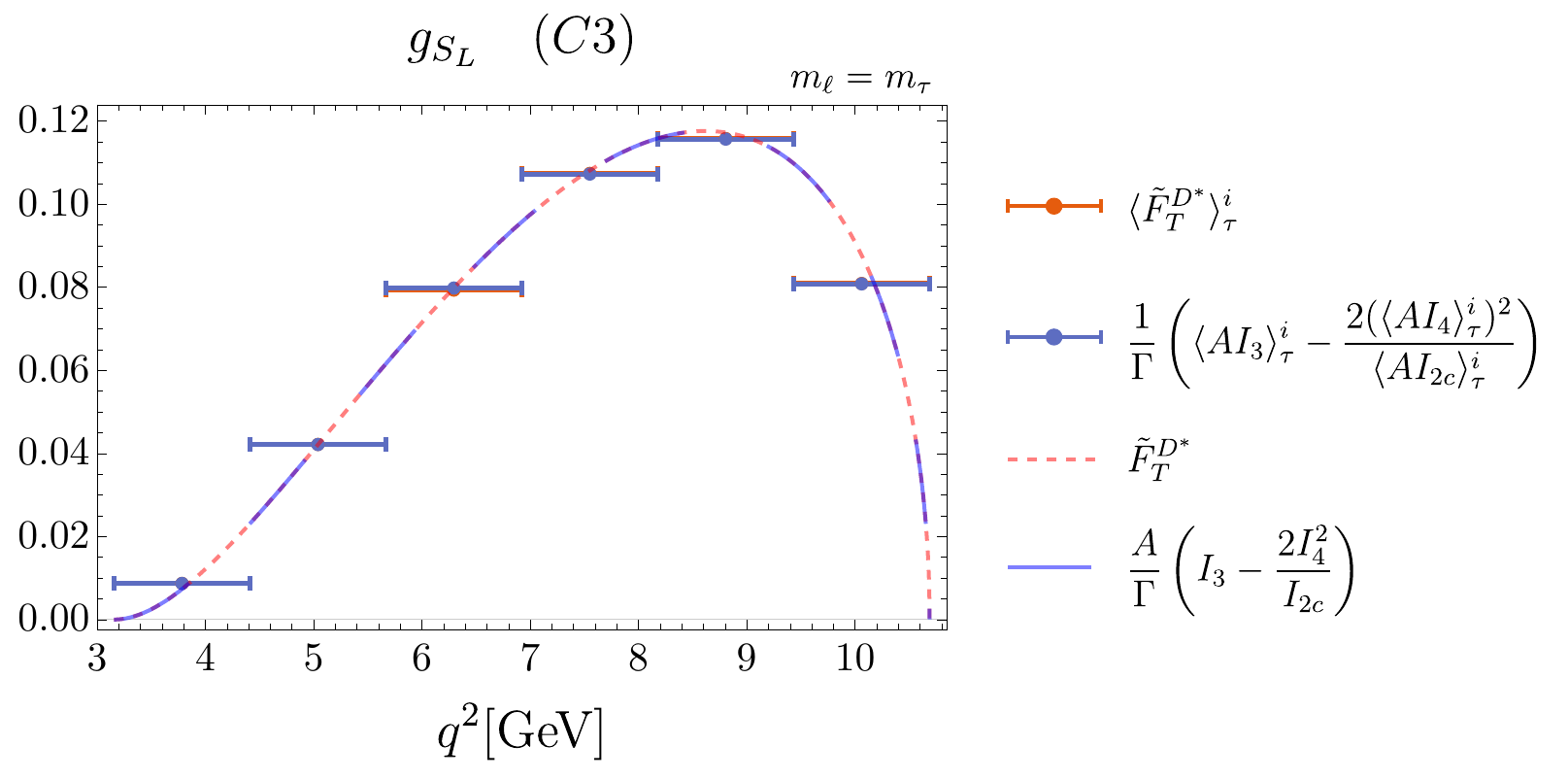}
  \label{fig:binningScalar}
 \end{subfigure}
 \caption{Same as Fig.~\ref{fig:binning} for Eq.~(\ref{binnedfLmass}).}
 \label{fig:binning2}
\end{figure}

\begin{figure}
 \centering
 \begin{subfigure}{\linewidth}
  \centering
  \includegraphics[height=.285\linewidth]{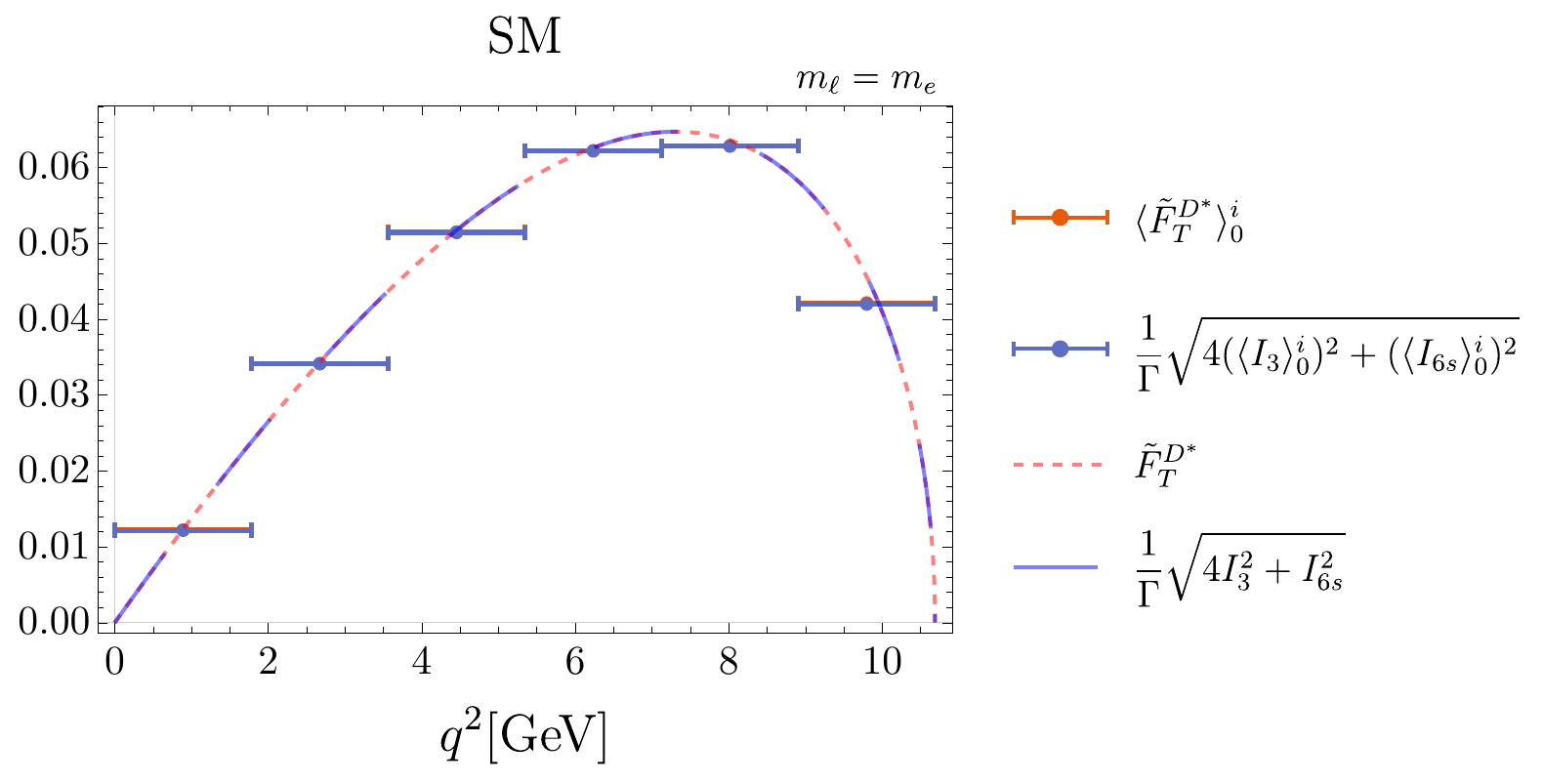}
  \label{fig:binning3Massless}
 \end{subfigure}

 \begin{subfigure}{\linewidth}
  \centering
  \includegraphics[height=.285\linewidth]{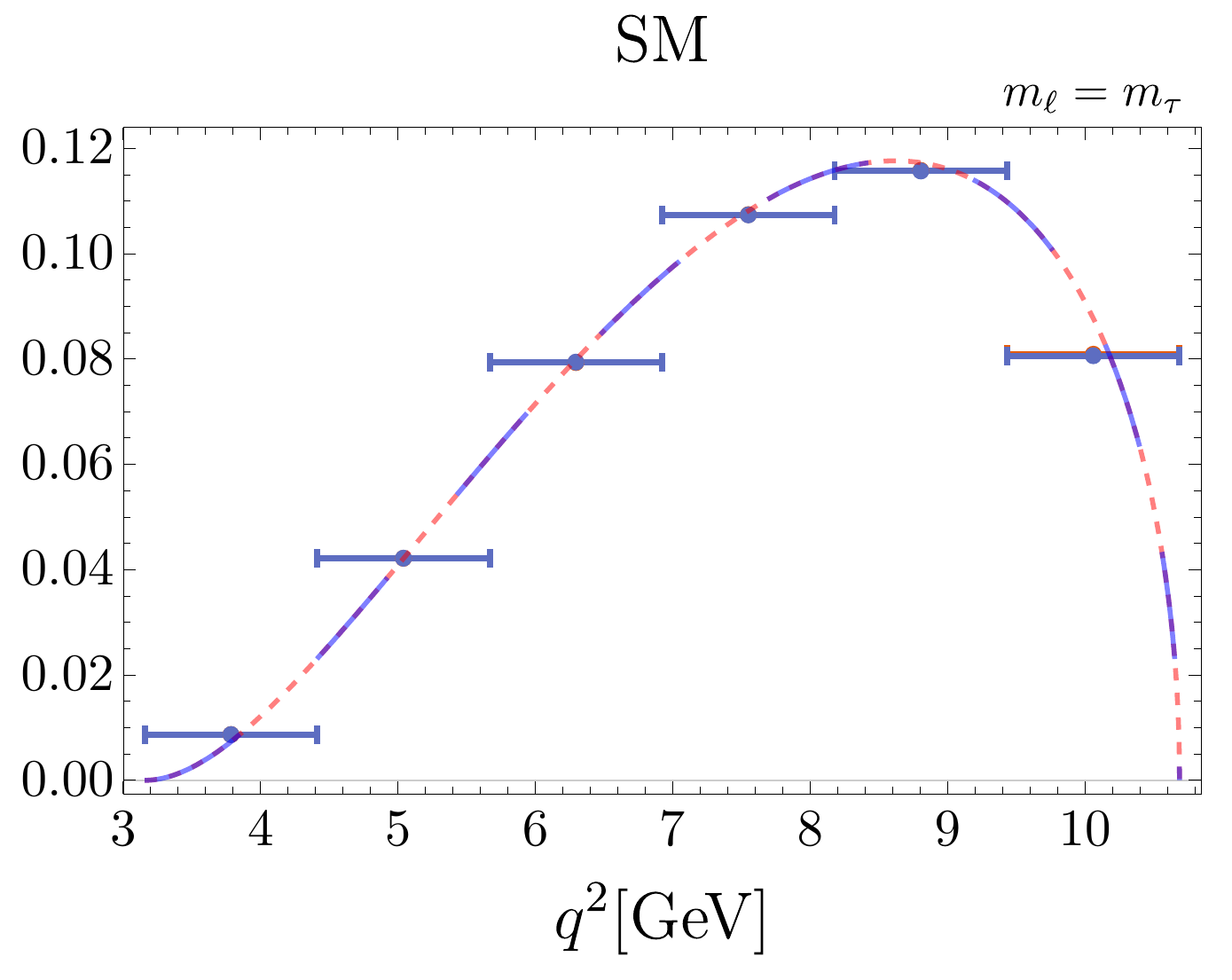}
  \hspace{10pt}
  \includegraphics[height=.285\linewidth]{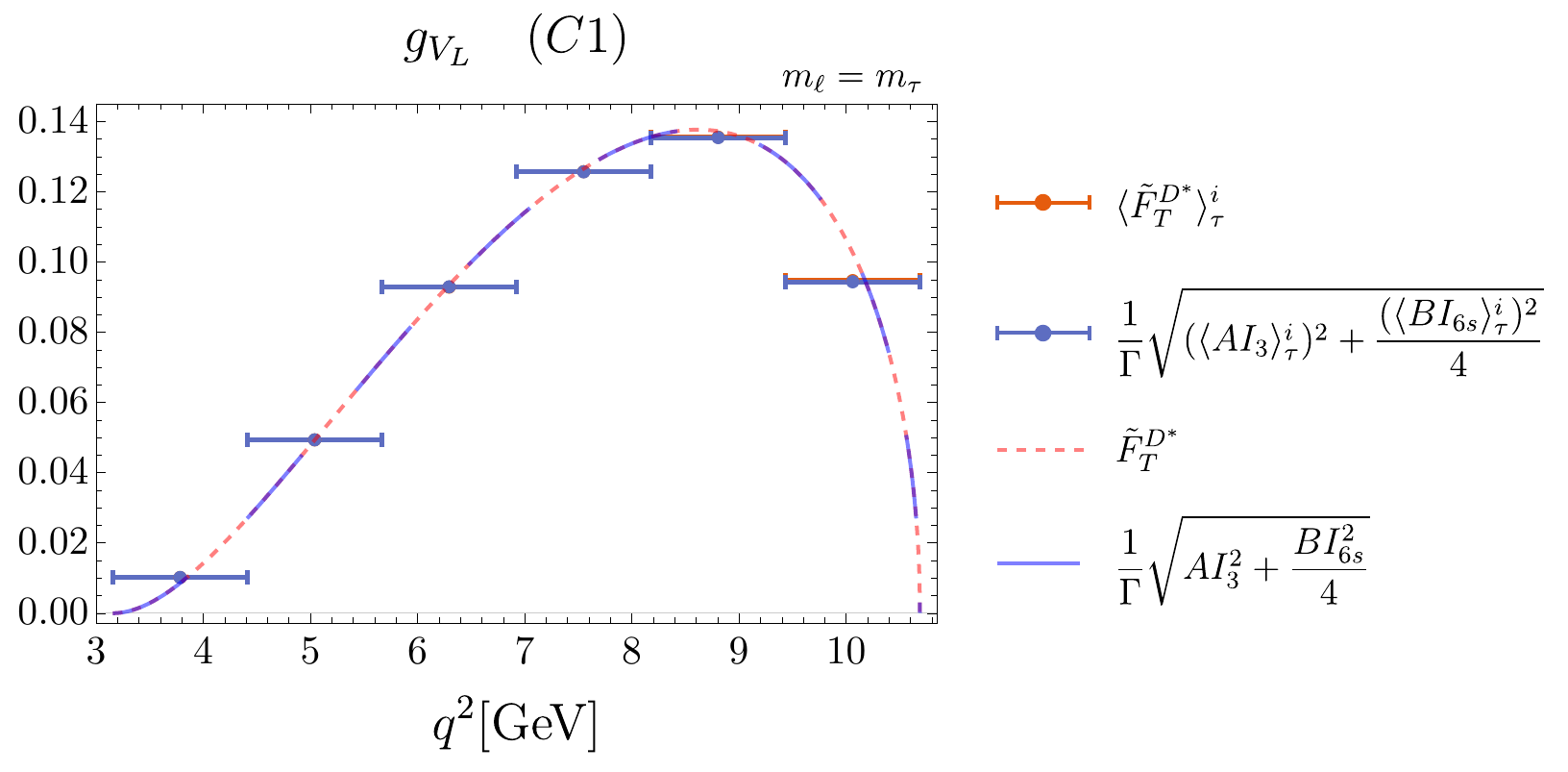}
  \label{fig:binning3Vector}
 \end{subfigure}

 \begin{subfigure}{\linewidth}
  \centering
  \includegraphics[height=.285\linewidth]{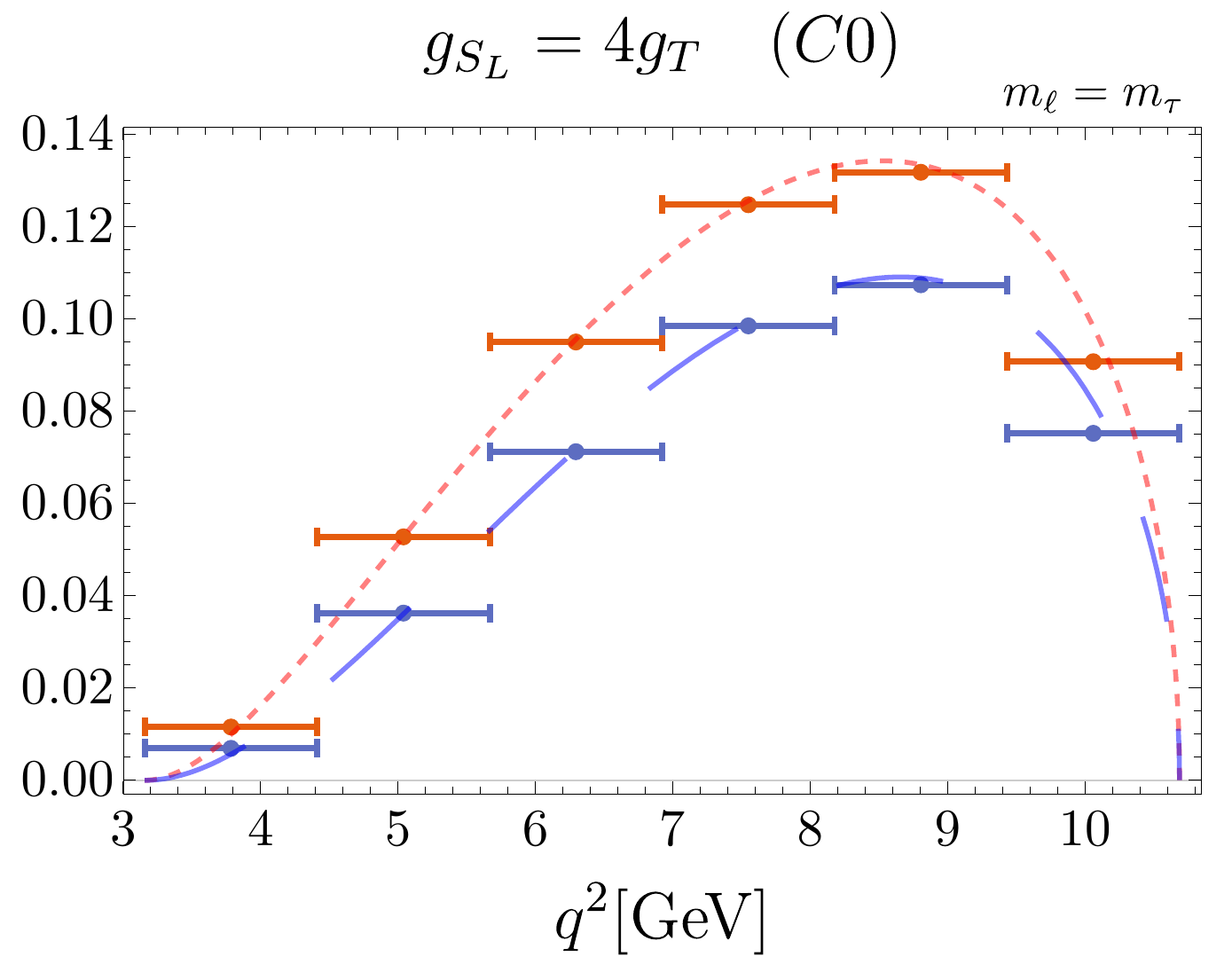}
  \hspace{10pt}
  \includegraphics[height=.285\linewidth]{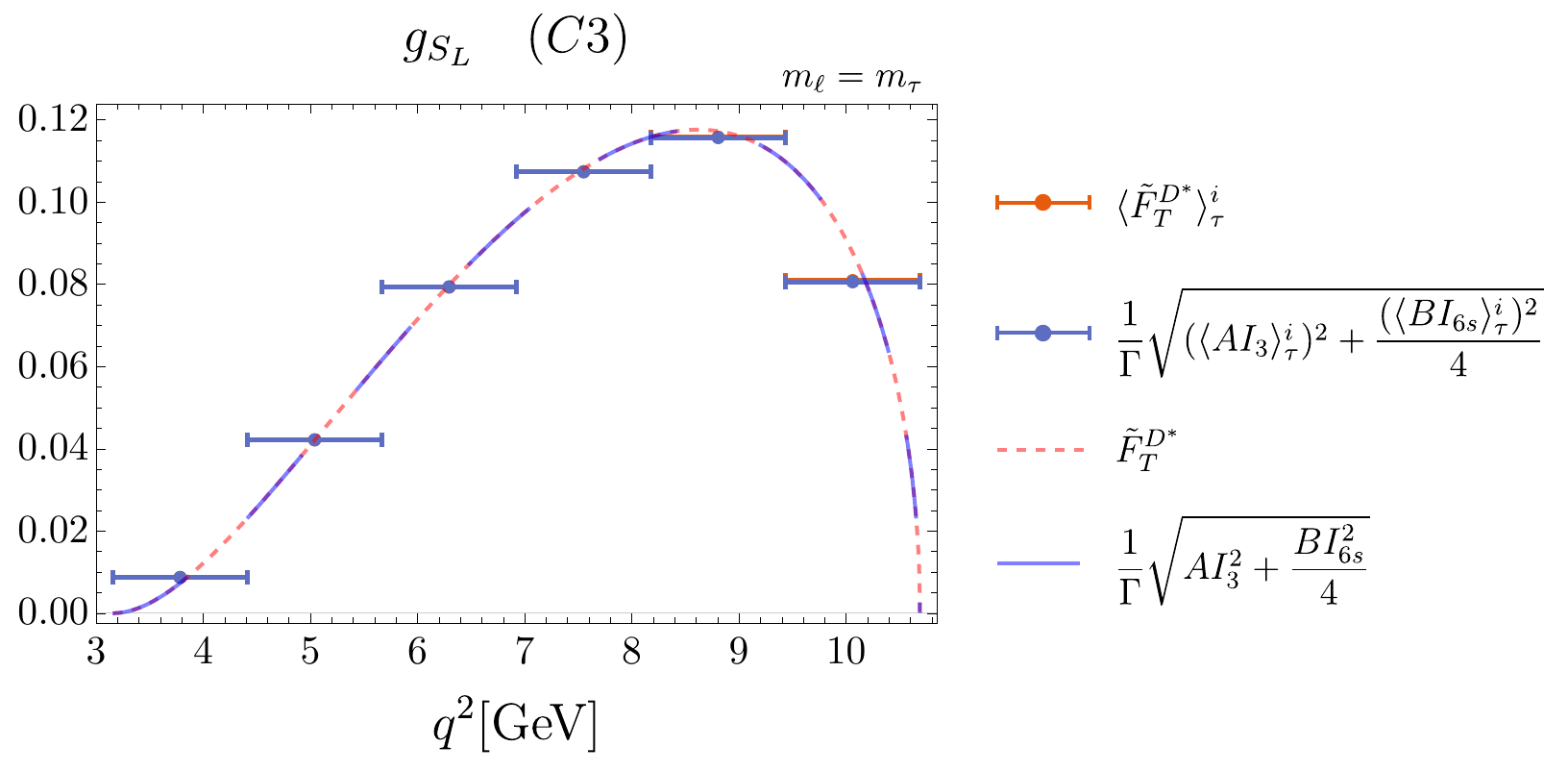}
  \label{fig:binning3Scalar}
 \end{subfigure}
 \caption{Same as Fig.~\ref{fig:binning} for Eq.~(\ref{binnedsecond}).}
 \label{fig:binning3}
\end{figure}

A sample of the results is shown in Figs.~\ref{fig:binning}, \ref{fig:binning2} and \ref{fig:binning3} to illustrate the accuracy of the determinations from Eqs.~(\ref{eq:FLbinnedmassive}) (taking into account the contribution from imaginary parts) and (\ref{binnedsecond}) or (\ref{binnedfLmass})  (neglecting this contribution). Additional scenarios are considered in App.~\ref{app:allplots}.
In order to be more precise, the relative errors of the approximate binned expression for $\tilde{F}_T^{D^* \rm alt}$ with respect to $\tilde{F}_T^{D^*}$ are given in Tabs. \ref{tab:numericalvalues1} and \ref{tab:numericalvalues2}.
Let us add that the $I_i$ are integrated with the kinematical weight $A$ or $B$ defined in Eq.~(\ref{eq:kinematicalWeights}) for the evaluation of the massive expressions whenever needed.
We obtain the following results for the benchmark points considered:
\begin{itemize}
 \item The binned approximation works very well in all cases when testing the relations in the case of scenarios where they are expected to hold. Conversely, when one considers a NP scenario with significant tensor contributions (like $(C0)$ or $(C5)$), the expressions are off by $\sim 70\%$ in the worst cases. Only when the NP contribution to the tensor coefficients is very small ($\left|g_T\right|\ll 1$), the expressions work quite well, for example $\sim 5\%$ for $(R4)$.

 \item When we consider NP scenarios for the $\tau$ lepton with complex values for the Wilson Coefficients but without tensor contributions, i.e. $(C1)-(C4)$ and $(C6)-(C9)$, the expressions hold with errors at the percent level. This occurs even when we consider the expressions meant for real coefficients (Sec.~\ref{sec:massivereal}). We stress again that this does not apply to scenarios with tensor contributions such as $(C0)$ and $(C5)$.

 \item We also tested the massless expressions in the case of NP scenarios affecting
       light leptons at the same level as the $\tau$ lepton. Such scenarios are ruled out by the current data, but they provide
       a further check of the robustness of our expressions. In these cases, the expressions that do not contain the angular coefficients containing imaginary parts of the amplitudes ($I_{7,8,9}$)
       (Sec.~\ref{sec:masslessreal}) are off by $\sim 20\%$ at worst. The agreement can be restored
       once we generalise the corresponding expressions so that they include these angular coefficients (Sec.~\ref{sec:imaginary}), where we find a perfect agreement.
 \item In the first bin of most of the massless expressions, the relations are not completely fulfilled, with a  difference up to 10\% due to binning effects enhanced at the endpoint of the massless distribution.

\end{itemize}

This study shows that the expressions derived above under the assumption of no imaginary NP contributions and no tensor contributions in Secs.~\ref{sec:masslessreal}  and \ref{sec:massivereal} work very well even in the binned approximation. They are  very accurate even in the presence of imaginary NP contributions.
Their simple generalization including imaginary parts  in Sec~\ref{sec:imaginary} are as expected to be even more accurate also in the binned approximation. Finally, all relations fail in the presence of large tensor contributions.

\subsection{Decision Tree}

We have proposed different ways of determining $F_L^{D^*}$ (or $F_T^{D^*}$) which can be compared to the usual definition, based on the existing symmetries if additional assumptions are made about the nature of NP (no tensors, real contributions). One may then wonder how to interpret the situation when the determination of $F_T^{D^*}$ in a narrow bin in the case of the tau lepton yields different results from Eq.~(\ref{second}) and from the traditional determination. While we have provided different possible determinations we will focus on Eq.~(\ref{second}) because it includes pseudoscalar contributions and it is easily generalised in the presence of
phases, see Eq.~(\ref{secondimag}). There are three possible conclusions:
\begin{itemize}
 \item[1)] Our first hypothesis is the absence (or negligible size) of tensors. In the presence of tensors, there are no dependencies among the angular observables, and we cannot use Eq.~(\ref{second}) to determine $F_T^{D^*}$. This first possibility seems to be in disagreement with the study in \cite{Becirevic:2019tpx} that shows that tensors tend to substantially worsen the situation reducing even further the value of $F_L^{D^*}$ (or increasing $F_T^{D^*}$).
       If needed, this question can be tested by probing the relationships shown in Sec.~\ref{sec:relationsangular} among the angular coefficients.
 \item[2)] The second hypothesis is the absence of large imaginary parts. In this case one can generalise the expression Eq.~(\ref{second}) to the presence of imaginary parts to get Eq.~(\ref{secondimag}), simply substituting:
       \begin{equation}
        {(A\, I_3)}^2 \to {(A\, I_3)}^2 + {(A\, I_9)}^2
       \end{equation}
       and similarly for the massless case.
       This simple substitution covers the presence of large phases but of course at the cost of measuring also $I_9$. Alternatively one can also measure $I_{7,8,9}$ which are sensitive to
       large imaginary parts and determine if they differ from zero in a significant way.
 \item[3)] The third option is the presence of an experimental issue in the determination of $F_L^{D^*}$ in the traditional way for $B \to D^* \tau \nu$. The alternative determination proposed here could help to determine the problem to be fixed and whether this second determination is also in disagreement not only with the SM but also with NP models.
\end{itemize}

\section{Impact of the presence of light right-handed neutrinos}\label{rhn}

We turn now to the analysis of a  case beyond the framework considered up to now, namely, the presence of light right-handed neutrinos (RHN) entering the decay $b \to c \tau {\bar \nu}$. The inclusion of light RHN was discussed in Refs.~\cite{Ligeti:2016npd,
 Asadi:2018wea,Greljo:2018ogz,Robinson:2018gza,Azatov:2018kzb, Heeck:2018ntp,Asadi:2018sym,Babu:2018vrl,
 Bardhan:2019ljo,Gomez:2019xfw,pich}  as a way to obey all phenomenological constraints as well as cosmological and astrophysical limits. Here we will follow closely the recent discussion in Ref.~\cite{pich} and we will use the results presented there  to generalise our expressions.

If one neglects neutrino masses, the $b \to c \tau {\bar \nu}$ decay probability is given by an incoherent sum of the contributions from left- and right-handed neutrinos. This introduces a substantial change in the structure of the angular distribution, requiring a separate discussion.

The inclusion of RHN leads to a more general dimension-six effective Hamiltonian (see Ref.~\cite{pich} for the definitions of the operators):
\begin{equation}
 {\cal H}_{\rm eff}=\frac{4 G_F V_{cb}}{\sqrt{2}} \left(O_{LL}^V+ \sum_{X=S,V,T}^{A,B=L,R} {\cal C}_{AB}^X {\cal O}_{AB}^X \right)
\end{equation}
The Wilson coefficients are defined  in such a way that ${\cal C}_{AB}^X=0$ in the SM.
Eq.~(17) of Ref.~\cite{pich} provides a translation table between our helicity basis and the transversity basis used in that reference.

The inclusion of RHN requires us to consider left and right chiralities of the leptonic current, while the hadronic current is not modified. Consequently the coefficients of the angular distribution get modified (see Ref.~\cite{pich}):
\begin{equation} \label{amp}
 { I}_j \to { I}_j(L) \pm I_j(R)
\end{equation}
where the relative sign depends on the angular observable considered, and
$I_j(L)$ and $I_j(R)$ involve different helicity amplitudes including ${\cal C}^L$ and ${\cal C}^R$ Wilson coefficients respectively. The total number of amplitudes entering the distribution gets thus enlarged from 7 to 14 (two of the helicity amplitudes always come in the same combination).

We can now discuss the impact of RHN on our previous discussion. Let us assume that there are neither tensor nor imaginary contributions, but that RHN are indeed present. We can compare the two determinations of
$\tilde{F}_T^{D^*}$:  the standard definition in  Eq.~(\ref{deffT}) and
the alternative determination in Eq.~(\ref{second}). The following relation holds:

\begin{equation} \label{dif}
 \frac{ \langle {(\tilde{F}_T^{D^* })}^2 - {(\tilde{F}_T^{D^* \, \rm alt, \,  I_9=0})}^2
 \rangle_\tau}{\langle\tilde{(B A_{6s})^2}\rangle_\tau}
 \!=\!\Delta^F\!\equiv\! \frac{64}{729}  \frac{(C^V_{LR}(1+ C^V_{LL})- C^V_{RL} C^V_{RR})^2}{ {((1+ C^V_{LL})}^2-{C^V_{LR}}^2-{C^V_{RL}}^2+{C^V_{RR}}^2)^2}
\end{equation}
where $\tilde{A}_{6s}$ refers to the observable  including left and right components defined by
\begin{equation}
 \langle{\tilde A}_{6s}\rangle_\tau=-\frac{27}{8} \frac{1}{\Gamma} \langle I_{6s}\rangle_\tau
 \label{difbin}
\end{equation}
In order that the previous expression becomes useful we  have checked that Eq.(\ref{dif}) still holds in the following binned form:\footnote{
We have scanned over a range of values of the RHN coefficients $C^V_{LL,LR,RL,RR}
$ to compare Eq.({\ref{dif})} and Eq.(\ref{dif2}). The result of this test clearly indicates that for combinations of RHN resulting in reasonably small values of $\Delta_F<1$, the two expressions agree up to ${\cal O}(10^{-3})$ corrections in all bins.}

\begin{equation} \label{dif2}
 \frac{  {(\langle\tilde{F}_T^{D^* }\rangle_\tau)}^2 - {(\langle\tilde{F}_T^{D^* \, \rm alt}\rangle^{\rm I_9=0}_\tau)}^2
 }{\langle\tilde{B A_{6s}}\rangle_\tau^2}
 \!\simeq\!\Delta^F
\end{equation}
Notice that given that $\Delta^F$ is always positive, Eq.(\ref{dif2}) implies that  an experimental determination using $\langle \tilde{F}_T^{D^* \, \rm alt}\rangle^{\rm I_9=0}_\tau$
should always be found equal or smaller than the ``standard'' $\langle \tilde{F}_T^{D^* } \rangle_\tau$ in absence of tensors and imaginary contributions.

We derived this expression assuming the hypotheses above and using the fact that Eq.~(\ref{eq:trivial}) is valid in presence of RHN while Eq.~(\ref{eq:dependency3}) 
holds if the constraint
\begin{equation} \label{constraint}
 C^V_{LR} (1 + C^V_{LL}) - C^V_{RL} C^V_{RR}=0
\end{equation}
is imposed.  In other words, only if this constraint is fulfilled, $\langle \tilde{F}_T^{D^* \, \rm alt}\rangle_\tau$ can be interpreted as the physical transverse polarization fraction.

In Ref.~\cite{pich} several interesting scenarios  are identified which are able to fulfill the constraints from ${\cal B}_{B_c \to \tau \bar{\nu}}$, ${\cal R}_{D,D^*}$, $F_L^{D^*}$ and ${\cal P}_\tau^{D^*}$:

\begin{itemize}
 \item[1)] The scenario with the highest pull$_{\rm SM}$ corresponds to scenario 3 ($V_\mu$) with NP only in $C^V_{RR}$. Since  $C^V_{LL}=C^V_{LR}=C^V_{RL}=0$  in this scenario, Eq.~(\ref{constraint}) is fulfilled and $\Delta^F=0$. However, in this scenario the NP contributions to $F_L^{D^*}$ cancel exactly and the tension with the experimental value is not relaxed.

 \item[2)] A second interesting scenario is called 4b ($\Phi_b$) in Ref.~\cite{pich}. This scenario can be generated by a two Higgs doublet model and it yields non-zero values for $C^S_{X}$ with $X=LL,LR,RL,RR$. Assuming ${\cal B}_{B_c \to \tau \bar{\nu}}<30\%$, this scenario is able to relax the tensions of all observables including $F_L^{D^*}$. Since this scenario yields NP contributions only in $C^S_i$ it fulfills automatically the constraint, leading to $\Delta^F=0$.

 \item[3)] In scenario 1 of Ref.~\cite{pich}, there are two solutions with non-vanishing values for $C_{LL,LR,RR}^V$ as well as $C^S_{LR,RR}$ and $C^T_{RR}$. One of the two solutions has a tensor contribution compatible with zero at 1$\sigma$. If we take this solution to remain under our initial hypothesis of the absence of tensor contributions we obtain $\Delta^F \sim  10^{-3}$ (central value of b.f.p)  if $C_{RL}^V=0$, which, obviously, cannot be detected. In Ref.~\cite{pich} the coefficient $C^V_{RL}$ is neglected because it is lepton-flavour universal within SMEFT and it cannot help to accommodate any of the deviations observed with LFUV observables.
 However, assuming the best-fit point of this scenario does not change when non-vanishing values of $C_{RL}^V$ are allowed, we find that $\Delta^F$ can be much larger when $C_{RL}^V$ approaches $\pm \sqrt{(1+C_{LL}^V)^2-{C_{LR}^{V}}^2+{C_{RR}^{V }}^2}$,  leading to a rather  visible effect.

\end{itemize}

In summary, a difference between the two measurements of $F_T^{D^*}$ (or $F_L^{D^*}$) in absence of tensors and imaginary contributions could be attributed, barring experimental issues, to contributions coming from RHN. For some  RHN scenarios, this would generate a non-zero value for $\Delta^F$.

\section{Experimental sensitivity} \label{expsensitivity}

 Our analysis is based on the possibility of performing a full angular analysis of the $B\to D^*\ell\nu$
 with a reasonable accuracy to check the relationships derived among angular observables.
There is a major experimental challenge associated to the difficulty of measuring angular distributions of semitauonic decays due to the loss of the two neutrinos, one from the B decay and the other from the subsequent $\tau$ decay, making it difficult to reconstruct the $\tau$ direction. This problem arises both when the $\tau$ decays into a pion or a lepton~\cite{Ligeti:2016npd,Alonso:2016gym}.
A novel approach~\cite{ Hill:2019zja} has been proposed using the three-prong $\tau^+\to \pi^+ \pi^+ \pi^- {\bar \nu}_\tau$ decay instead of the muonic $\tau$ decay and a multidimensional template fit able to measure the coefficients of the angular distribution.
We can use the numerical results from Ref.~\cite{Hill:2019zja} to compare the expected experimental sensitivity of $F_L^{D^*}$ using the standard definition in  Eq.~(\ref{deffT}) with the one using
the alternative determination in Eq.~(\ref{second}).\footnote{We refrain from using the more complete alternative definition in Eq~(\ref{secondimag})  because the ratio ${\langle A\, I_{9}\rangle}/{\langle I_{9}\rangle}$ necessary to get the rough estimate described in the text is not properly defined in the SM.}

Taking the results of
the template fit for the $50\ {\rm fb}^{-1}$ collider scenario given in Tab.~11 and Fig.~10 of Ref.~\cite{Hill:2019zja} and applying the transformation described in Eq.~(\ref{errorbin}) we can obtain a rough estimate of the sensitivity of $\langle \tilde{F}_L^{D*\, \rm alt}\rangle^{\rm I_9=0}_\tau$.
Obtaining this estimate is not straightforward since $\langle \tilde{F}_L^{D*\, \rm alt}\rangle^{\rm I_9=0}_\tau$ includes not only the angular observables $I_3$ and $I_{6s}$ but also the kinematic factors $A$ and $B$. As mentioned in Sec.~\ref{sec:binning}, experimentalists can measure directly $A\, I_3$, and $B\, I_{6s}$  following the same binning as the angular observables arising in the differential branching ratio.
In order to get a rough idea of these quantities in the absence of a dedicated experimental study including estimates of $A I_3$ and $B I_{6s}$, we study the ratios ${\langle A\, I_{3}\rangle}/{\langle I_{3}\rangle}$ and ${\langle B\, I_{6s}\rangle}/{\langle I_{6s}\rangle}$ and how they change in the presence of NP.
Scanning the parameter space, we find these ratios to be rather independent of the NP considered.  We find that ${\langle A\, I_{3}\rangle}/{\langle I_{3}\rangle}\approx 4.1$ and ${\langle B I_{6s}\rangle}/{\langle I_{6s}\rangle}\approx2.4$, leading to our approximate determination of the binned observables
\begin{equation}
 \langle A\, I_3\rangle_{\rm exp}  \approx 4.1 \langle I_3\rangle_{\rm exp} \qquad
 \langle B\, I_{6s}\rangle_{\rm exp}  \approx 2.4 \ \langle I_{6s}\rangle_{\rm exp}
\end{equation}
It is important to emphasise that this approximation would not be needed for future experimental measurements as long as $A\, I_3$, $A\, I_9$ and $B\, I_{6s}$ are measured directly.

Under these approximations and considering the uncertainties and correlations given for the $50\ {\rm fb}^{-1}$ collider scenario in Ref.~\cite{Hill:2019zja}, we obtain the following rough estimate for the alternative determination for the SM case considered in this reference
\begin{equation}
 \langle \tilde{F}_L^{D*\, \rm alt}\rangle_{50\ {\rm fb}^{-1}}^{\rm I_9=0}= 0.47\pm 0.12
\end{equation}
to be compared with the standard determination
\begin{equation}
 \langle \tilde{F}_L^{D*}\rangle_{50\ {\rm fb}^{-1}}= 0.45\pm 0.01
\end{equation}
The alternative determination suffers from the larger errors of the angular observables involved in its definition, in comparison
with the standard determination which is dominated by $I_{1s}$ with a smaller uncertainty than the other angular observables, as show in Fig. 10 of Ref.~\cite{Hill:2019zja}.

These uncertainties would be enough to identify discrepancies coming from tensor contributions, such as our scenario C5. The  smaller differences between the two determinations coming from other types of scenarios (such as Wilson coefficients with imaginary parts) could not be distinguished and the two determinations should yield similar results. Conversely, it means that our relations will provide a non-trivial experimental cross-check of the angular analyses projected in Ref.~\cite{Hill:2019zja}, unless large tensor contributions are present.

\section{Conclusions}\label{sec:conclusions}

The charged-current $B \to D^*\ell\nu$ transition has been under scrutiny recently, as it exhibited
a deviation from the SM in the LFUV ratio $R_{D^*}$ comparing the branching ratios $\ell=\tau$ and lighter leptons. Moreover, the polarisation of both the $D^*$ meson and the $\tau$ lepton have been measured for $B \to D^*\tau\nu$.
If the latter agrees with the SM within large uncertainties, the Belle measurement  of $F_L^{D^*}$ yields a rather high value compared to the SM prediction, which appears difficult to accommodate with NP scenarios.

We could understand better this situation by considering in more detail the angular observables that could be extracted from the differential decay rate, as described in Ref.~\cite{Becirevic:2019tpx}.
We applied the formalism of amplitude symmetries of the angular distribution of the decays $B \to D^* \ell \nu$ for $\ell=e,\mu,\tau$. We showed that the set of angular observables used to describe the distribution of this class of decays are not independent in absence of NP contributing to tensor operators. We derived sets of relations among the angular coefficients of the decay distribution for the massless and massive lepton cases.
These relations  can be used to probe in a very general way the consistency among the angular observables and the underlying NP at work, and in particular whether it involves tensor operators or not.

We used these relations to access the integrated longitudinal polarisation fraction of the $D^*$ using different angular coefficients from the ones used by Belle experiment. This in the near future can provide
an alternative strategy to measure $F_L^{D^*}$ for $B\to D^*\tau\nu$ and  to understand the relatively high value measured by Belle. We presented expressions in Eqs.~(\ref{firstimag}) and (\ref{secondimag}) for the massless and massive case that cover the most general NP scenario including also pseudoscalars and imaginary contributions, with the only exception of tensor contributions.

We then studied the accuracy of these expressions if only binned observables are available, or if they are used in the case of scenarios beyond the assumptions made in their derivation (imaginary contributions, tensor contributions). We used several benchmark points corresponding to best-fit points from global fits to $b\to c\tau\nu$ observables, relying on a simple quark model for the hadronic form factors for this exploratory study.
The expressions derived under the assumption of no imaginary NP contributions and no tensor contributions work very well even in the binned approximation. They are very  accurate even in the presence of imaginary NP contributions. As expected, their generalisations, derived assuming the presence of imaginary contributions, are very well behaved also in the binned approximation. All relations fail in the presence of large tensor contributions, where no dependencies can be found among the angular observables.

Besides presenting the most general expressions for $F_L^{D^*}$  in the massless and massive case, we also derived  a relation among observables ($\tilde{A}_{3,9,6s}$ and $F_L^{D^*}$) that are potentially  interesting  from the NP point of view if the deviation in $F_L^{D^*}$ is confirmed. Having specific model building predictions for these observables would be highly interesting.
We also discussed the impact of the presence of light right-handed neutrinos. We showed that we could test their presence in some specific cases under the hypothesis that there are no tensor nor imaginary contributions, by comparing our two determinations of $F_L^{D^*}$. Moreover, under this hypothesis, the sign of the difference between the two determinations is fixed.

We have explored alternative determinations of $F_L^{D^*}$ based on our symmetries. In the absence of tensor contributions, these determinations based on other angular observables are fulfilled very accurately. This provides an important cross check for the experimental measurements: if our relations are not fulfilled by the experimental measurements, this would mean either a problem on the experimental side or the presence of large tensor contributions.
Using recent projections on the experimental prospects for the measurements of angular observables, we find that these relations could be checked with an accuracy of 0.1 in the scenario of a 50 fb$^{-1}$ hadron collider, which would be enough to spot a scenario with tensor contributions and would provide an interesting cross-check of the determination of the angular observables.

These additional measurements needed for this extraction make obviously this determination more challenging experimentally, but they can help to corner the kind of NP responsible for this high value or to understand the experimental problem responsible for this unexpected value of the $D^*$ polarisation.
We hope that our results will be of particular interest once the LHCb and Belle II experiments are able to analyse the $B\to D^*\ell\nu$ decays in more detail and thus to provide us with a more detailed picture of the intriguing deviations currently observed in $b\to c\ell\nu$ transitions.

\section*{Acknowledgments}

This work received financial support from European Regional Development Funds under the Spanish Ministry of Science, Innovation and Universities (project FPA2017-86989-P) and from the Research Grant Agency of the Government of Catalonia (project SGR 1069) [MA, JM] and from European Commission (Grant Agreements 690575, 674896 and 69219) [SDG]. JM acknowledges TH-Division at CERN where part of this work was done and
gratefully acknowledges the financial support by ICREA under the ICREA Academia programme.

\appendix

\section{Explicit dependencies in the massive case}
\label{app:massivecase}

In this appendix we provide the detailed methodology followed and the full expressions of the dependencies among the angular coefficients in the massive case with no tensor contributions. It is useful to define the following four combinations in order to obtain compact expressions:

\begin{equation}
 R_{s,d}={\rm Re}(H_{+})\pm {\rm Re}(H_{-}) \, , \quad  I_{s,d}={\rm Im}(H_{+})\pm {\rm Im}(H_{-}) \label{eqRsd}
\end{equation}

One can solve the system of equations in terms of the variables defined above and find a twofold solution:
\begin{eqnarray}
 R_s&=&\frac{1}{ H_0} \frac{I_4 q^2}{q^2-m_\ell^2} \\
 I_d&=&\frac{1}{ H_0} \frac{I_8 q^2}{q^2-m_\ell^2} \\
 R_d&=&(-1)^n \frac{ q^2 \left(I_4 I_8 q^2 + H_0^2 I_9 (q^2-m_\ell^2)\right)}{\sqrt{H_0^2 (q^2-m_\ell^2)^2}\sqrt{-I_4^2 q^4+H_0^2 (m_\ell^2-q^2)\big[(|H_-|^2+|H_+|^2)(m_\ell^2-q^2)+I_3 q^2)\big]}} \qquad \\
 I_s&=&(-1)^n \frac{\sqrt{-I_4^2 q^4+H_0^2 (m_\ell^2-q^2)\big[(|H_-|^2+|H_+|^2)(m_\ell^2-q^2)+I_3 q^2)\big]}}{\sqrt{H_0^2 (q^2-m_\ell^2)^2}}
\end{eqnarray}
with $n=0,1$. However, this sign ambiguity product of the twofold nature of the solution can be fixed, since physical combinations prevent interference terms that could be problematic. This set of solutions can be used to determine the square of the four amplitudes once $H_0$ is fixed to be real and positive through the symmetry of the angular distribution.
One can also rewrite the real and imaginary parts of $H_t$ in terms of the variables in Eq.~(\ref{eqRsd}) and $H_0$:
\begin{eqnarray}
 {\rm Re}(H_{t})&=&-\frac{q^2 \left[ I_7\, I_s+ I_5 R_d -2 H_0 (I_s^2+R_d^2) \right]}{2 m_\ell^2 \left(I_d I_s + R_sR_d \right)}  \\
 {\rm Im}(H_{t})&=&\frac{q^2 \left[ -I_5I_d+ I_7 R_s +2  H_0 (I_d R_d-R_s I_s) \right]}{2 m_\ell^2 \left(I_dI_s + R_sR_d \right)}
\end{eqnarray}
With these definitions, one can find the whole set of dependencies among angular coefficients. Besides the trivial dependency Eq.~(\ref{eq:trivial}), there are four more relations which are obtained by taking combinations of the modulus of $H_+$, $H_-$ and ${\rm Re}(H_t)$, ${\rm Im}(H_t)$.

The first non-trivial relation can be derived from the sum $|H_+|^2+|H_-|^2$:
\begin{eqnarray}
 0&=& \frac{m_{\ell}^2-q^2}{2a}\Big\{-4 I_{1s}^2 I_{2c} (m_\ell^2 - q^2)^2 + 4 I_{1s}(I_4^2+I_8^2)(m_\ell^2-q^2)(m_\ell^2+3q^2) \nonumber \\
 &+& \left[-2 I_{3} I_{4}^2+ 2 I_{3} I_8^2-4 I_4 I_8 I_9 + I_{2c} (I_3^2+I_9^2) \right] (m_\ell^2+3 q^2)^2\Big\}
\end{eqnarray}
where
\begin{equation}
 a=(m^2_{\ell}-q^2)^2(m^2_{\ell}+3q^2)\big[2 I_{1s}I_{2c}(m^2_{\ell}-q^2) + (I_{2c}I_3-2 I^2_4) (m^2_{\ell}+3q^2)\big] \label{eqa}
\end{equation}

From $|H_+|^2|H_-|^2$ one can obtain the second dependency:
\begin{equation} \label{seconddep}
 0=-I_3^2-I_9^2+\left(1-\frac{m_\ell^2}{q^2}\right)^2 \left[ \left(\frac{2 I_{1s}}{3 + m_\ell^2/q^2} \right)^2-\frac{I_{6s}^2}{4}\right]
\end{equation}

The third one follows from $[{\rm Re}(H_t)]^2$:
\begin{eqnarray}
 0&=&\frac{8 q^4}{a}\bigg[ 2 I_{1s}I_{2c}I_7 (m_\ell^2-q^2)+(I_{2c} I_{3} I_7 - 2 I_4^2 I_7 + 2 I_4 I_5 I_8 - I_{2c} I_5 I_9) (m_\ell^2+3 q^2)\bigg]^2 \nonumber \\
 &-&\left[\frac{I_{6s} I_{6c}}{2} -\frac{4 q^4}{a} \Big(4 I_{1s}^2 I_{2c}^2 (m_\ell^2-q^2)^2+4 I_{1s} I_{2c} (I_{2c} I_3-2 I_4^2)(m_\ell^2-q^2)(m_\ell^2+3 q^2)\right. \nonumber \\
  &+&(4 I_4^2 (I_4^2 + I_8^2) - 4 I_{2c} I_4 (I_3 I_4 + I_8 I_9) + I_{2c}^2 (I_3^2 + I_9^2)) (m_\ell^2 + 3 q^2)^2\Big)\bigg]^2 \label{eq:reHt}
\end{eqnarray}
with $a$ defined in Eq.~(\ref{eqa}).

Finally, the last dependency is related to $[{\rm Im}(H_t)]^2$:
\begin{eqnarray}
 0&=& 256 I_{6s}^2 (I_4 I_7 - I_5 I_8)^2 (m_\ell^2 - q^2)^4 q^{12} \nonumber\\
 &&\qquad\qquad \times \big[I_{6c}^2 (m_\ell^2 - q^2)^2 + 8 I_{1c} I_{2c} m_\ell^2 (-m_\ell^2 + q^2) + 8 I_{2c}^2 m_\ell^2 (m_\ell^2 + q^2)\big]\nonumber \\
 &&+ \big[64b-64 (I_4 I_7 - I_5 I_8)^2 (m_\ell^2 - q^2)^2 q^8 +I_{6s}^2 (m_\ell^2 - q^2)^2 q^4 (I_{6c}^2 (m_\ell^2 - q^2)^2 \nonumber\\
 &&\qquad\qquad+ 8 I_{1c} I_{2c} m_\ell^2 (-m_\ell^2 + q^2)+8 I_{2c}^2 m_\ell^2 (m_\ell^2 + q^2))\big]^2
\end{eqnarray}
with
\begin{equation}
 b=\frac{ 2 q^{12} (2 I_{1s} I_{2c} I_4 (m_\ell^2 - q^2) + (-2 I_4 (I_4^2 + I_8^2) + I_{2c} (I_3 I_4 + I_8 I_9)) (m_\ell^2 + 3 q^2))^2}{(m_\ell^2 + 3 q^2) (2 I_{1s} I_{2c} (m_\ell^2 - q^2) + (I_{2c} I_3 - 2 I_4^2) (m_\ell^2 + 3 q^2))}
\end{equation}

As a final comment, let us remark that these dependencies among angular coefficients yield Eqs.~(\ref{eq:dependency3})-(\ref{eq:dependency5}) when one considers only real Wilson coefficients, so that all imaginary contributions and $I_{7,8,9}$ can be neglected.

\section{Comparison of the binned expressions in benchmark NP scenarios} \label{app:allplots}

\begin{figure}
 \centering
 \begin{subfigure}{\linewidth}
  \centering
  \includegraphics[height=.285\linewidth]{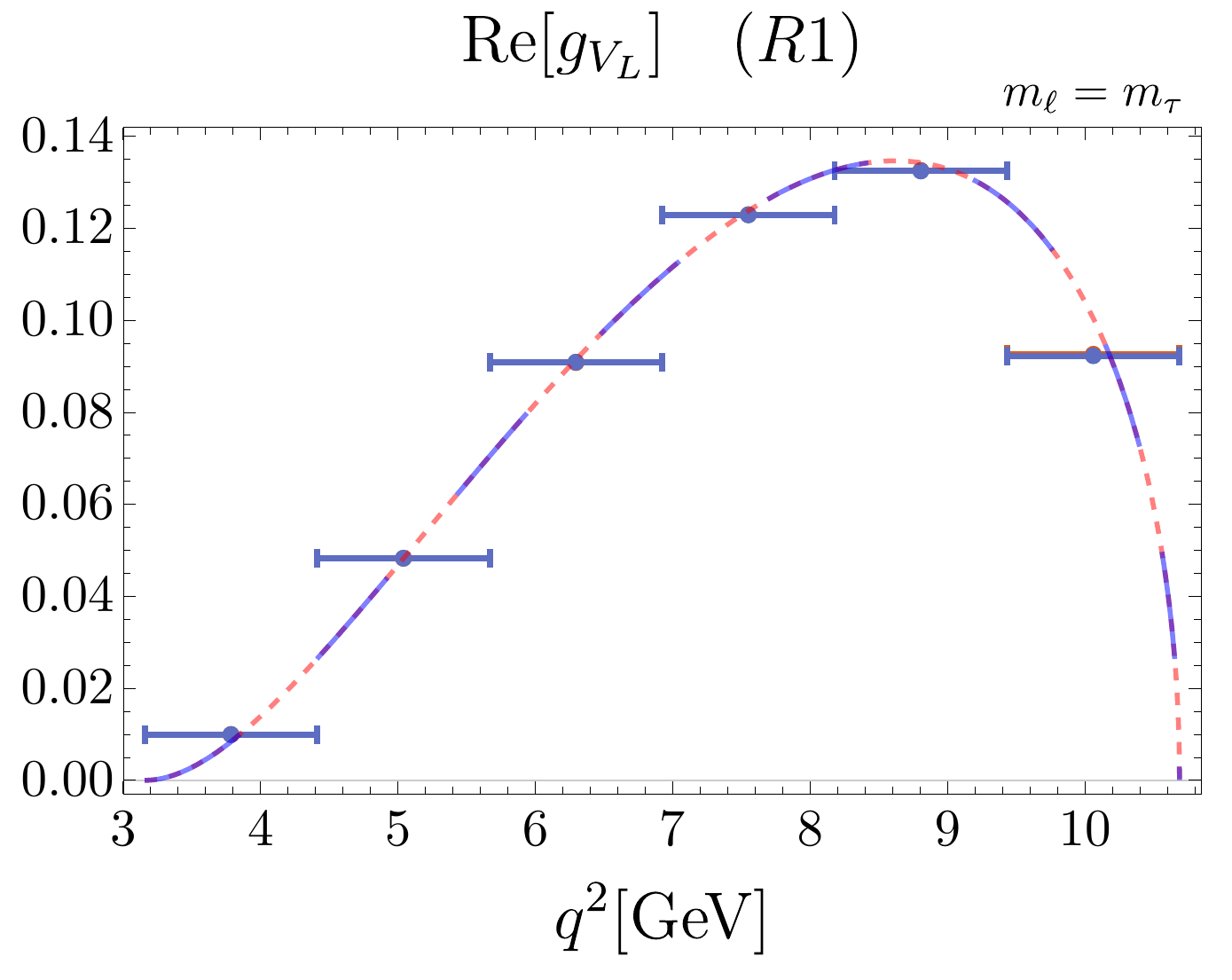}
  \hspace{10pt}
  \includegraphics[height=.285\linewidth]{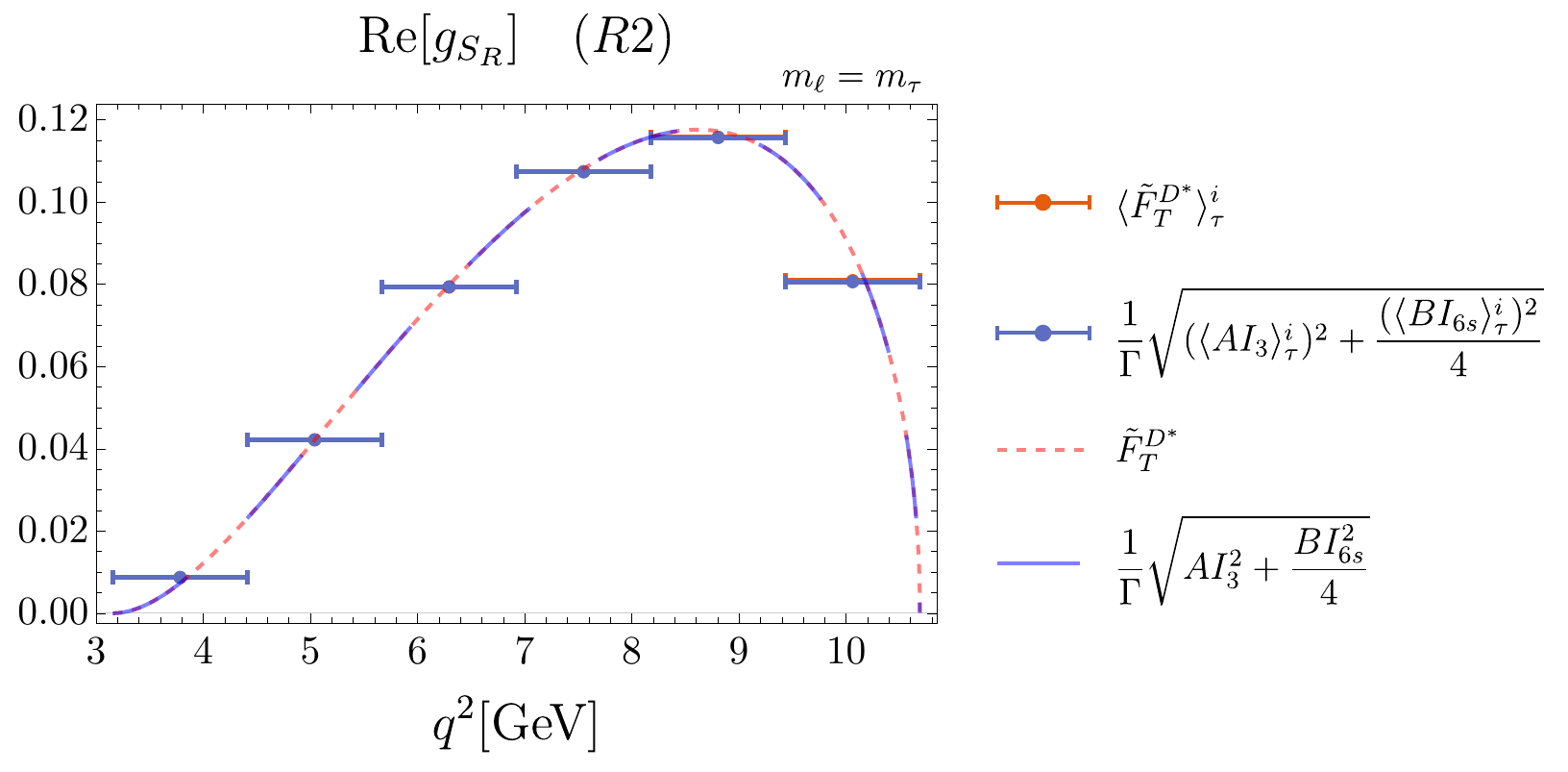}
 \end{subfigure}

 \begin{subfigure}{\linewidth}
  \centering
  \includegraphics[height=.285\linewidth]{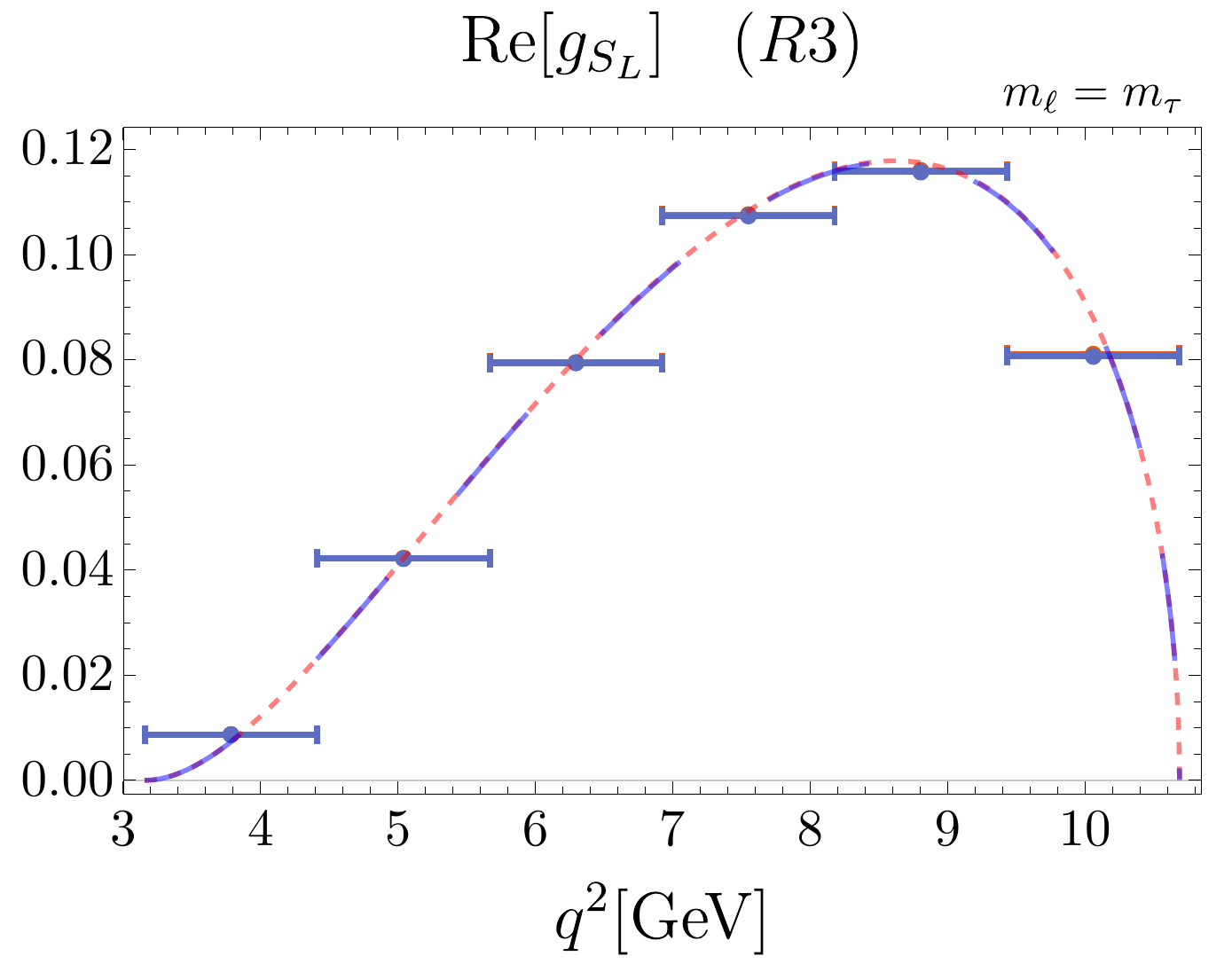}
  \hspace{10pt}
  \includegraphics[height=.285\linewidth]{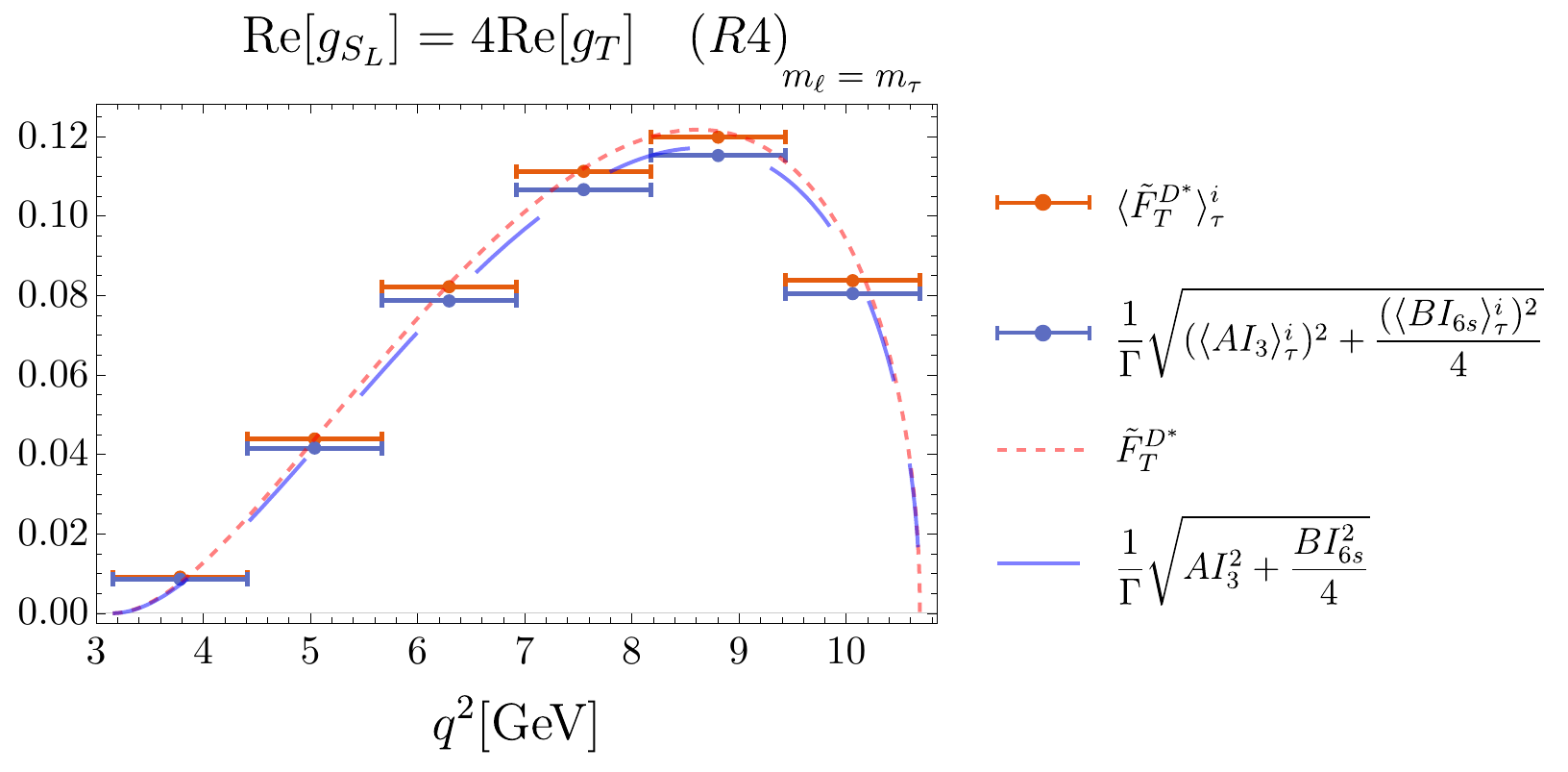}
 \end{subfigure}

 \begin{subfigure}{\linewidth}
  \centering
  \includegraphics[height=.285\linewidth]{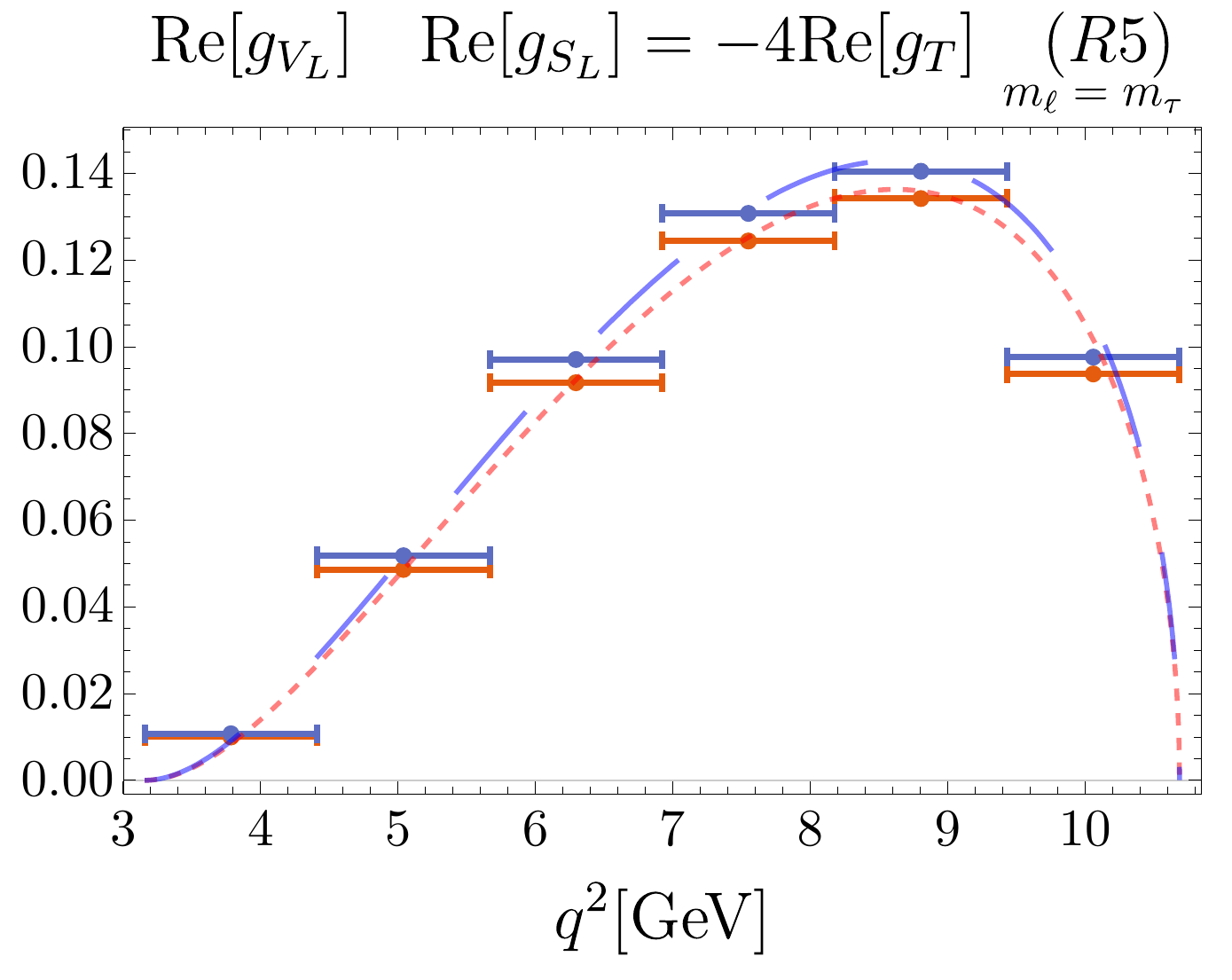}
  \hspace{10pt}
  \includegraphics[height=.285\linewidth]{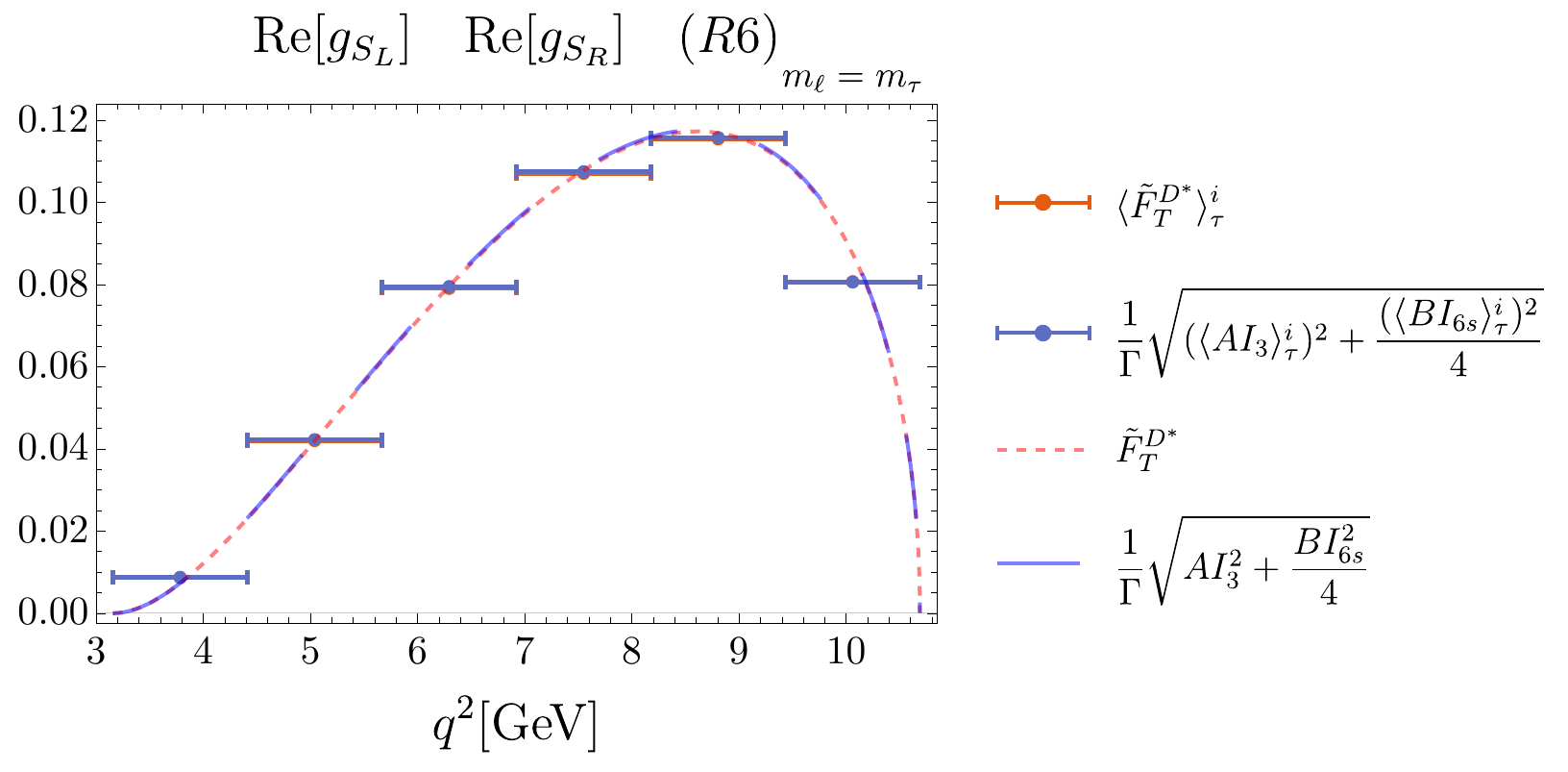}
 \end{subfigure}
 \begin{subfigure}{\linewidth}
  \centering
  \includegraphics[height=.285\linewidth]{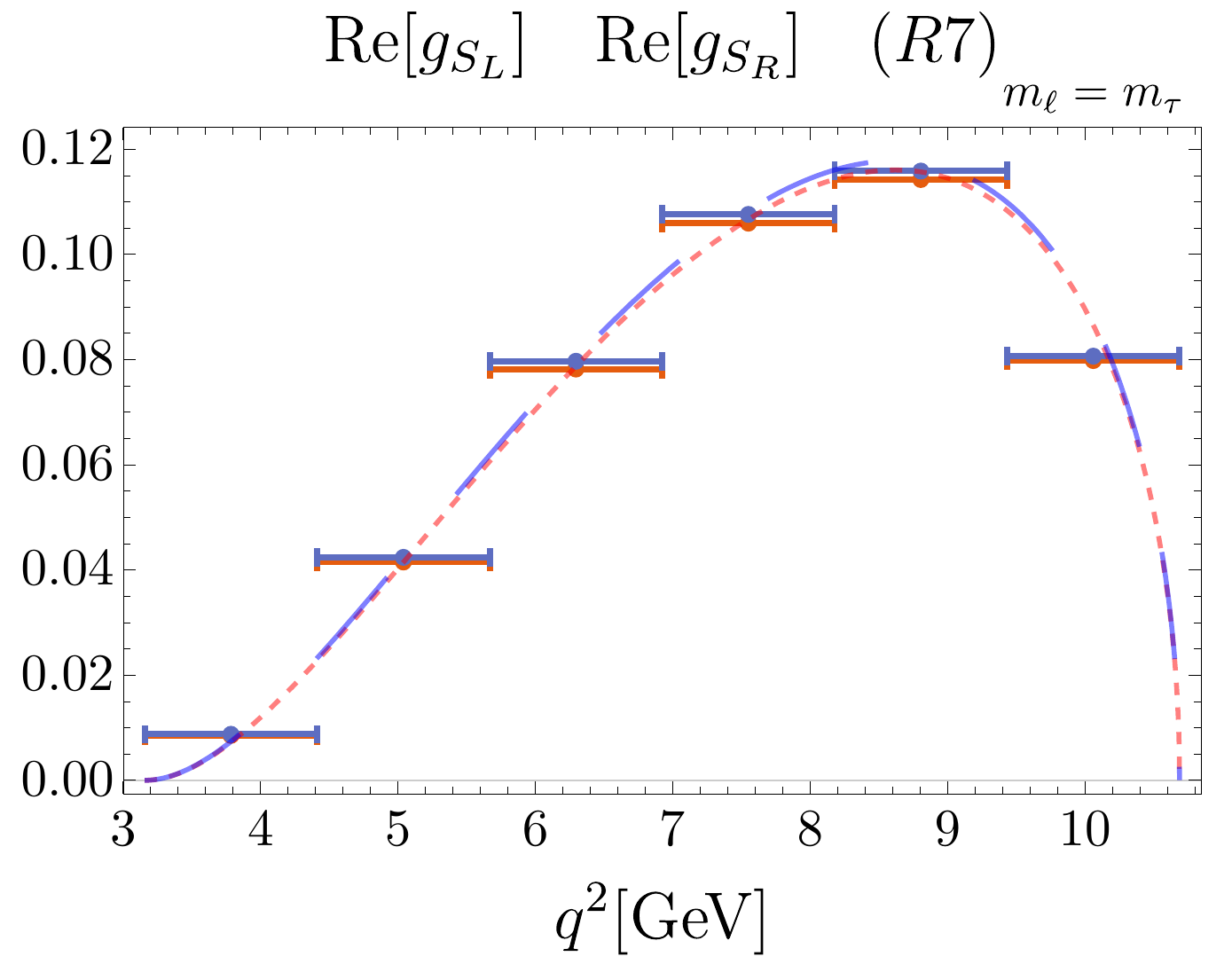}
  \hspace{10pt}
  \includegraphics[height=.285\linewidth]{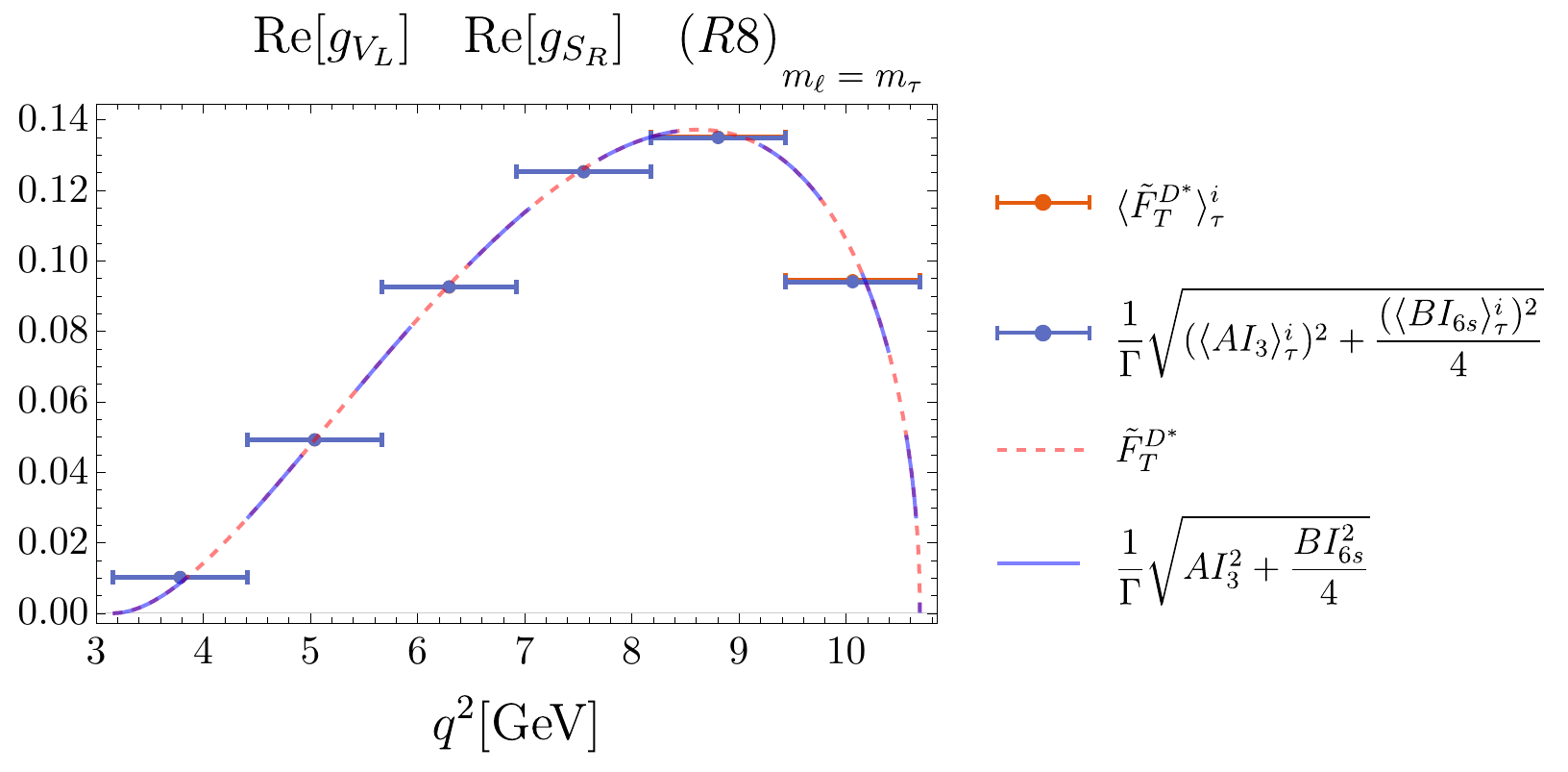}
 \end{subfigure}

 \caption{ Study of binning effects for Eq.~(\ref{binnedsecond}) for benchmark NP scenarios with real contributions.}
 \label{fig:appendix1}
\end{figure}

\begin{figure}
 \centering

 \begin{subfigure}{\linewidth}
  \centering
  \includegraphics[height=.285\linewidth]{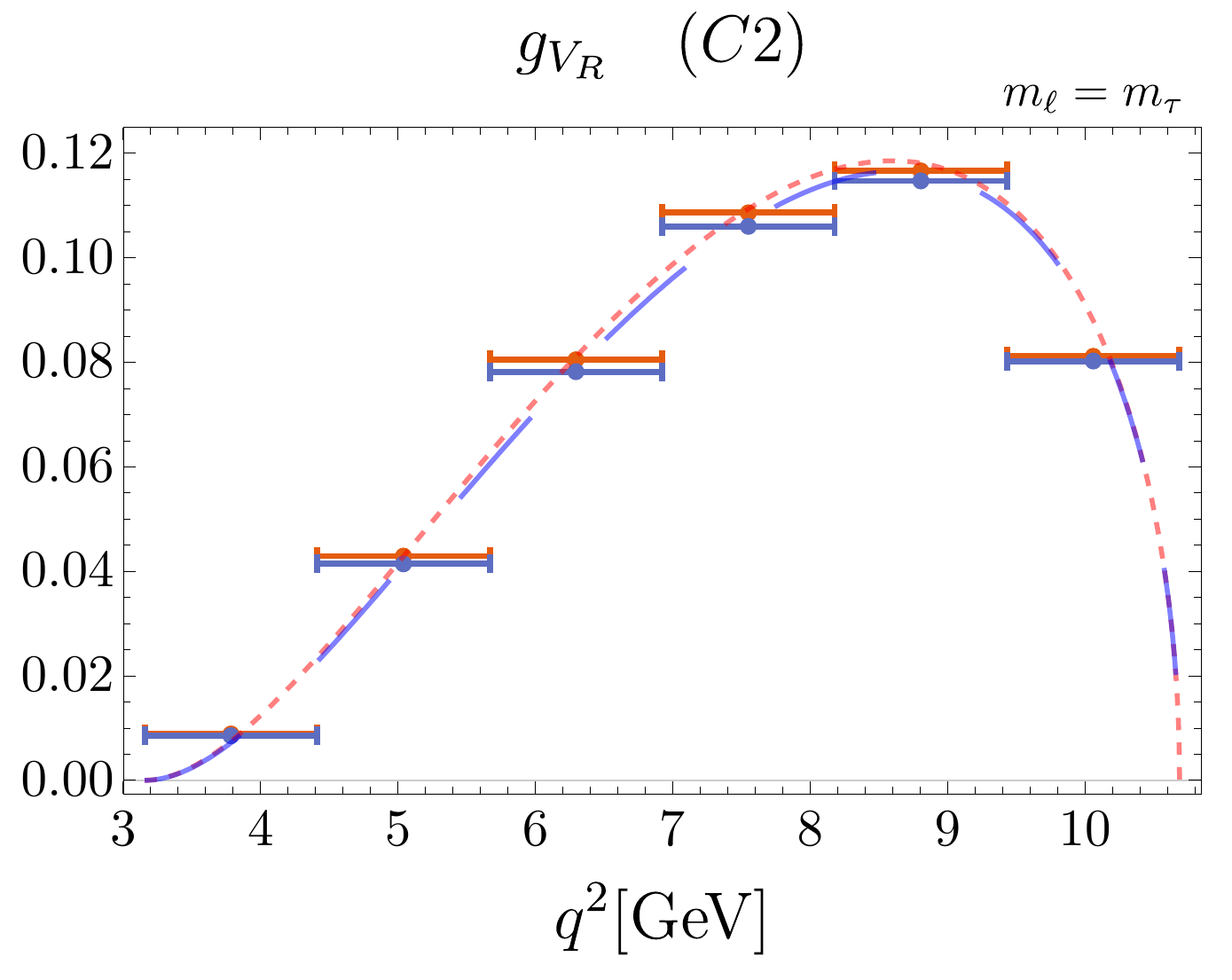}
  \hspace{10pt}
  \includegraphics[height=.285\linewidth]{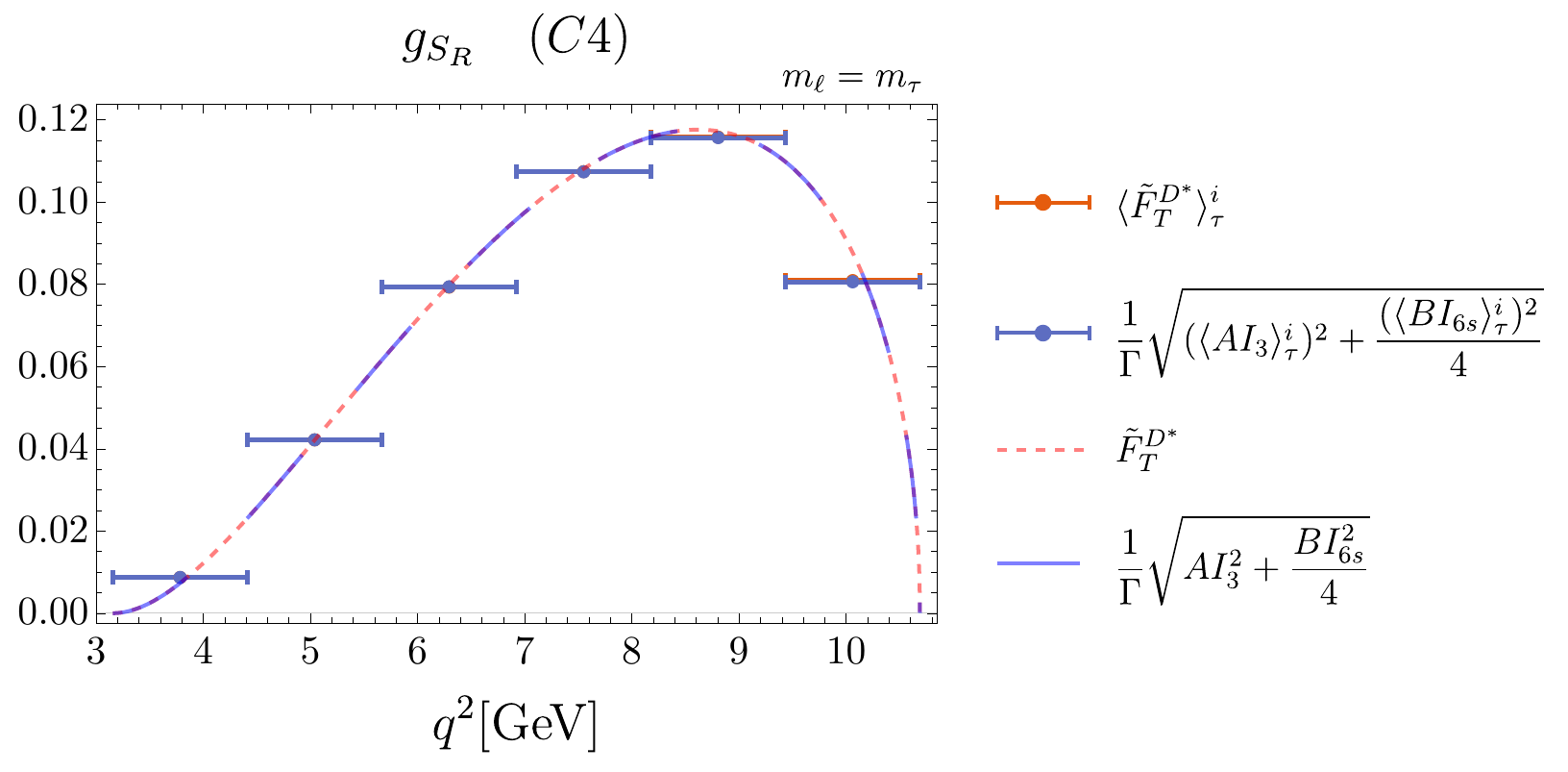}
 \end{subfigure}
 \begin{subfigure}{\linewidth}
  \centering
  \includegraphics[height=.285\linewidth]{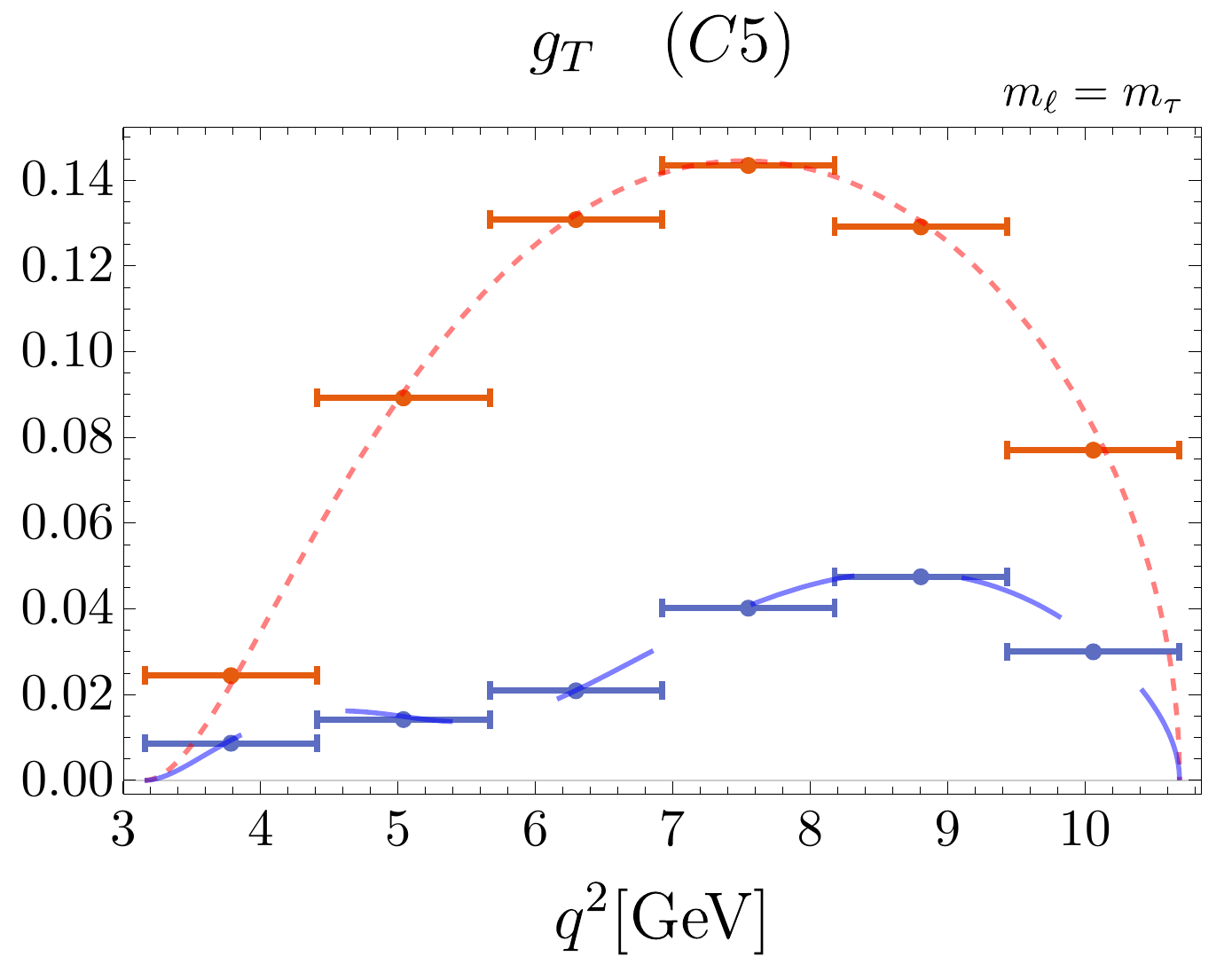}
  \hspace{10pt}
  \includegraphics[height=.285\linewidth]{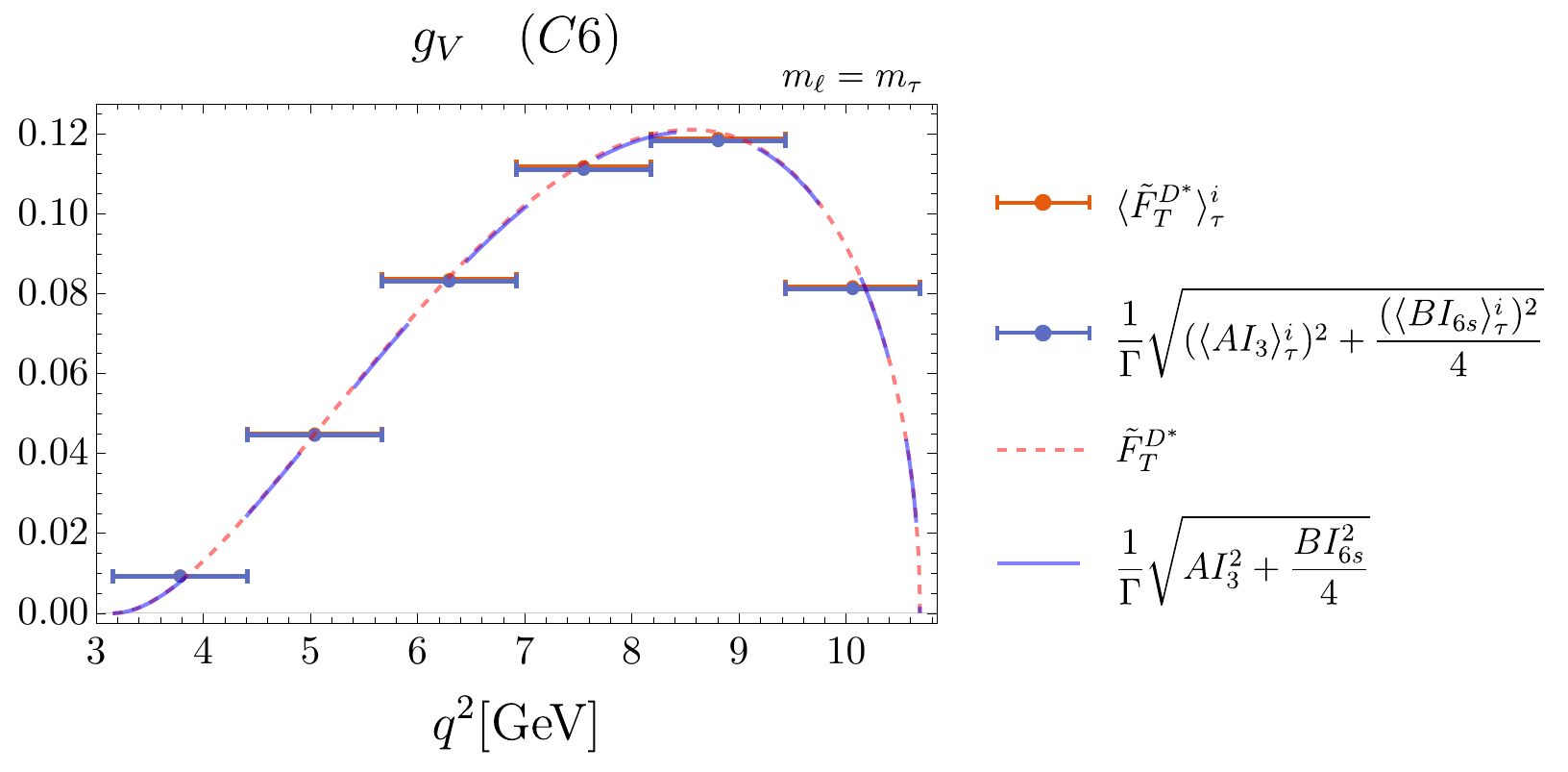}
 \end{subfigure}
 \begin{subfigure}{\linewidth}
  \centering
  \includegraphics[height=.285\linewidth]{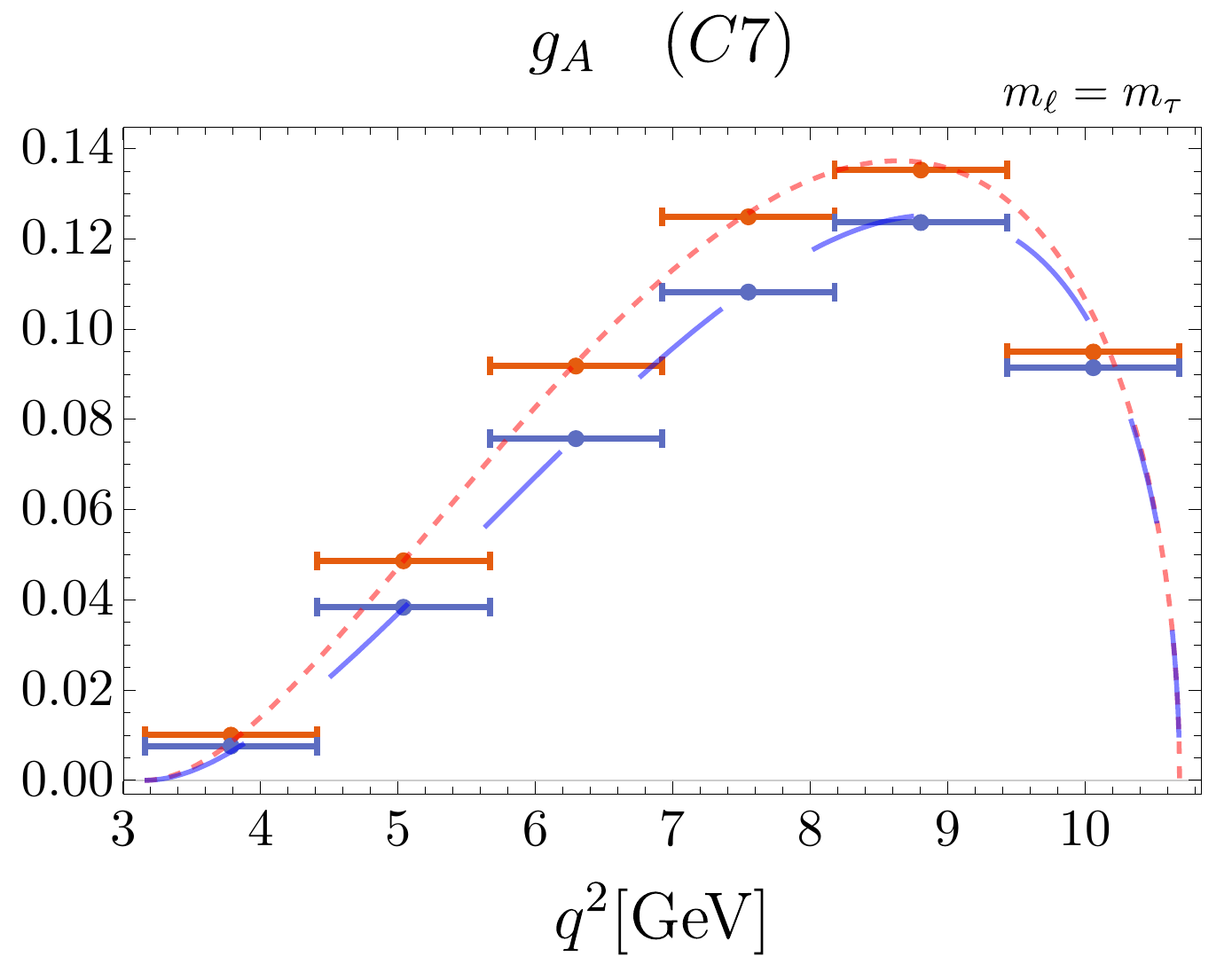}
  \hspace{10pt}
  \includegraphics[height=.285\linewidth]{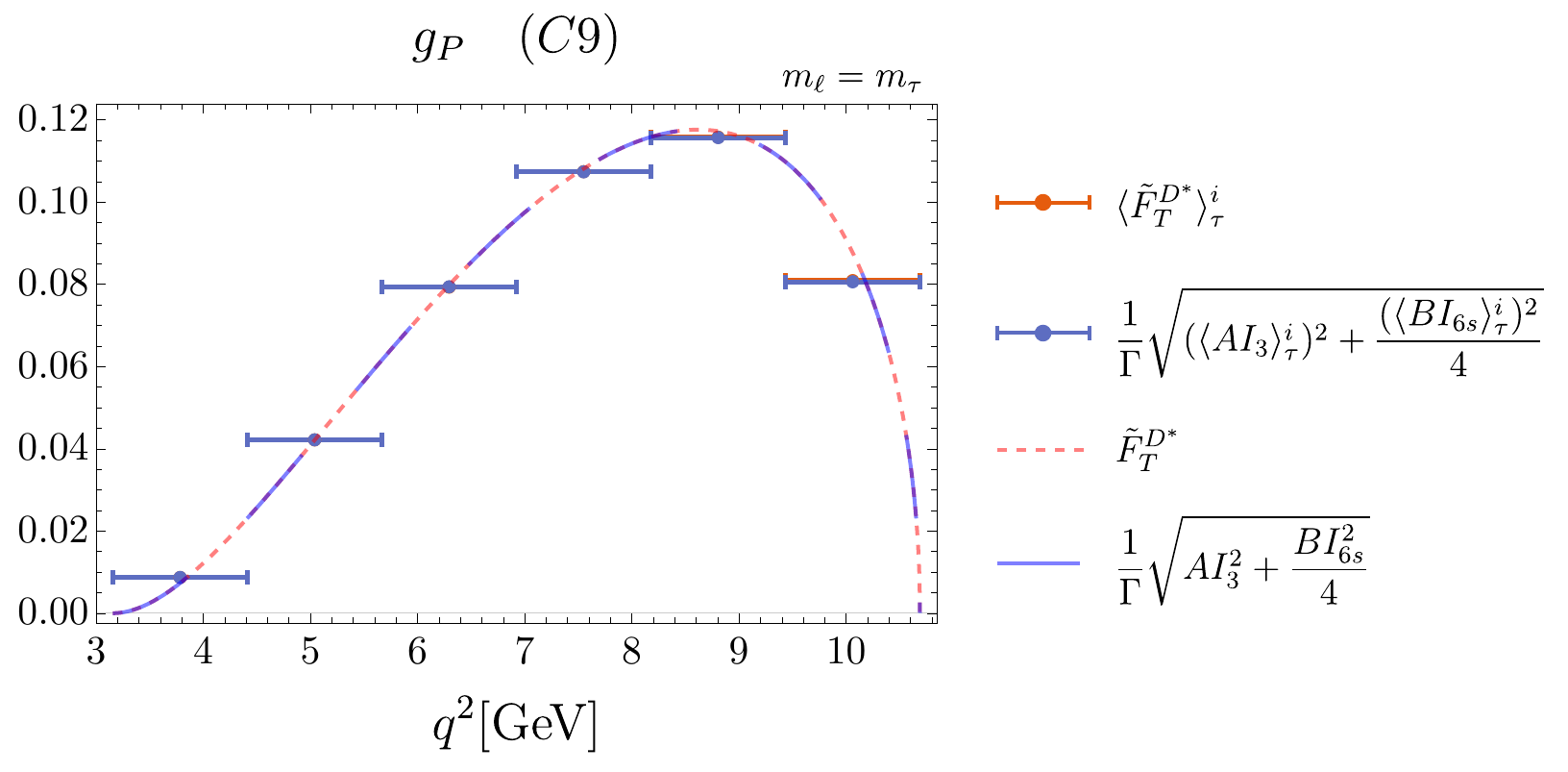}
 \end{subfigure}

 \caption{ Study of binning effects for Eq.~(\ref{binnedsecond}) for benchmark NP scenarios with complex contributions.}
 \label{fig:appendix2}
\end{figure}

\begin{figure}
 \centering
 \begin{subfigure}{\linewidth}
  \centering
  \includegraphics[height=.285\linewidth]{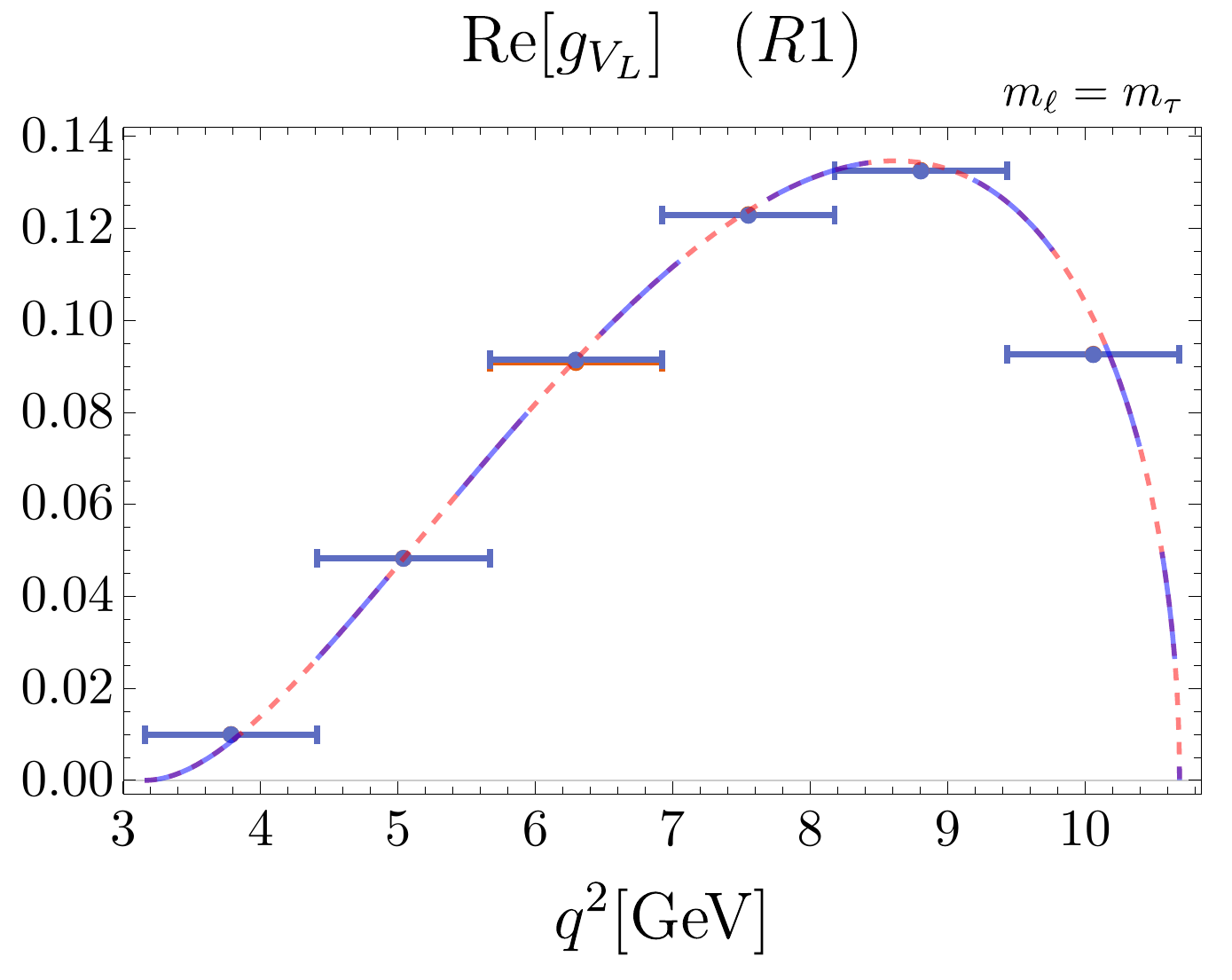}
  \hspace{10pt}
  \includegraphics[height=.285\linewidth]{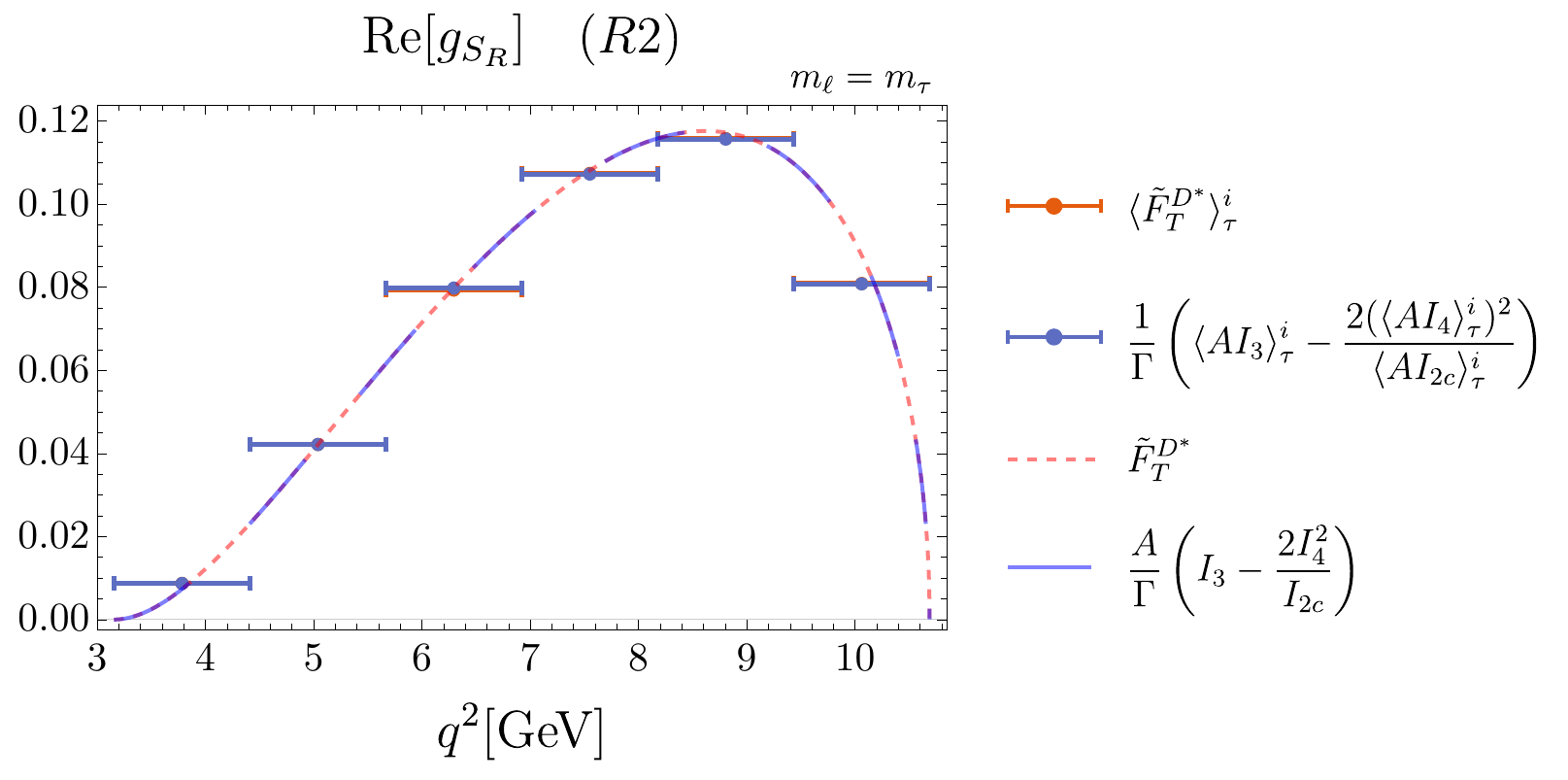}
 \end{subfigure}

 \begin{subfigure}{\linewidth}
  \centering
  \includegraphics[height=.285\linewidth]{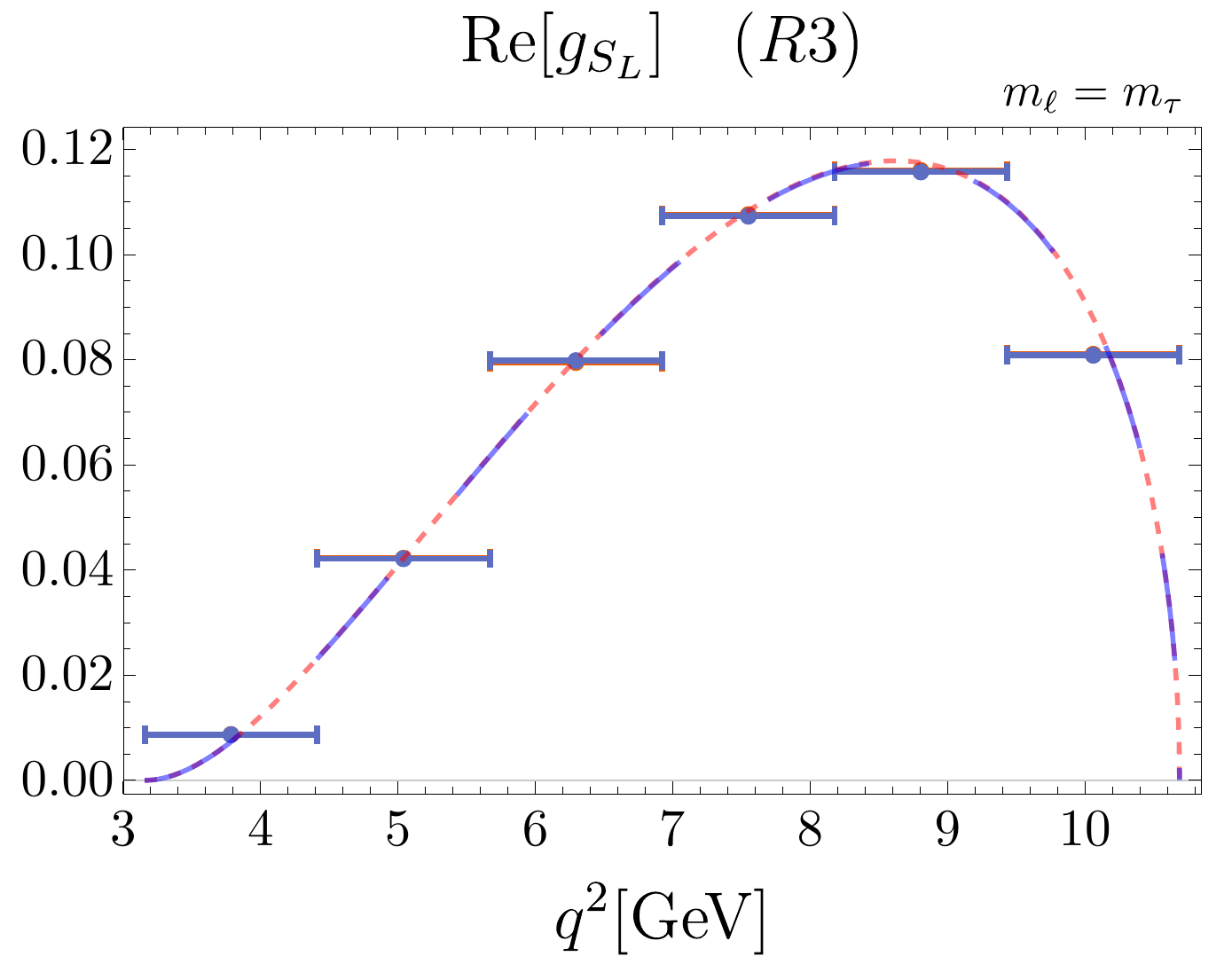}
  \hspace{10pt}
  \includegraphics[height=.285\linewidth]{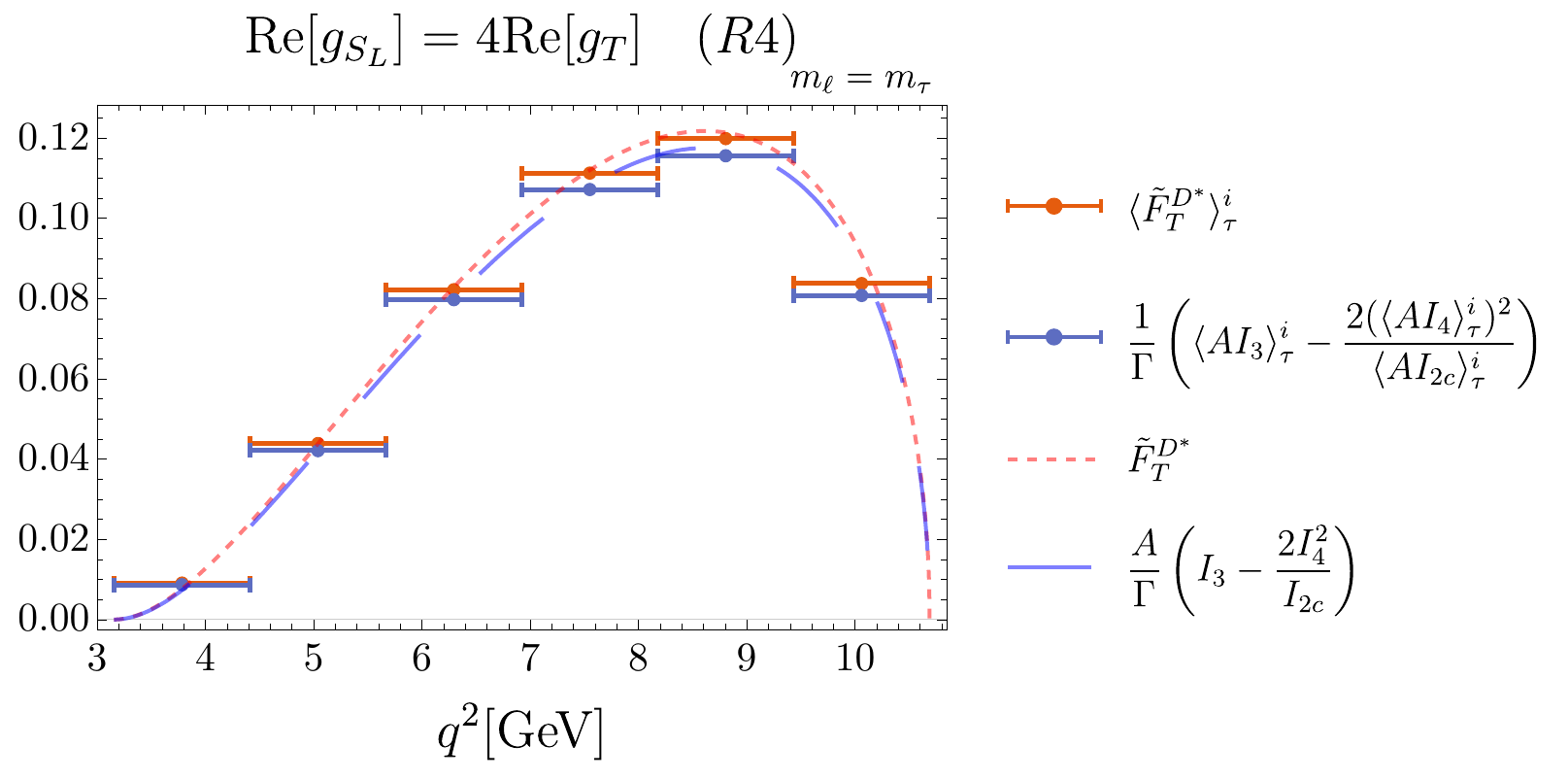}
 \end{subfigure}

 \begin{subfigure}{\linewidth}
  \centering
  \includegraphics[height=.285\linewidth]{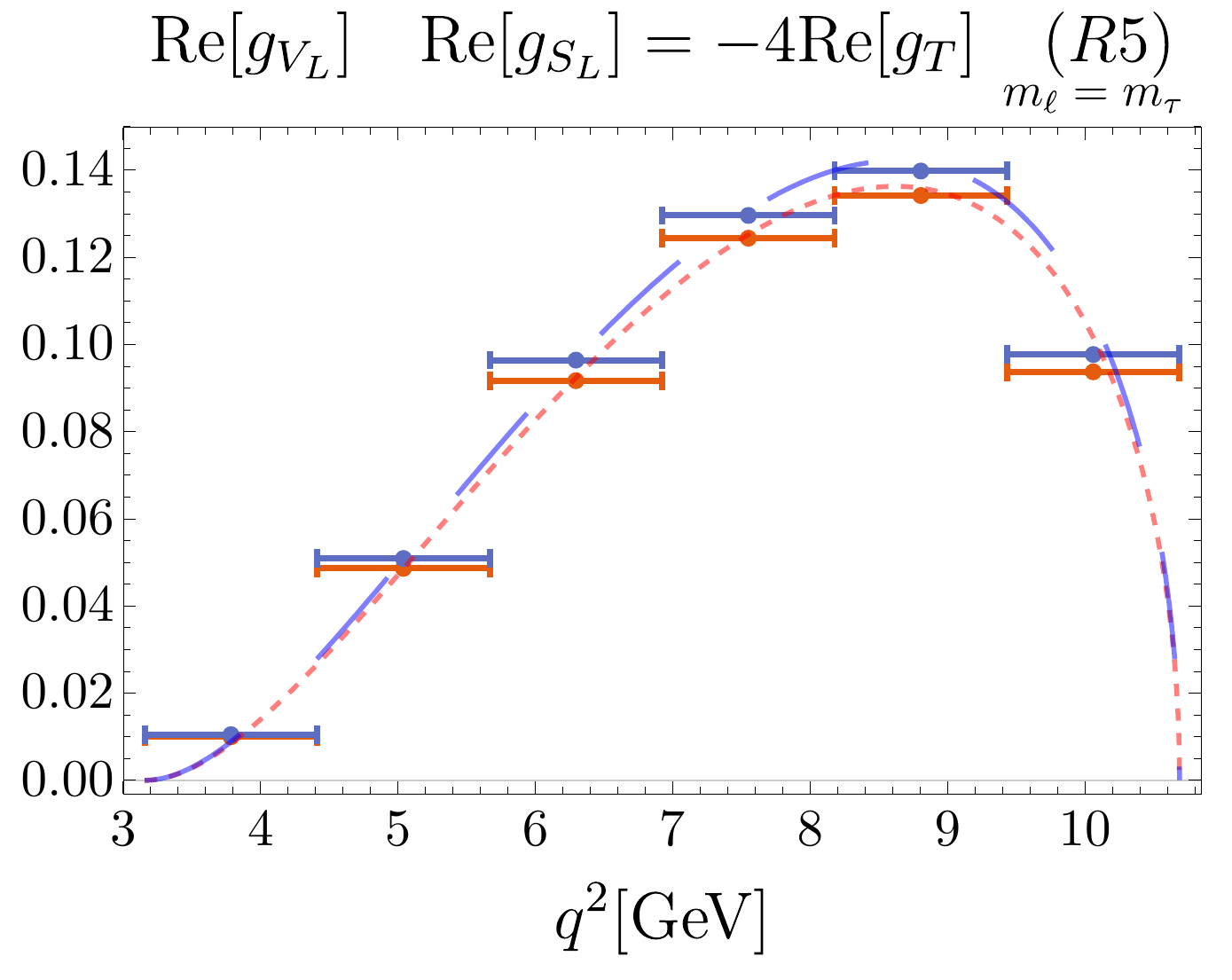}
  \hspace{10pt}
  \includegraphics[height=.285\linewidth]{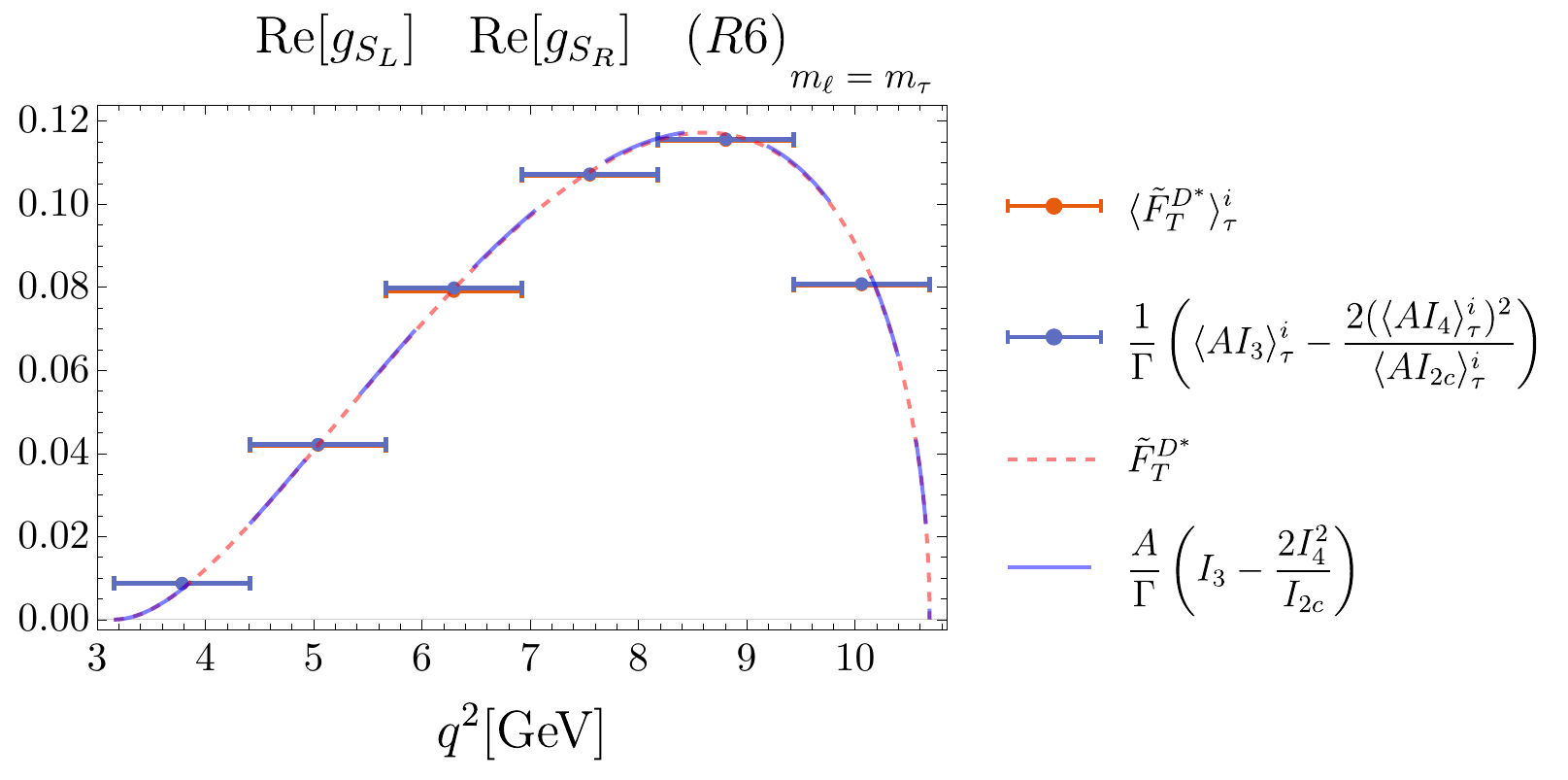}
 \end{subfigure}
 \begin{subfigure}{\linewidth}
  \centering
  \includegraphics[height=.285\linewidth]{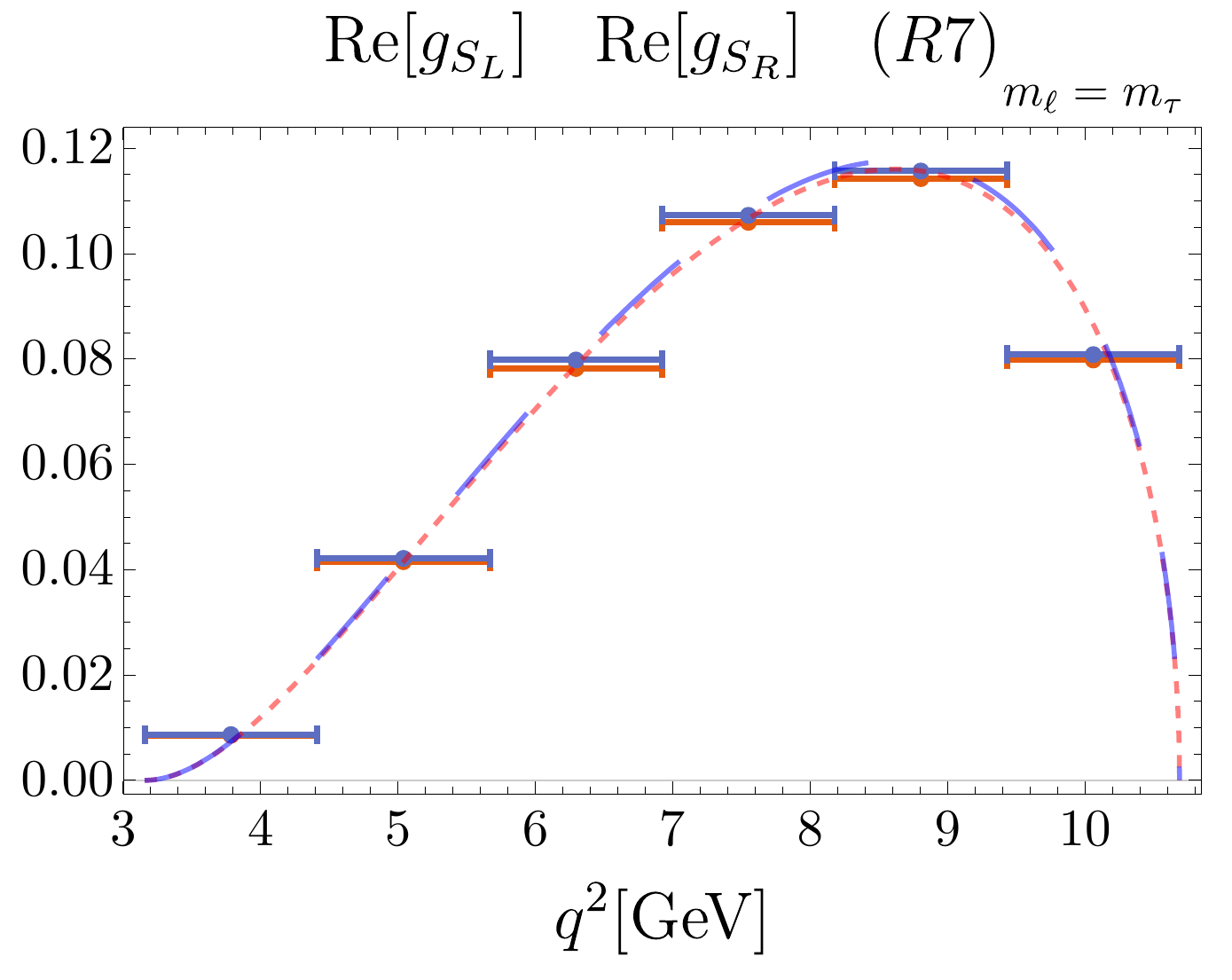}
  \hspace{10pt}
  \includegraphics[height=.285\linewidth]{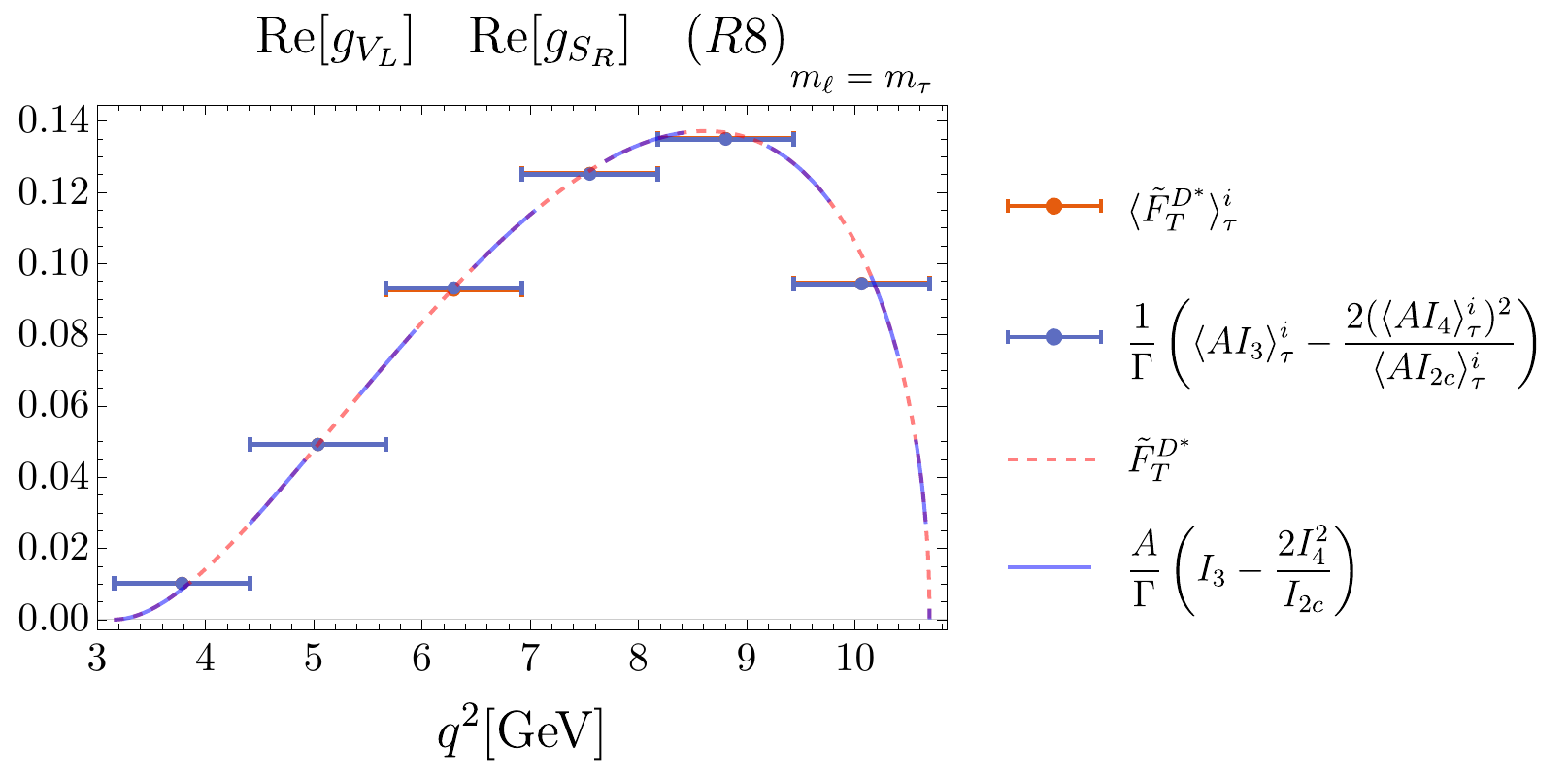}
 \end{subfigure}

 \caption{ Study of binning effects for Eq.~(\ref{binnedfLmass}) for benchmark NP scenarios with real contributions.}
 \label{fig:appendix3}
\end{figure}

\begin{figure}
 \centering

 \begin{subfigure}{\linewidth}
  \centering
  \includegraphics[height=.285\linewidth]{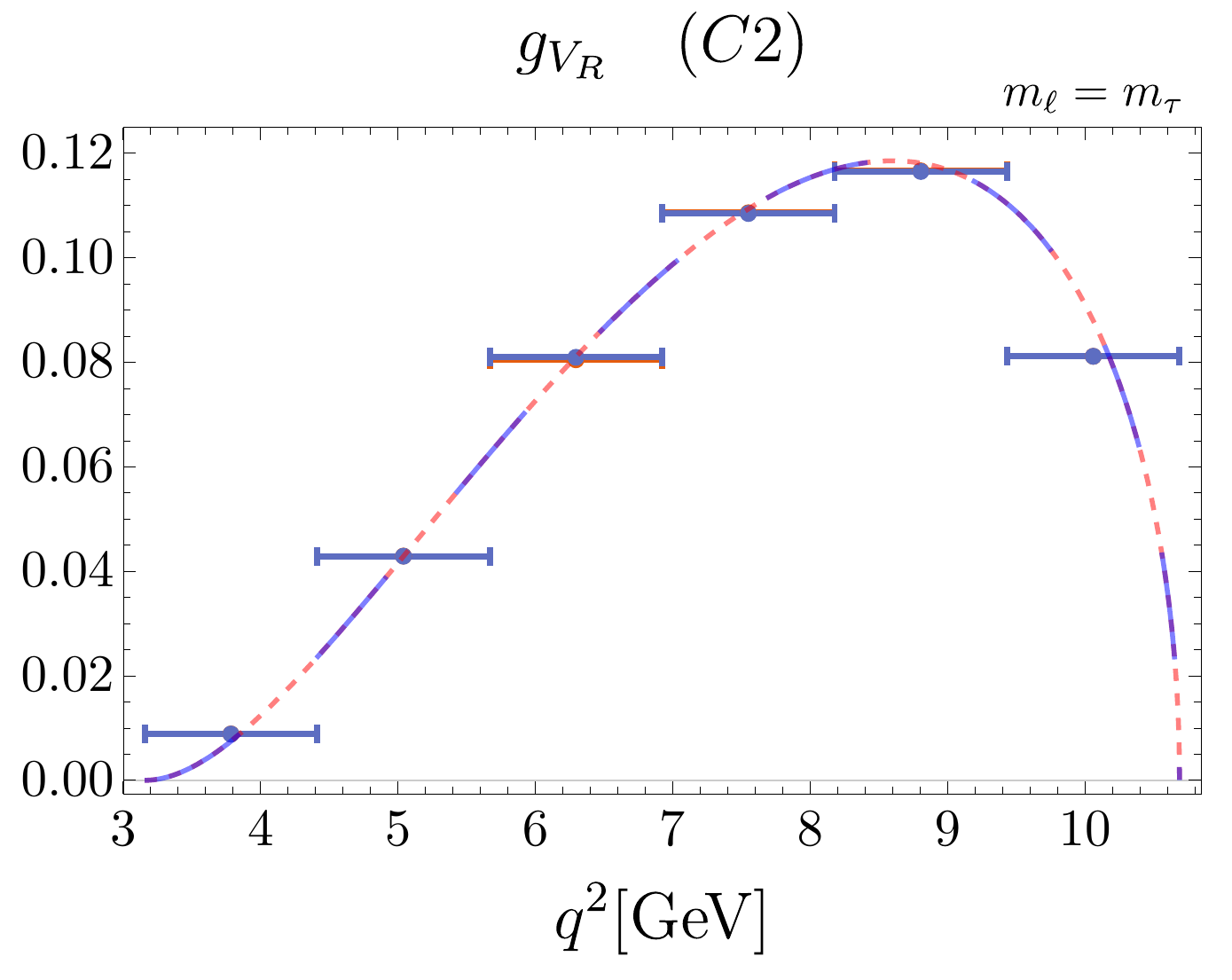}
  \hspace{10pt}
  \includegraphics[height=.285\linewidth]{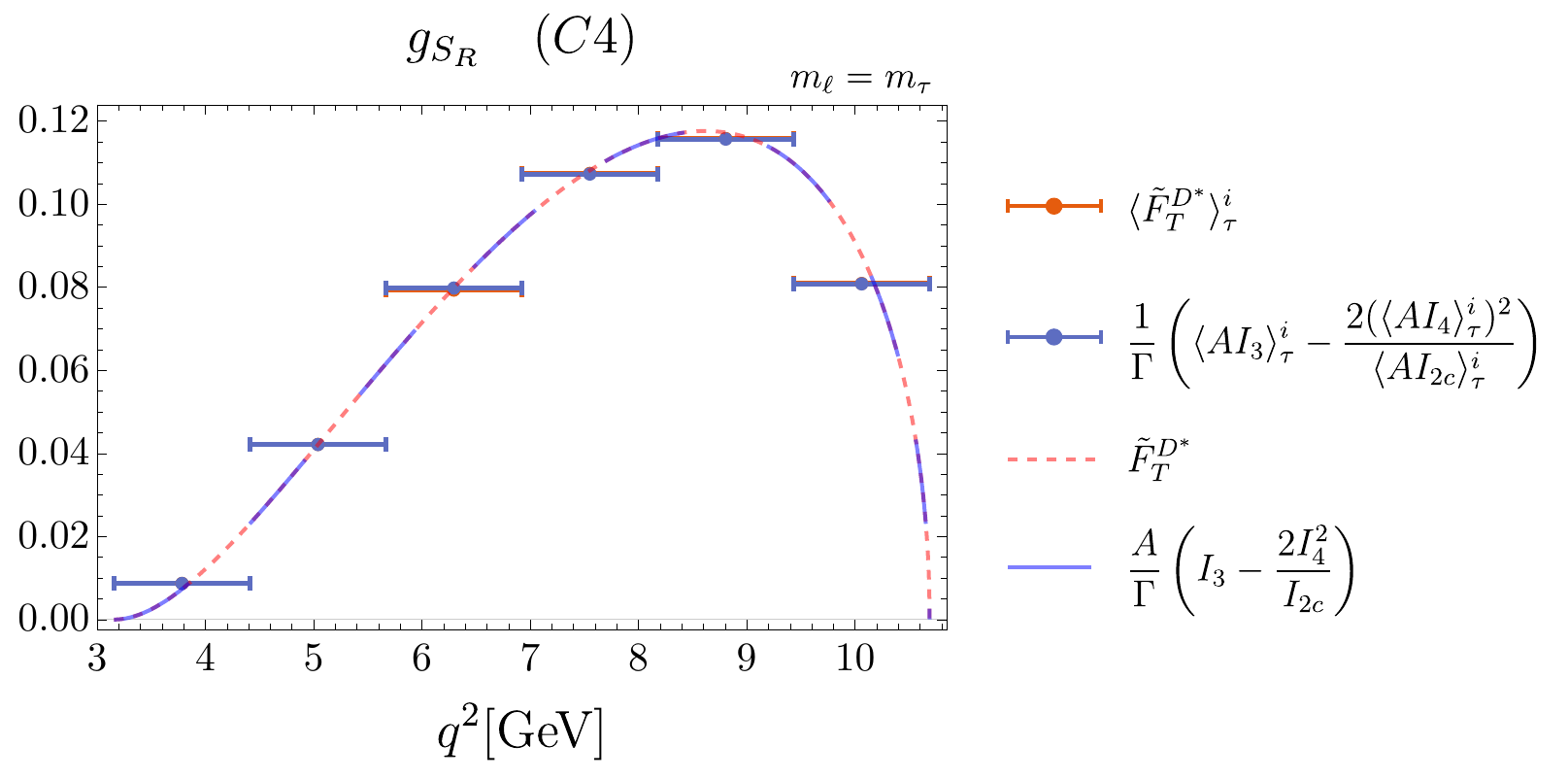}
 \end{subfigure}
 \begin{subfigure}{\linewidth}
  \centering
  \includegraphics[height=.285\linewidth]{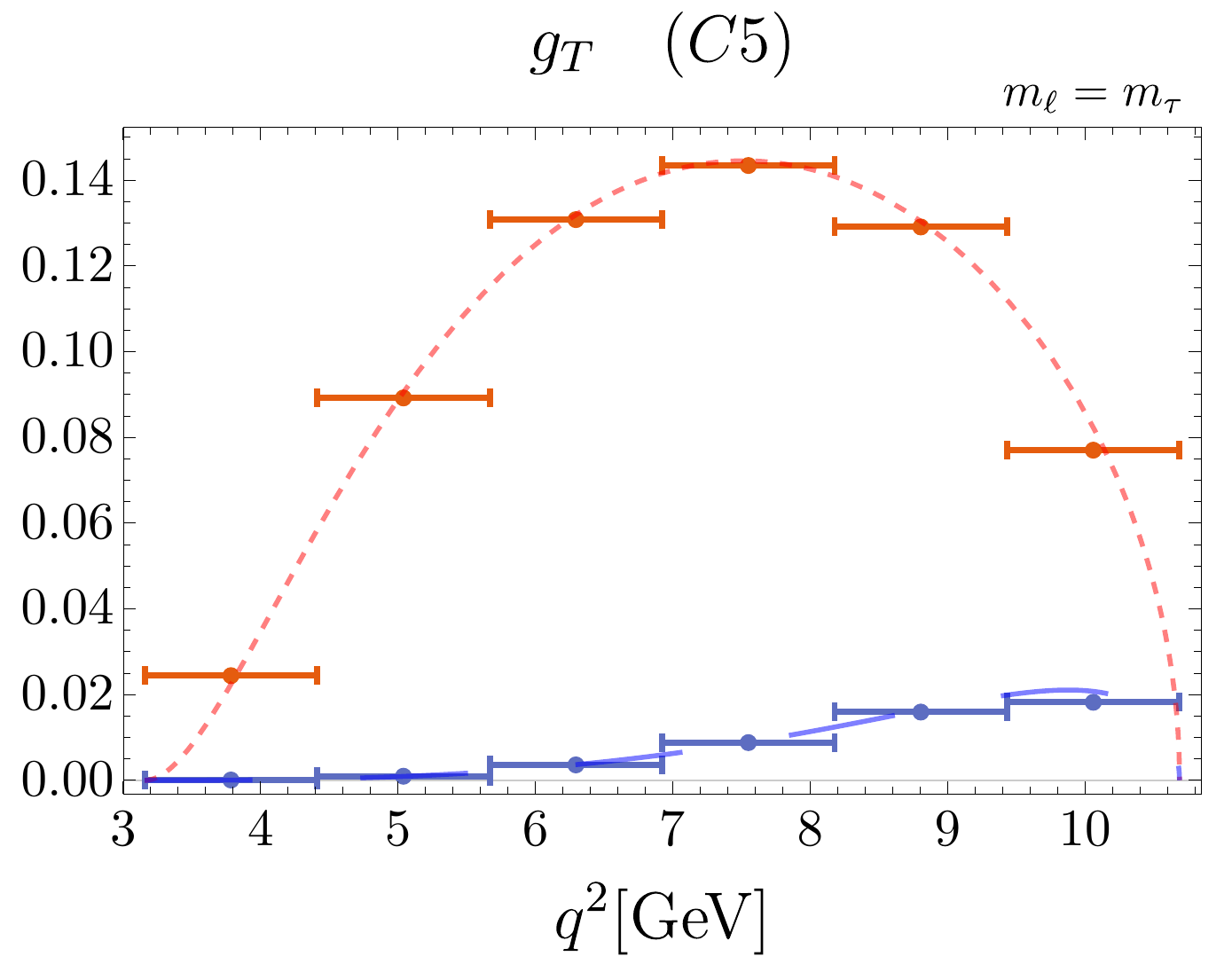}
  \hspace{10pt}
  \includegraphics[height=.285\linewidth]{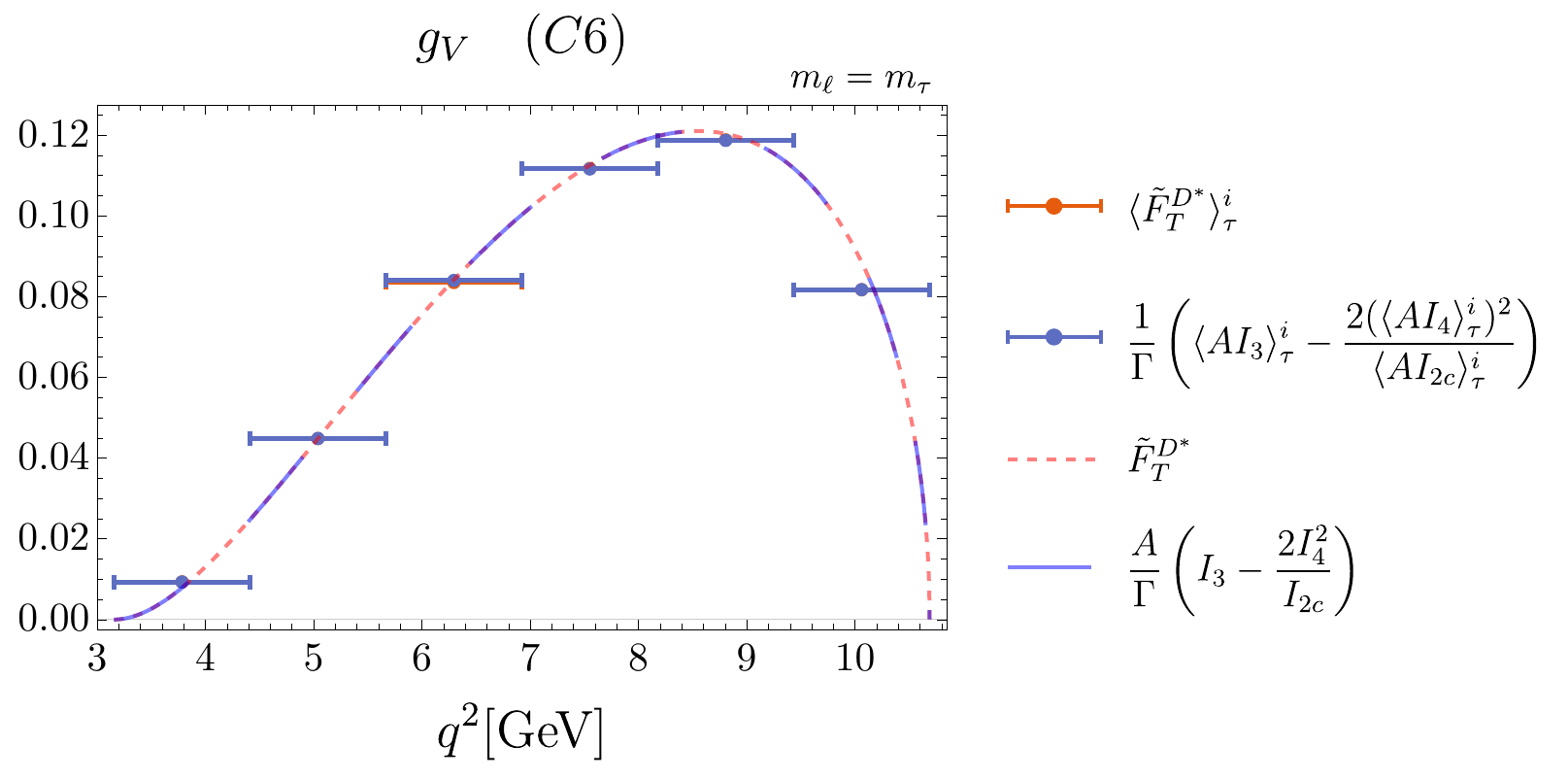}
 \end{subfigure}
 \begin{subfigure}{\linewidth}
  \centering
  \includegraphics[height=.285\linewidth]{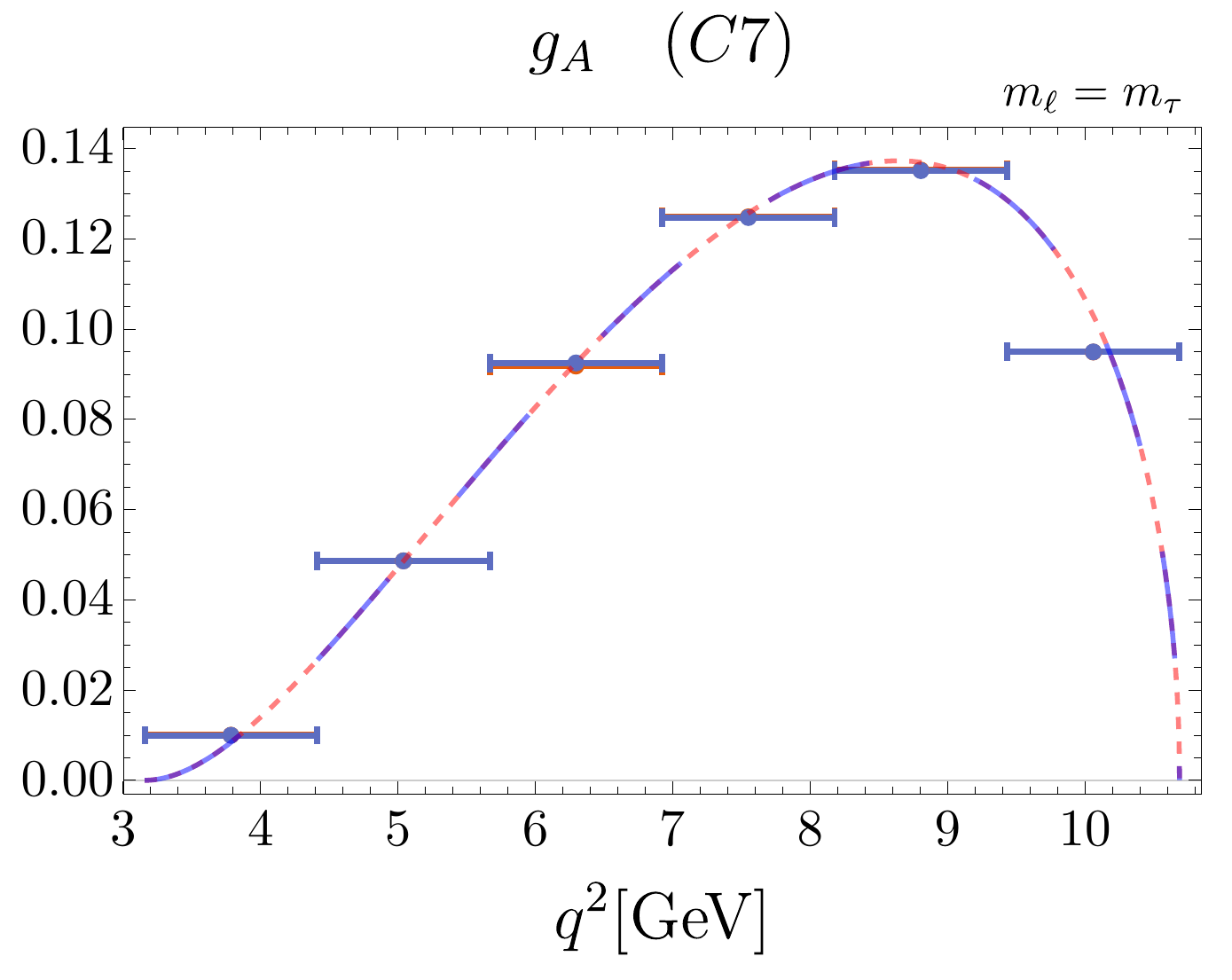}
  \hspace{10pt}
  \includegraphics[height=.285\linewidth]{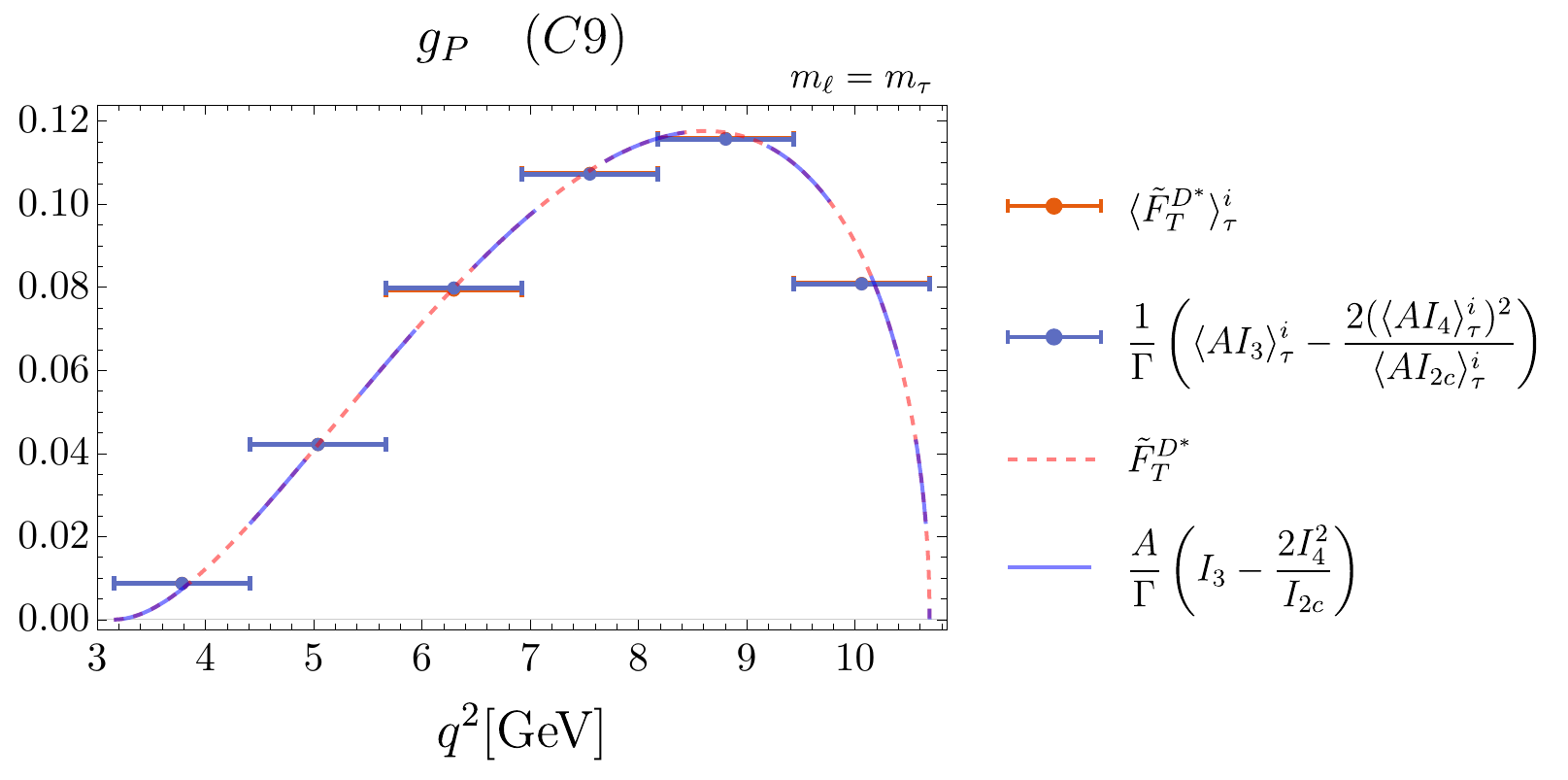}
 \end{subfigure}

 \caption{ Study of binning effects for Eq.~(\ref{binnedfLmass}) for benchmark NP scenarios with complex contributions.}
 \label{fig:appendix4}
\end{figure}

\begin{figure}
 \centering
 \begin{subfigure}{\linewidth}
  \centering
  \includegraphics[height=.27\linewidth]{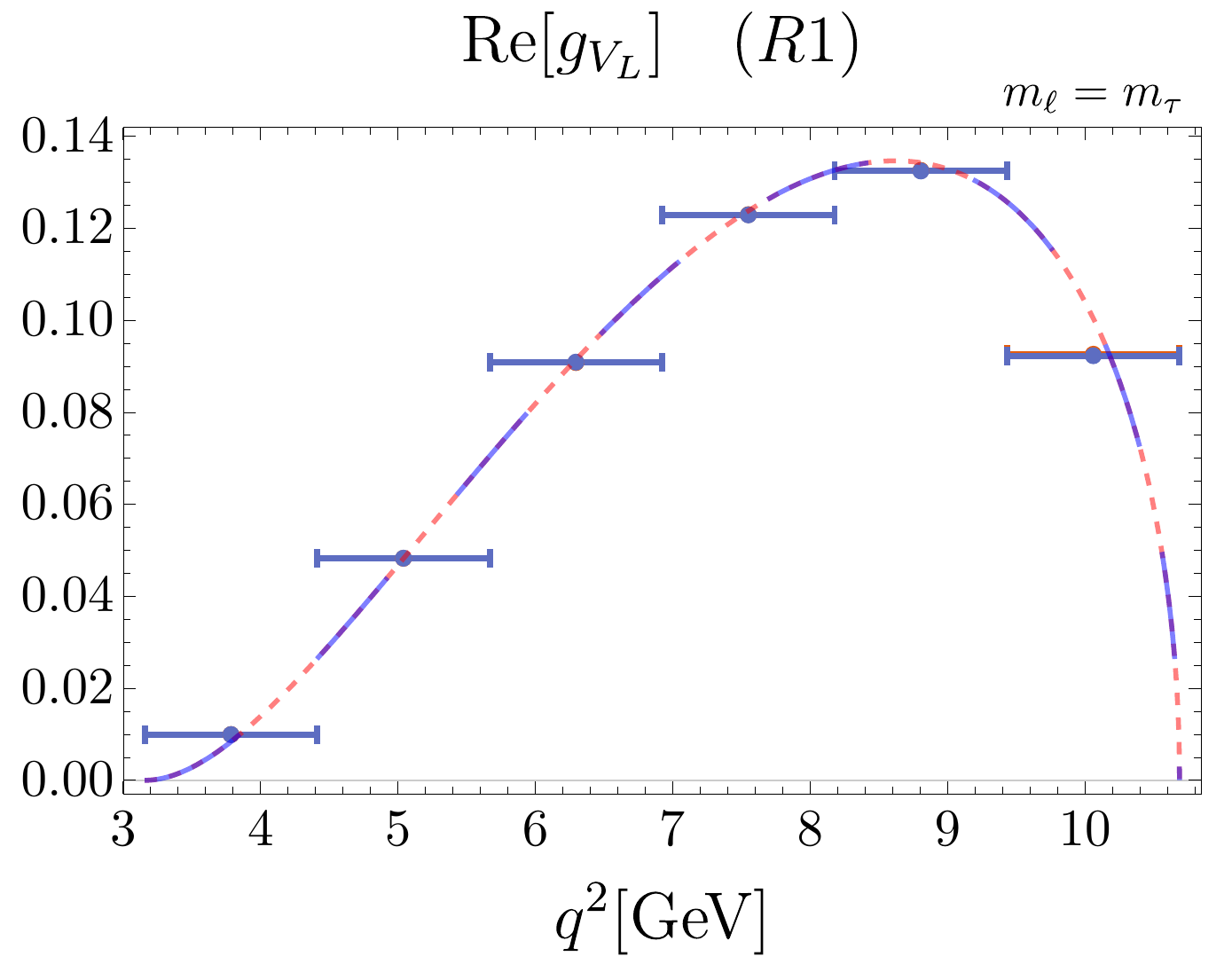}
  \hspace{10pt}
  \includegraphics[height=.27\linewidth]{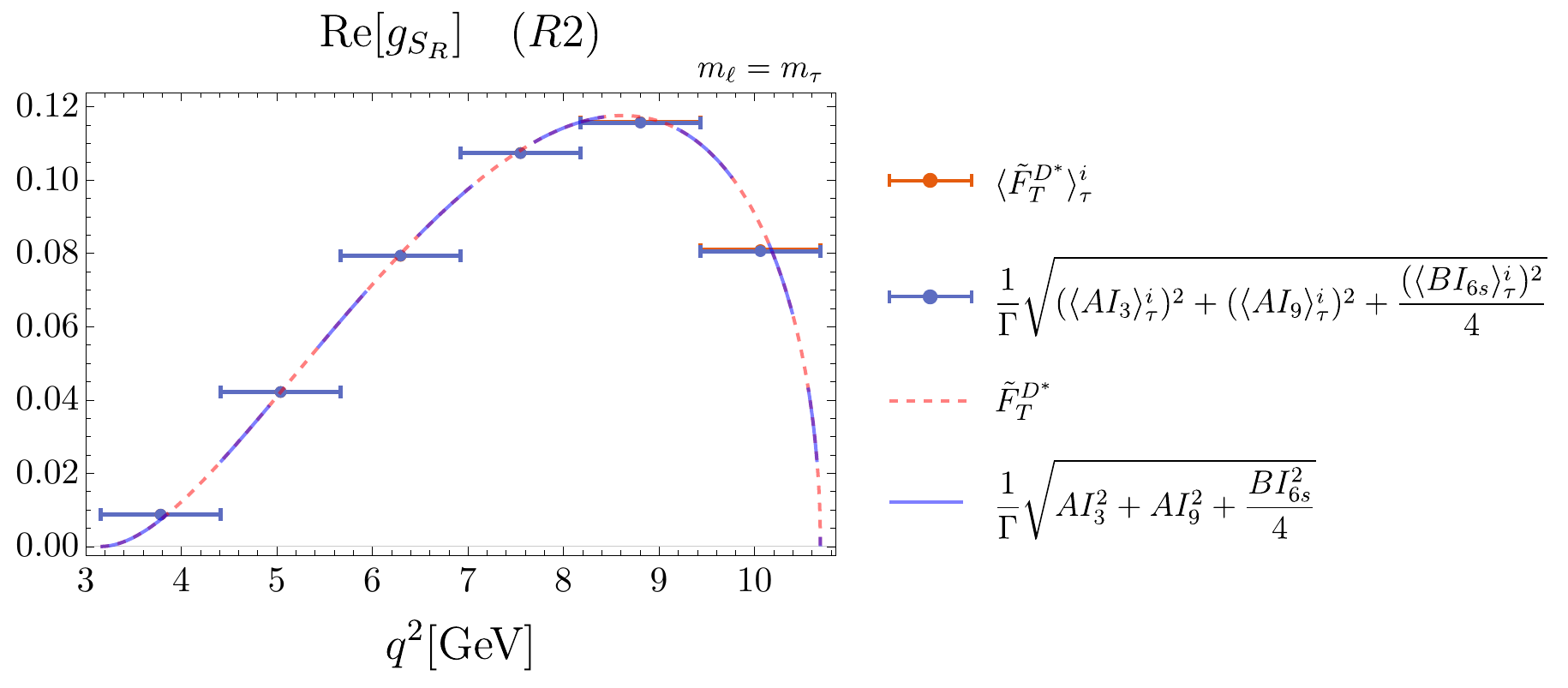}
 \end{subfigure}

 \begin{subfigure}{\linewidth}
  \centering
  \includegraphics[height=.27\linewidth]{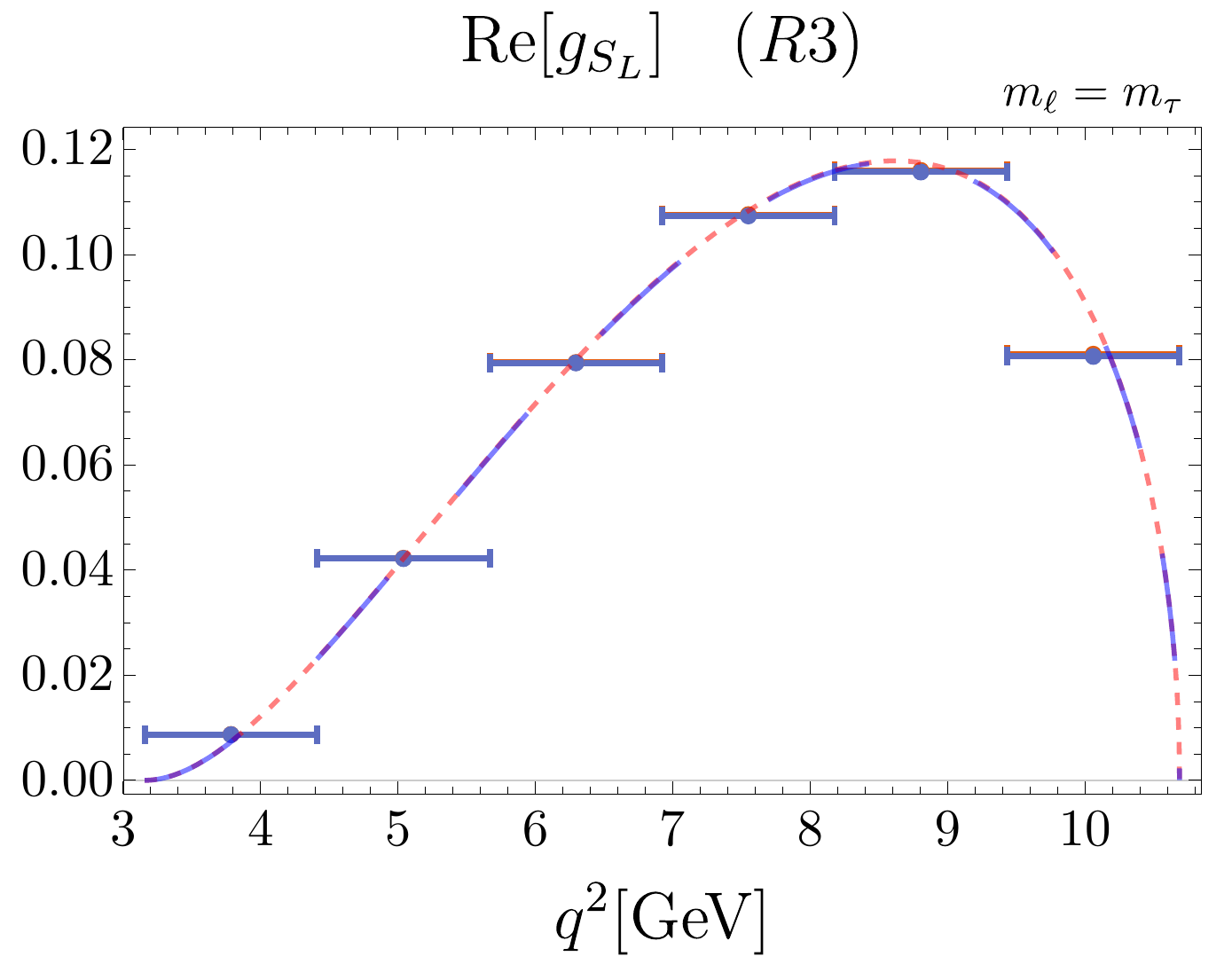}
  \hspace{10pt}
  \includegraphics[height=.27\linewidth]{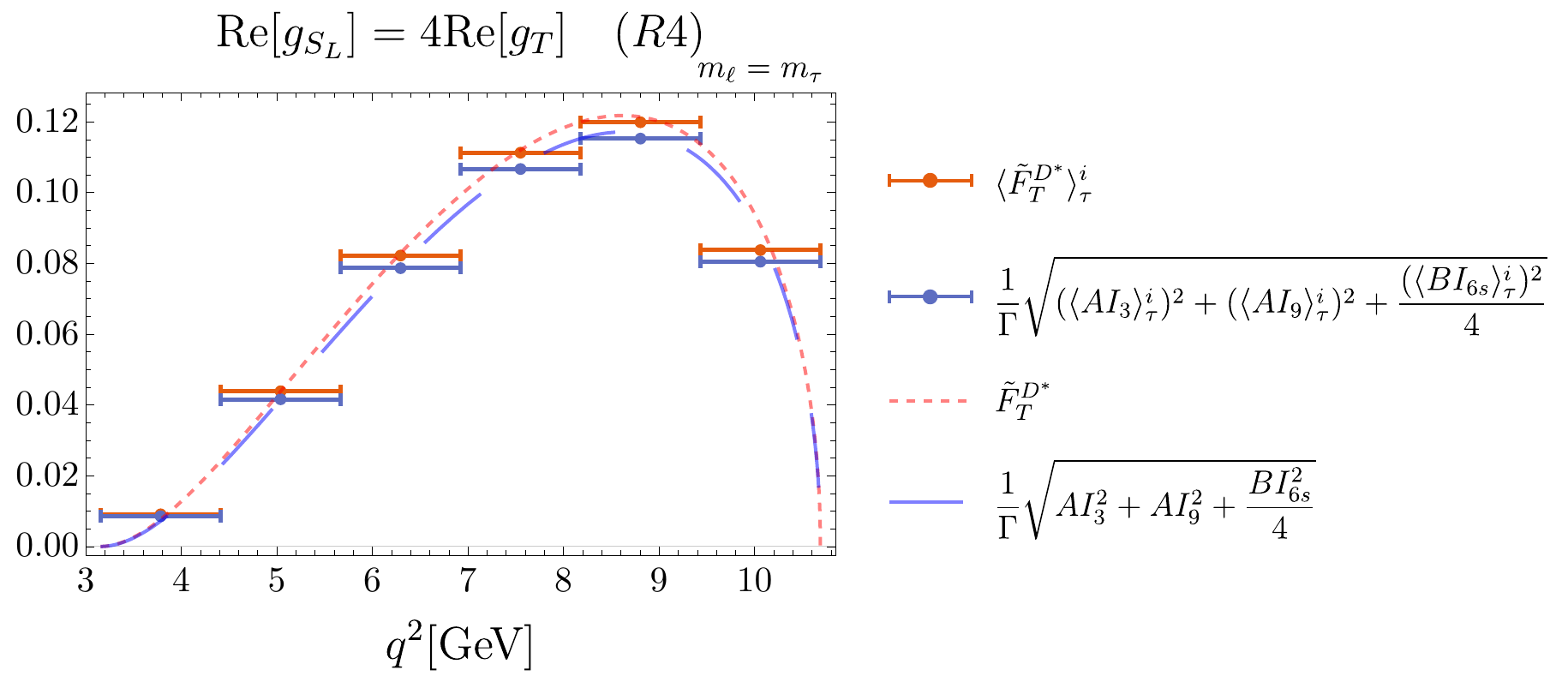}
 \end{subfigure}

 \begin{subfigure}{\linewidth}
  \centering
  \includegraphics[height=.27\linewidth]{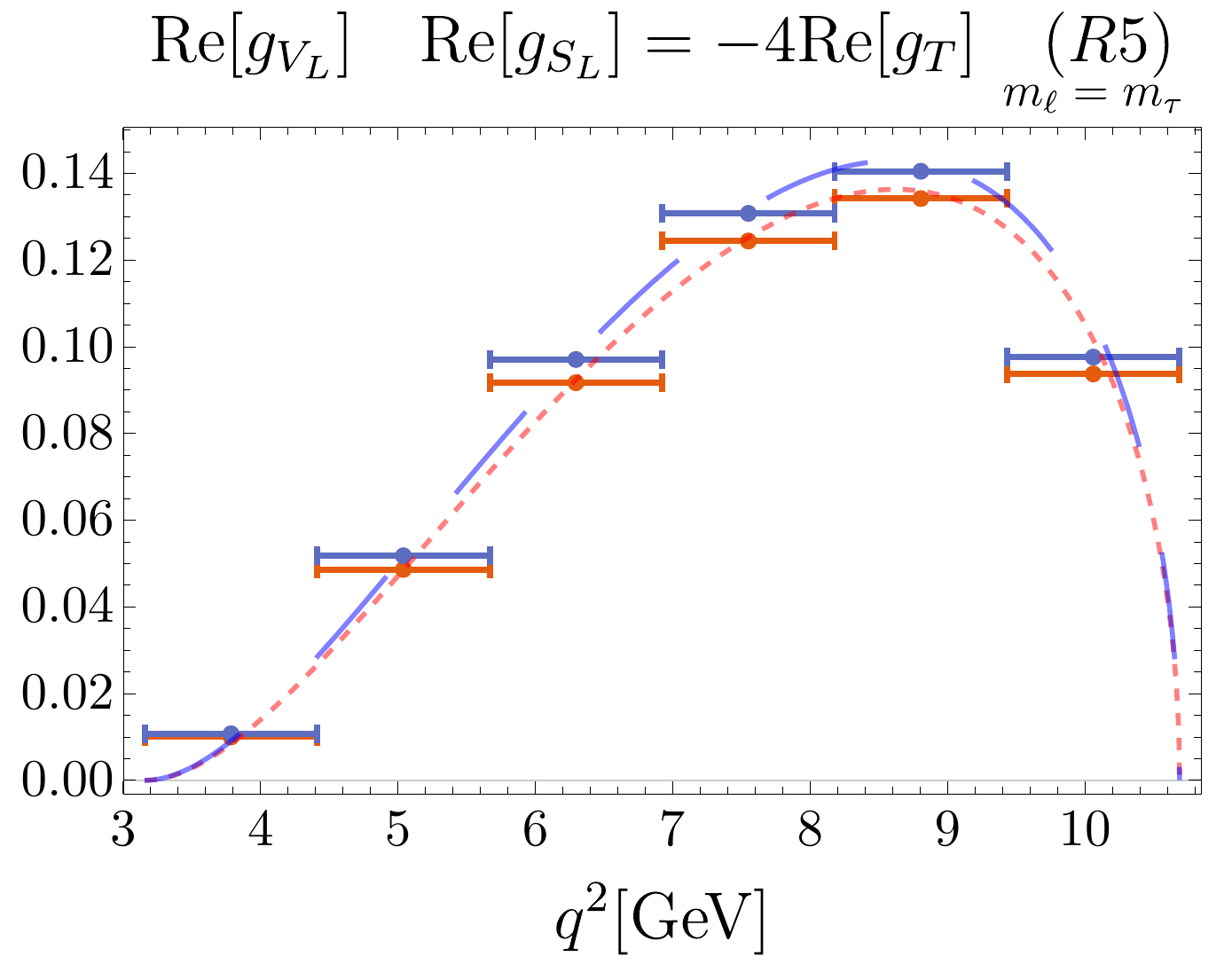}
  \hspace{10pt}
  \includegraphics[height=.27\linewidth]{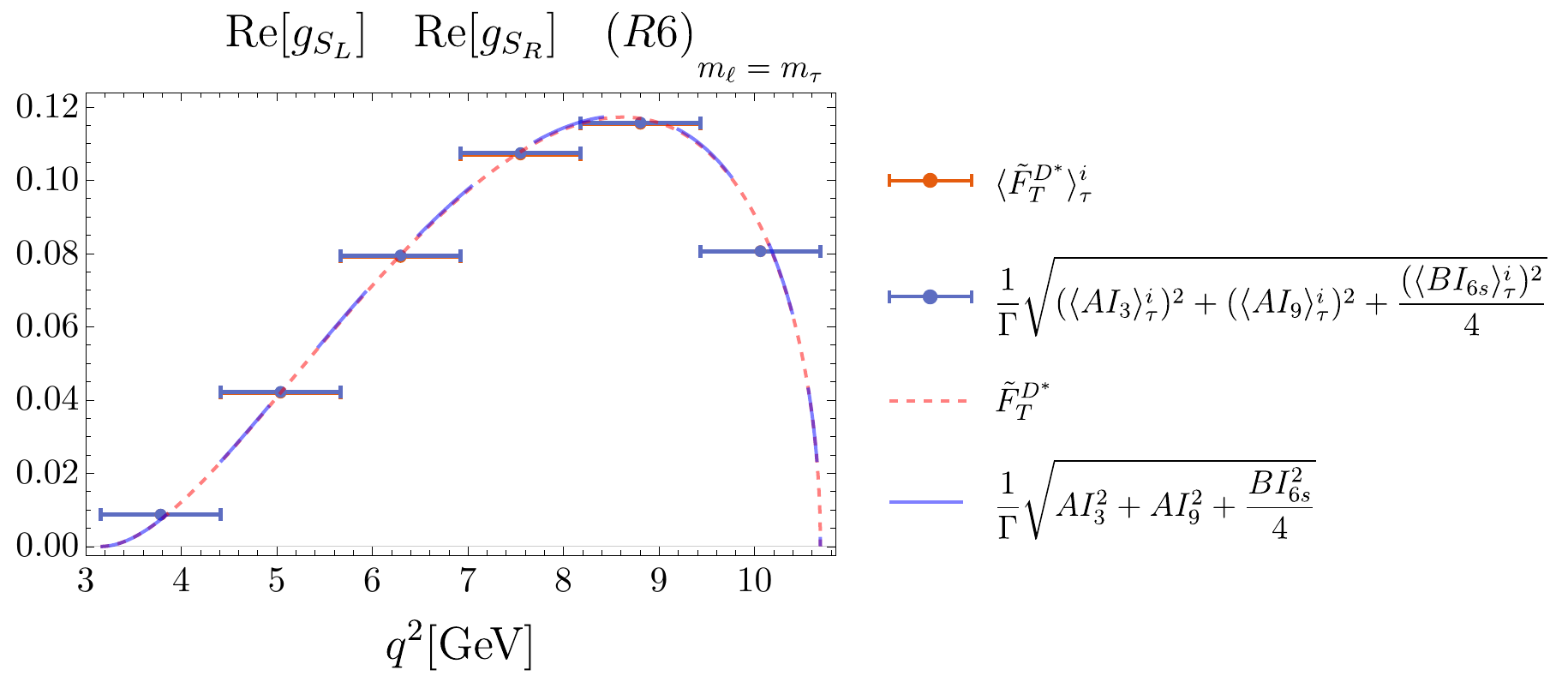}
 \end{subfigure}
 \begin{subfigure}{\linewidth}
  \centering
  \includegraphics[height=.27\linewidth]{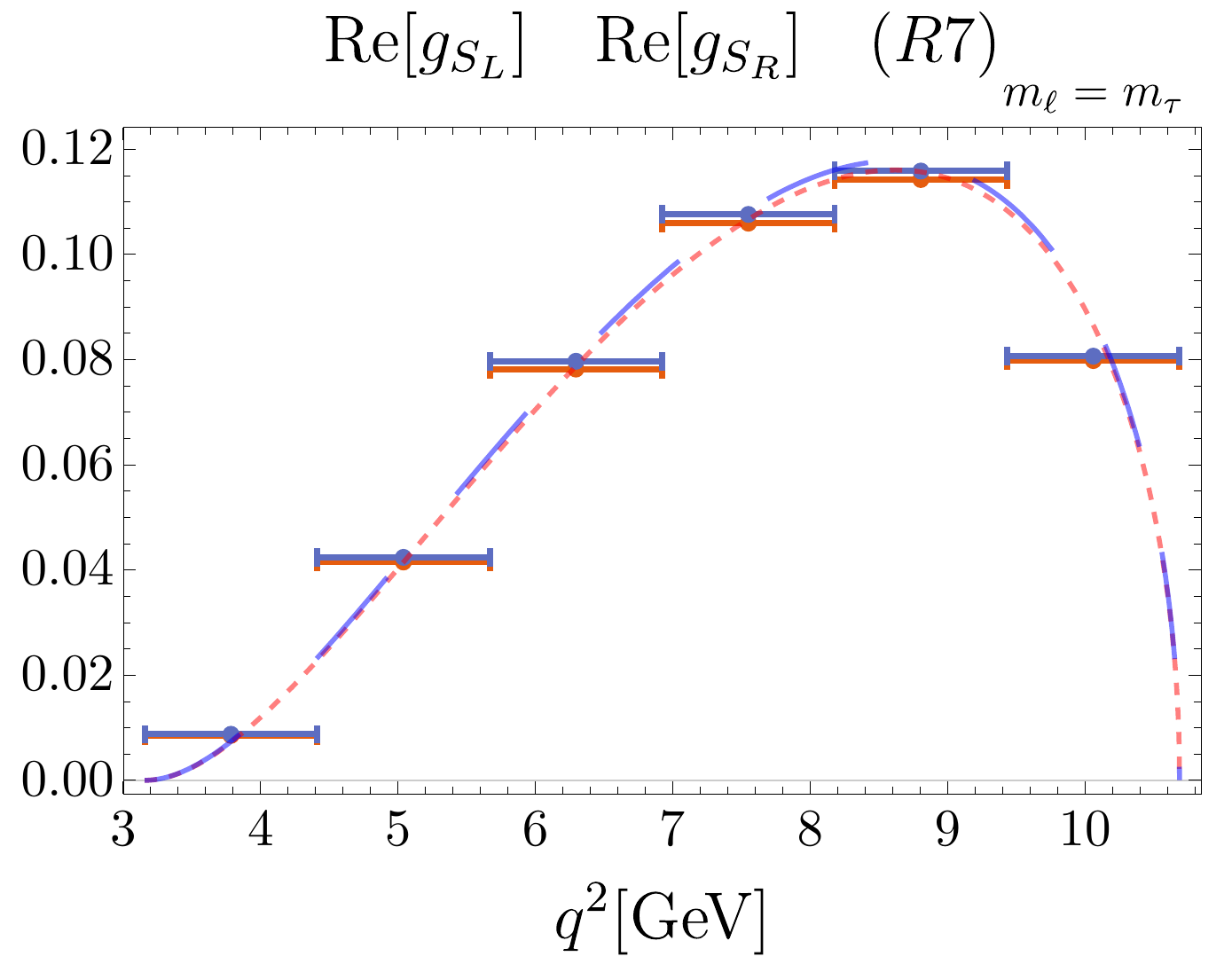}
  \hspace{10pt}
  \includegraphics[height=.27\linewidth]{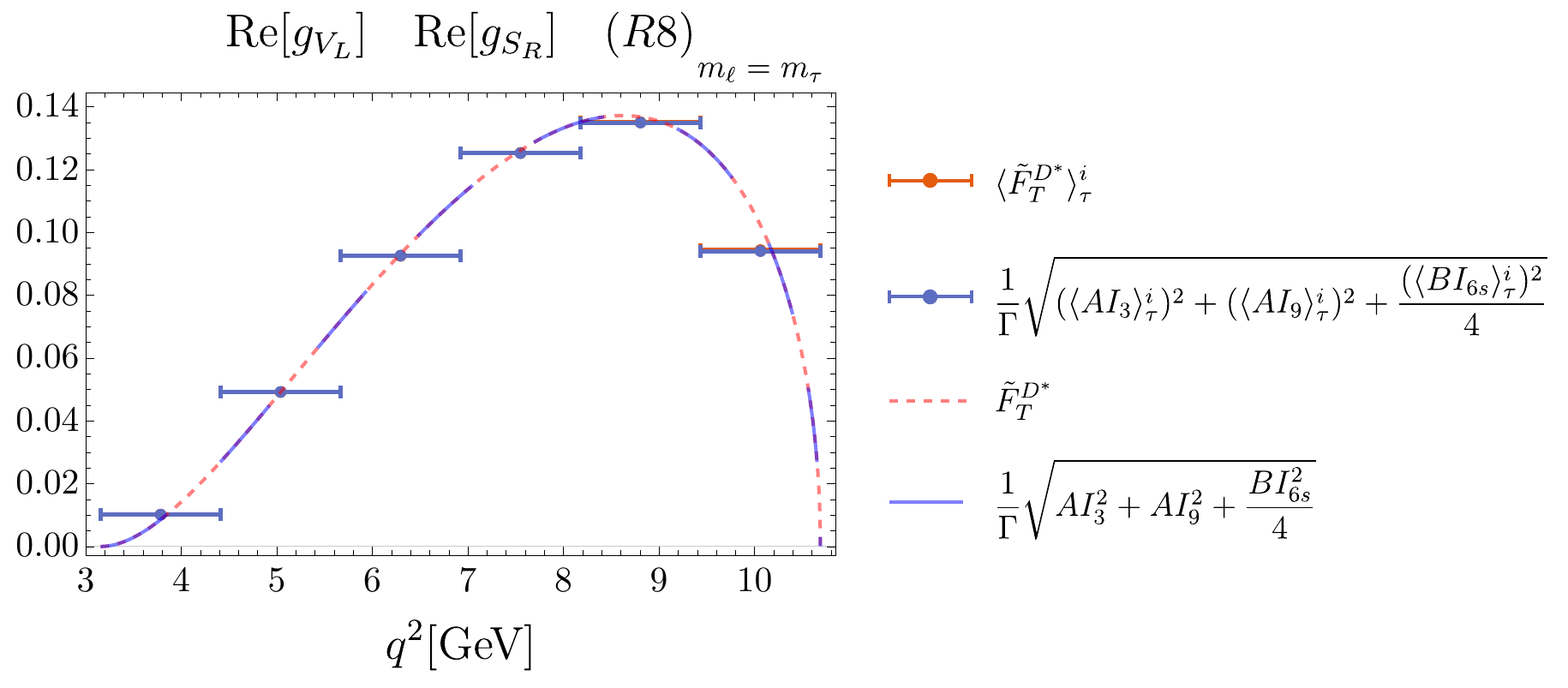}
 \end{subfigure}

 \caption{ Study of binning effects for Eq.~(\ref{eq:FLbinnedmassive}) for benchmark NP scenarios with real contributions.}
 \label{fig:appendix5}
\end{figure}

\begin{figure}
 \centering

 \begin{subfigure}{\linewidth}
  \centering
  \includegraphics[height=.27\linewidth]{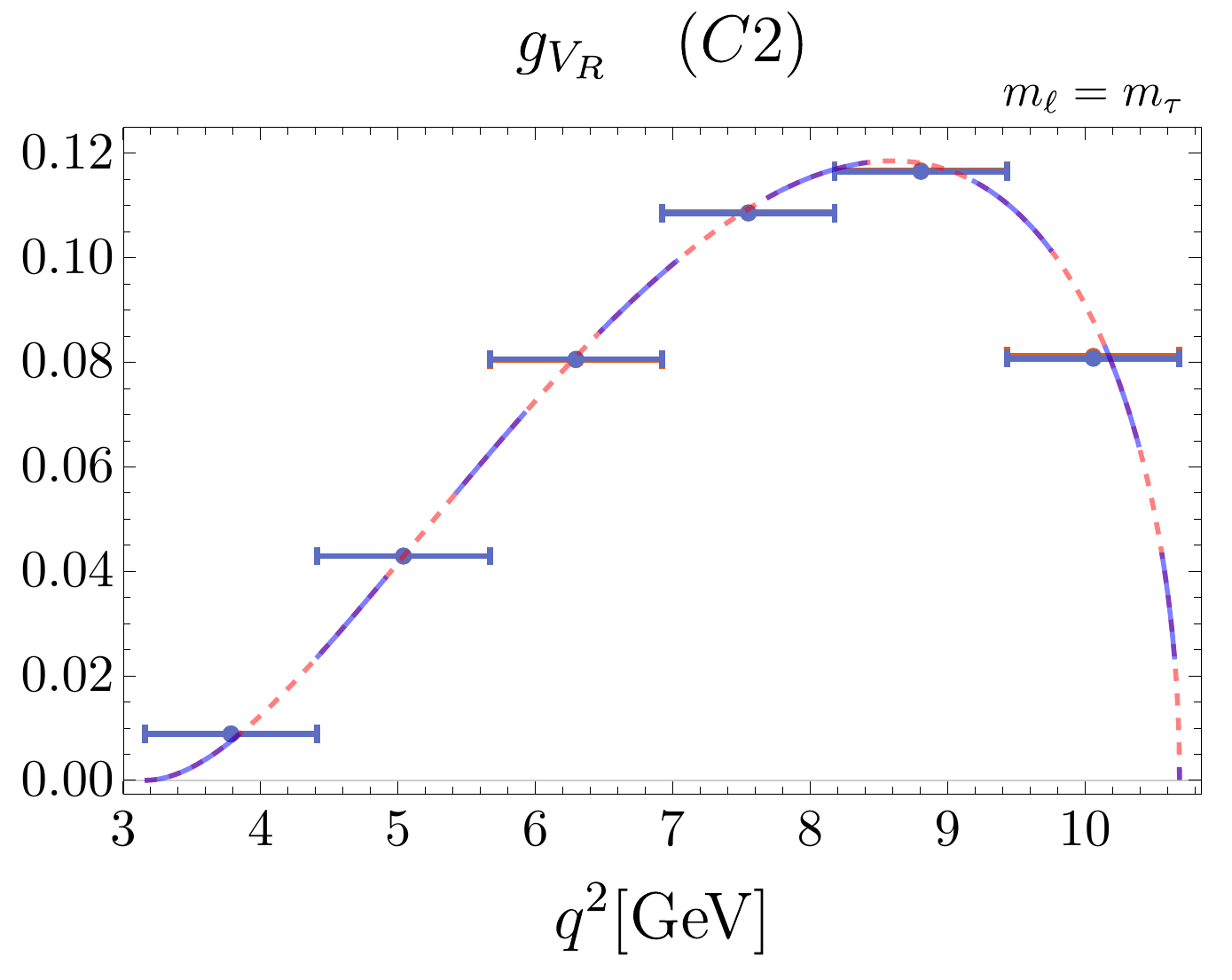}
  \hspace{10pt}
  \includegraphics[height=.27\linewidth]{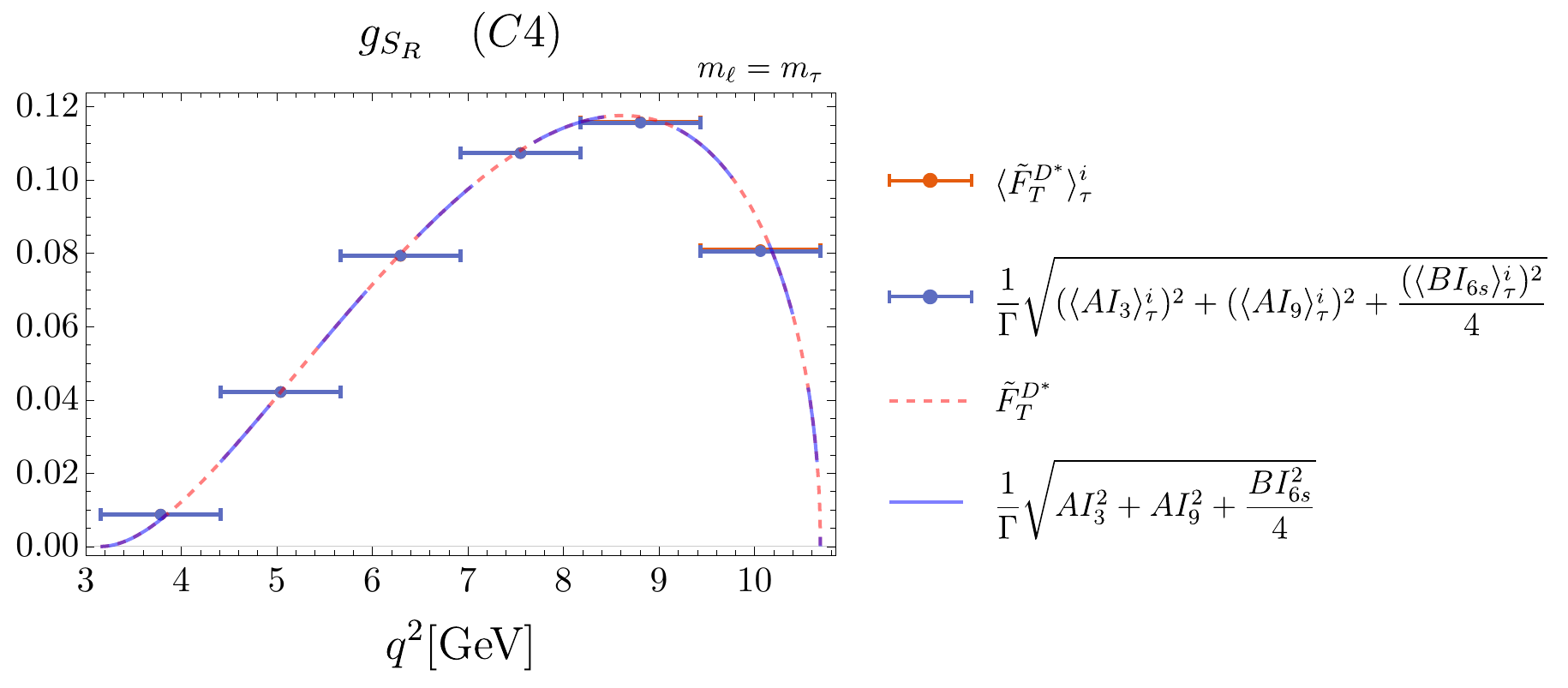}
 \end{subfigure}
 \begin{subfigure}{\linewidth}
  \centering
  \includegraphics[height=.27\linewidth]{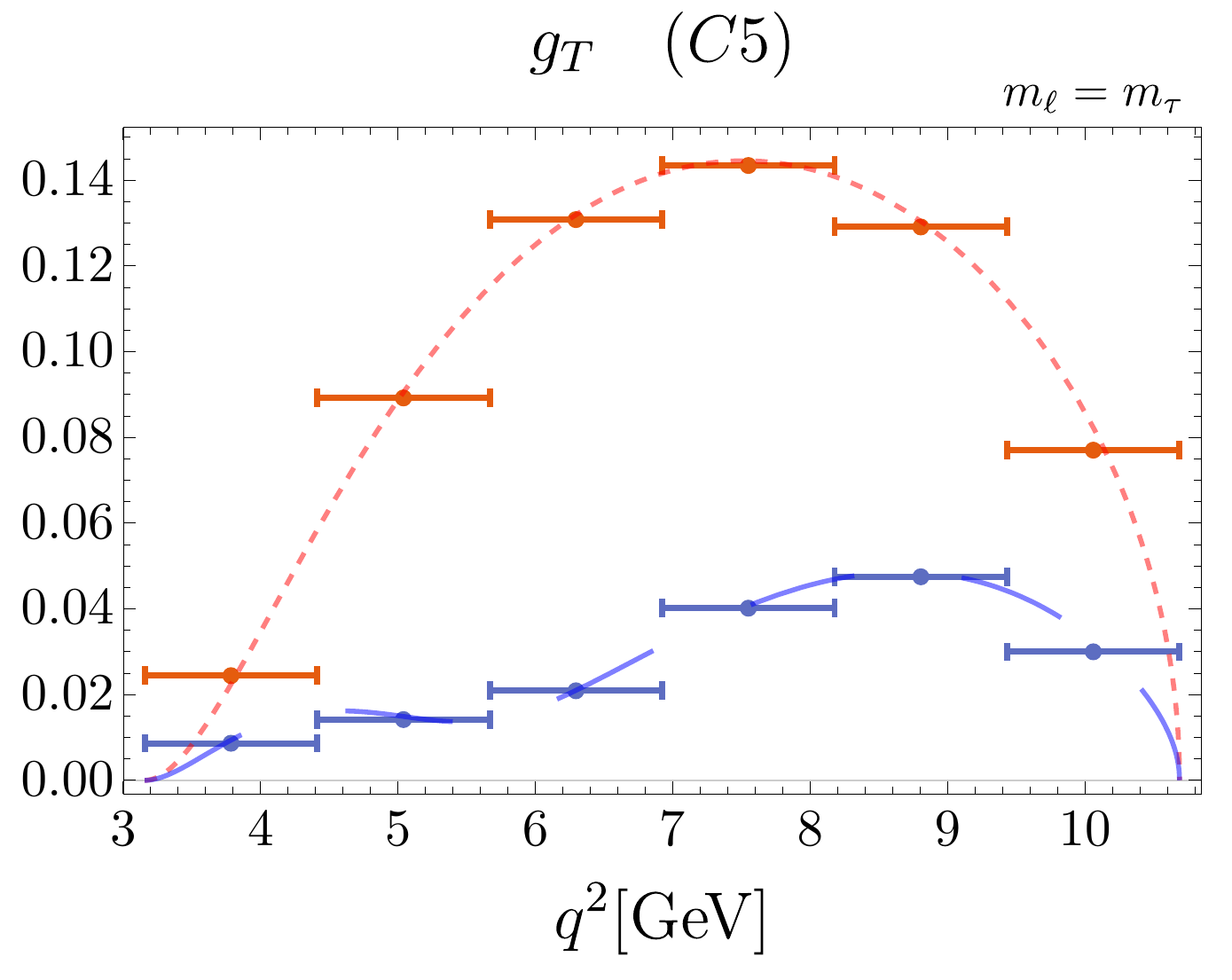}
  \hspace{10pt}
  \includegraphics[height=.27\linewidth]{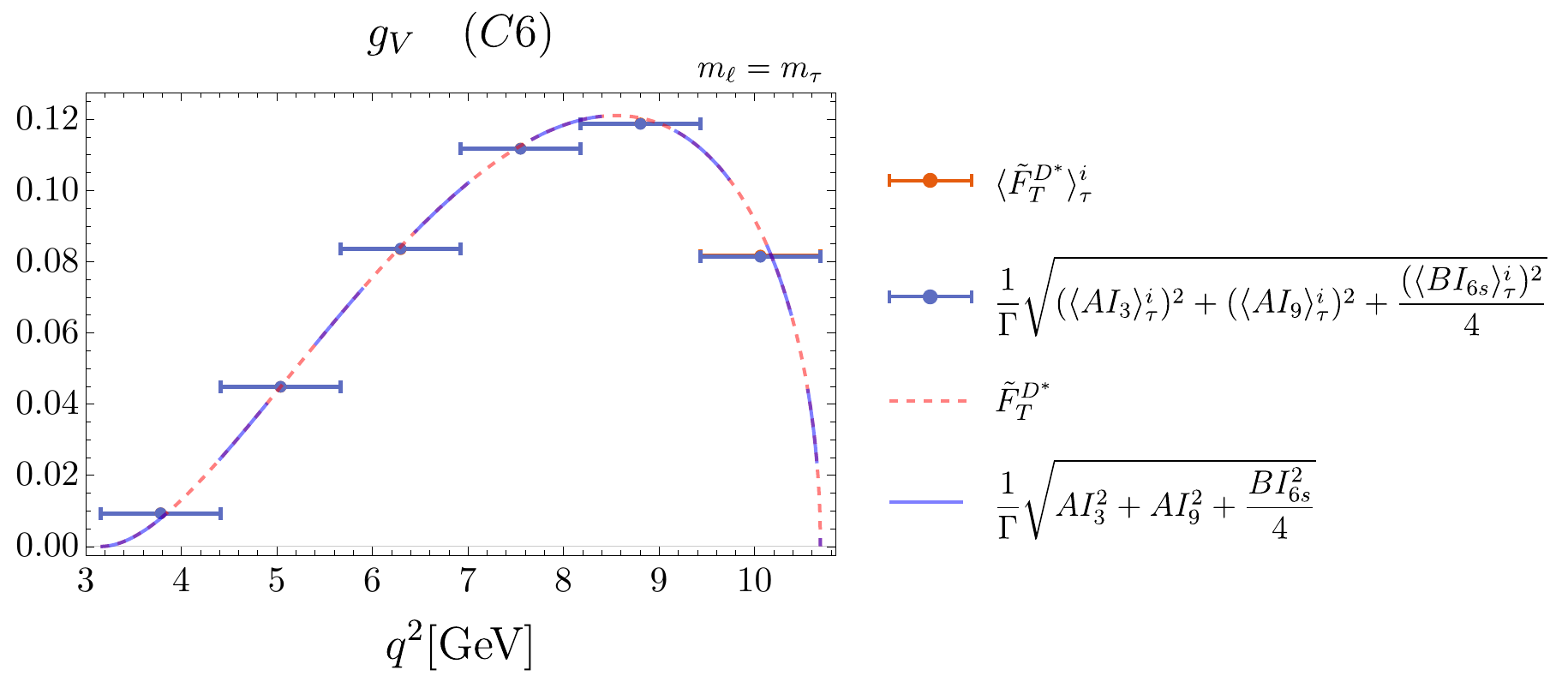}
 \end{subfigure}
 \begin{subfigure}{\linewidth}
  \centering
  \includegraphics[height=.27\linewidth]{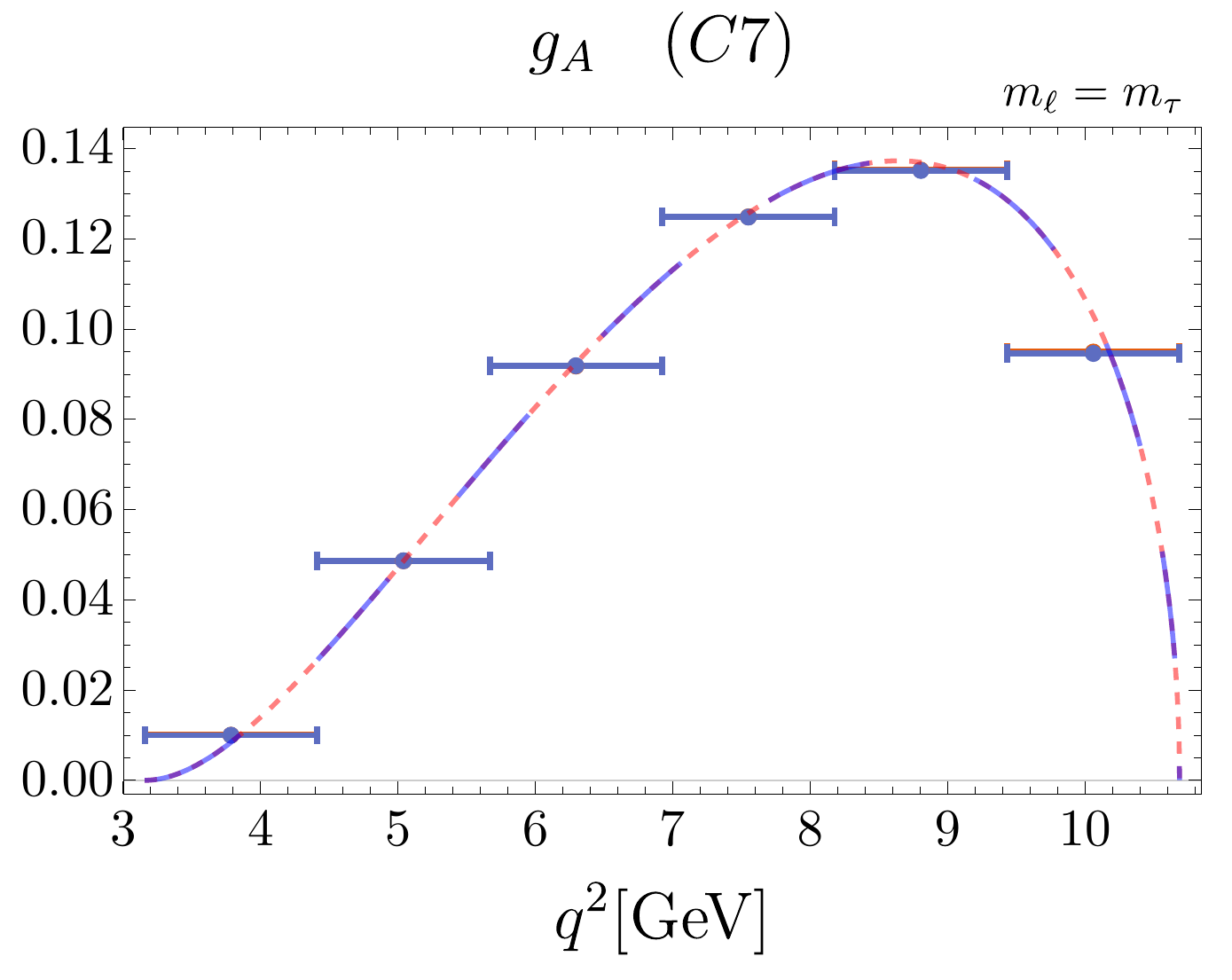}
  \hspace{10pt}
  \includegraphics[height=.27\linewidth]{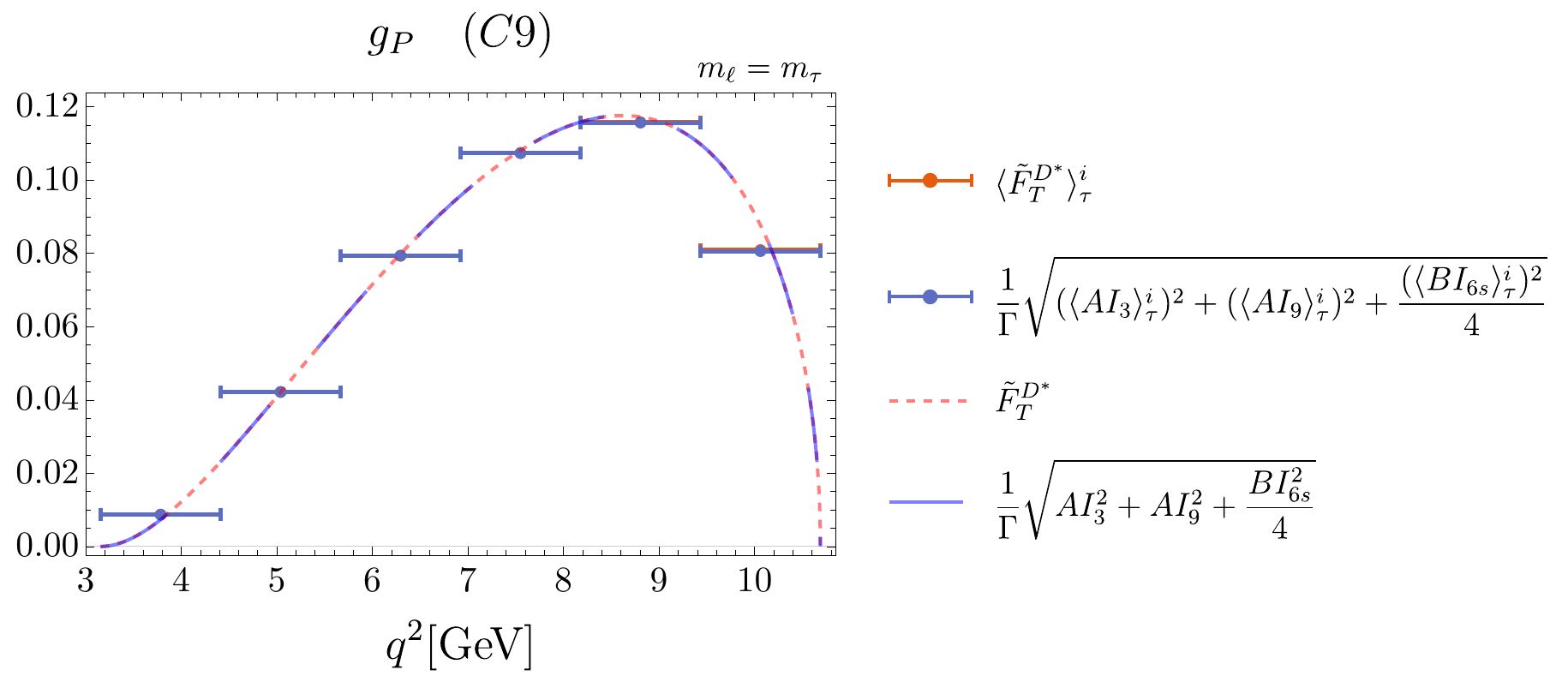}
 \end{subfigure}

 \caption{ Study of binning effects for Eq.~(\ref{eq:FLbinnedmassive}) for  benchmark NP scenarios with complex contributions.}
 \label{fig:appendix6}
\end{figure}

Following the setup of Sec.~\ref{sec:binning}, we illustrate in Fig.~\ref{fig:appendix1} to Fig.~\ref{fig:appendix6} the errors induced on the binning by the approximation Eq.~(\ref{errorbin}) on relations derived using the amplitude symmetries under various assumptions
on the NP scenario in the $\tau$ lepton case. We follow same convention as in Fig.~\ref{fig:binning}.

We provide the relative errors for selected scenarios in Tables~\ref{tab:numericalvalues1} and \ref{tab:numericalvalues2}.


\clearpage
\begin{sidewaystable}
 \centering
 \begin{tabular}{|c|c|l|l|l|l|l|l|l|}
  \hline

  \diagbox{Scenario}{Bin} & $ [3.2, 4.4]$ & $ [4.4, 5.7]$ & $ [5.7, 6.9]$ & $ [6.9, 8.2]$ & $ [8.2, 9.4]$ & $ [9.4, 10.7]$ & $[m_\tau^2,(m_B-m_{D^*})^2]$ \\
  \hline
  SM                      & 0.03\%        & 0.03\%        & 0.1\%         & 0.04\%        & 0.09\%        & 0.4\%          & 1\%                          \\
  \hline
  C1                      & 0.03\%        & 0.03\%        & 0.1\%         & 0.04\%        & 0.09\%        & 0.4\%          & 1\%                          \\
  \hline
  C0                      & 40\%          & 30\%          & 30\%          & 20\%          & 20\%          & 20\%           & 20\%                         \\
  \hline
  C3                      & 0.03\%        & 0.03\%        & 0.1\%         & 0.04\%        & 0.09\%        & 0.4\%          & 1\%                          \\
  \hline
 \end{tabular}
 \caption{Relative difference in percent of the approximate binned expression of $\langle\tilde{F}_T^{D*\, \rm alt}\rangle_\tau $ with respect to the ``standard'' $\langle\tilde{F}_T^{D*}\rangle_\tau$ for the SM and different NP scenarios. It corresponds to the relative difference in between the orange and blue bins displayed in Fig.~\ref{fig:binning} (normalised by the ``standard'' $\langle\tilde{F}_T^{D*}\rangle_\tau$ i.e. the orange bins).
 The bins in the first 6 columns correspond to the division of the kinematic range ($[m_\tau^2,(m_B-m_{D^*})^2]$) in 6 equally sized intervals. The last column corresponds to the whole kinematic range (not displayed in Fig.~\ref{fig:binning}). Notice that, as expected, this approximation works better for smaller bins. The scenario $C0$ is displayed as an example of a scenario with tensor contributions where, as expected, the two determinations should yield different results.}\label{tab:numericalvalues1}

 \vspace{2\baselineskip}

 \centering\begin{tabular}{|c|c|l|l|l|l|l|l|l|}
  \hline

  \diagbox{Scenario}{Bin} & $ [0, 1.8]$ & $ [1.8, 3.6]$ & $ [3.6,5.3]$ & $ [5.3,7.1]$ & $ [7.1,8.9]$ & $ [8.9,10.7]$ & $[0,(m_B-m_{D^*})^2]$ \\
  \hline
  SM                      & 0.08\%      & 0.04\%        & 0.1\%        & 0.05\%       & 0.1\%        & 0.4 \%        & 2\%                   \\
  \hline
 \end{tabular}
 \caption{Relative difference in percent of the approximate binned expression of $\langle\tilde{F}_T^{D*\, \rm alt}\rangle_0 $ with respect to the ``standard'' $\langle\tilde{F}_T^{D*}\rangle_0$ for the SM. It corresponds to the relative difference in between the orange and blue bins displayed in Fig.~\ref{fig:binning} (normalised by the ``standard'' $\langle\tilde{F}_T^{D*}\rangle_0$ i.e. the orange bins).
 The bins in the first 6 columns correspond to the division of the kinematic range ($[m_\tau^2,(m_B-m_{D^*})^2]$) in 6 equally sized intervals. The last column corresponds to the whole kinematic range (not displayed in Fig.~\ref{fig:binning}). Notice that, as expected, this approximation works better for smaller bins. The scenario $C0$ is displayed as an example of a scenario with tensor contributions where, as expected, the two determinations should yield different results.}
 \label{tab:numericalvalues2}
 \end{sidewaystable}

\end{document}